%% file: main.tex
\documentclass[a4paper,11pt]{article}
\usepackage{jinstpub} 
\usepackage[table]{xcolor}
\usepackage{lineno}
\usepackage{adjustbox}
\newcolumntype{P}[1]{>{\centering\arraybackslash}p{#1}}

\title{On the Feasibility of Future Colliders:\\ Report of the Snowmass'21 Implementation Task Force}

\author[1]{Thomas Roser,}
\author[2]{Reinhard Brinkmann,}
\author[3]{Sarah Cousineau,}
\author[1]{Dmitri Denisov,}
\author[4]{Spencer Gessner,}
\author[5,6]{Steve Gourlay,}
\author[7]{Philippe Lebrun,}
\author[8]{Meenakshi Narain,}
\author[9]{Katsunobu Oide,}
\author[4]{Tor Raubenheimer,}
\author[4]{John Seeman,} 
\author[6]{Vladimir Shiltsev,} 
\author[5,6]{Jim Strait,} 
\author[5]{Marlene Turner,}
\author[10]{Lian-Tao Wang.}

\affiliation[1]{Brookhaven National Laboratory, Upton, NY 11973, USA}
\affiliation[2]{DESY, 22607 Hamburg, Germany}
\affiliation[3]{Oak Ridge National Laboratory, Oak Ridge, TN 37830, USA}
\affiliation[4]{SLAC National Laboratory, Menlo Park, CA 94025, USA}
\affiliation[5]{Lawrence Berkeley National Laboratory, Berkeley, CA 94720, USA}
\affiliation[6]{Fermi National Accelerator Laboratory, Batavia, IL 60510, USA}
\affiliation[7]{ESI Archamps, 74160 Archamps, France}
\affiliation[8]{Brown University, Providence, RI, 02912, USA}
\affiliation[9]{KEK, Tsukuba, Ibaraki 305-0801, Japan}
\affiliation[10]{University of Chicago, Chicago, IL 60637, USA}

\emailAdd{roser@bnl.gov}

\abstract{Colliders are essential research tools for particle physics. Numerous future collider proposal were discussed in the course of the US high energy physics community strategic planning exercise {\it Snowmass'21}. The Implementation Task Force (ITF) has been established to evaluate the proposed future accelerator projects for performance, technology readiness, schedule, cost, and environmental impact. Corresponding metrics has been developed for uniform comparison of the proposals ranging from Higgs/EW factories to multi-TeV lepton, hadron and $ep$ collider facilities, based on traditional and advanced acceleration technologies. This article describes the metrics and approaches, and presents evaluations of future colliders performed by the ITF.}

\arxivnumber{2208.06030} 

\begin{document}
\maketitle
\flushbottom

\clearpage

\section{Introduction}
\label{sec:Intro}

While being arguably among the largest and the most complex scientific instruments, colliders are essential research tools for particle physics \cite{shiltsev2021modern}. Numerous future collider proposal were discussed in the course of the US high energy physics community strategic planning exercise {\it Snowmass'21} \cite{Snowmass21}. 
As part of the the Snowmass'21, the Accelerator Frontier (AF) group established an Implementation Task Force (ITF) to evaluate the proposed future accelerator projects for performance, technology readiness, schedule, cost, and environmental impact. Part of the work of ITF builds on the recently published report "European Strategy for Particle Physics - Accelerator R\&D Roadmap"\cite{Mounet:European_roadmap}.

One of the key goals of the Accelerator Frontier is to address the question {\it “…What are the time and cost scales of the R\&D and associated test facilities as well as the time and cost scale of the facility?”} \cite{Snowmass21AF}. A large number of accelerator projects are being considered and/or developed as part of the Snowmass'21 effort. One of the challenges for the AF topical groups is to compare the expected cost scales, schedule, and R\&D status for the projects in various stages of development and utilizing different accounting rules for costing. The collider Implementation Task Force is charged with developing metrics and processes to facilitate such a comparison between projects.    
More specifically, the Snowmass’21 ITF charge includes:
\begin{itemize}
\item Development of the metrics to compare projects’ cost, schedule/timeline, technical risks (readiness), operating cost and environmental impact, and R\&D status and plans; 
\item 
Select the accelerator projects to be evaluated; 
\item Work with the proponents of the selected accelerator projects to evaluate them against the metrics;
\item 
Consider the ultimate limits of various types of colliders: $e^+/e^-$, $p/p$, $\mu^+/\mu^-$;
\item 
Consider limits and timescales due to accelerator technology for various types of colliders: $e^+/e^-$, $p/p$, $\mu^+/\mu^-$; 
\item Lead the evaluation of the different HEP accelerator proposals and inform and communicate with the Snowmass’21 Frontiers - Accelerator, Energy, Neutrino, and Theory (AF, EF, NF and TF) -  in the course of the Snowmass'21 activities; 
\item Document the metrics, processes, and conclusions for the Snowmass'21 Community Summar Study (July, 2022); write and submit a corresponding report. 
\end{itemize}

To make the evaluations of the ITF most useful to the Snowmass'21 exercise it was decided on four categories of colliders that address similar physics:
\begin{itemize}
\item Higgs factory colliders
\item Lepton colliders with up to 3 TeV COM energy
\item Colliders with 10 TeV or higher parton COM energy
\item Lepton-hadron colliders
\end{itemize}

A separate group consists of versions of the proposals from these categories that could be located at Fermilab.


The ITF comparative evaluations are organized along four topics: 
\begin{itemize}
\item Physics Reach of Collider Proposals
\item Size, Complexity, and Impact on Environment
\item Technical Risk and Technical Readiness
\item Cost and Schedule
\end{itemize}
Each topic is covered in a section below.


\input{PhysicsRequirements}


\input{TechnicalReadiness}

\input{PowerComplexity}

\input{Cost}

\clearpage
\input{Summary}

\clearpage
\section*{Acknowledgements}

Deliberations and analysis of the Snowmass'21 Implementation Task Force have required various factual information and expertise opinions on numerous subjects related to the ITF charge. The Task Force approached many experts world-wide and got tons of very helpful input which is greatly appreciated. Our particular thanks go those colleagues who have shared their knowledge in written, often in documented form, including : on SC magnets - G.Apollinari, G.Ambrosio, R.Carcagno, S.Feher, and A.Zlobin (Fermilab); on NC magnets - T.Shaftan (BNL), H.Piekarz (Fermilab) and T.Zickler (CERN); on the US project accounting practices - J.Kerby (ANL); on high-power lasers - R.Assmann (DESY) and C.Schroeder, E.Esarey and C.Gededes (LBNL). 

While we were working on this article, our dear friend, colleague and one of the most active contributors to the Snowmass'21 in general and to this report in particular, Professor Meenakshi Narain (1964-2023) of Brown University has passed away. We will all miss her more than words can express and will long remember Meenakshi’s energy, tenacity and her many important contributions to our field. 

This work was supported by the Fermi National Accelerator Laboratory, managed and operated by Fermi Research Alliance, LLC under Contract No. DE-AC02-07CH11359 with the U.S. Department of Energy. The U.S. Government retains and the publisher, by accepting the article for publication, acknowledges that the U.S. Government retains a non-exclusive, paid-up, irrevocable, world-wide license to publish or reproduce the published form of this manuscript, or allow others to do so, for U.S. Government purposes.

\clearpage
\input{Appendices}

\clearpage
\newpage

\bibliographystyle{JHEP}
\bibliography{biblio.bib}

\end{document}

%% file: PhysicsRequirements.tex
\section{Energy and Luminosity Reach, and Achievable Science}
\label{sec:Physics}

\newcommand{\LTW}[1]{{\bf [LW: #1] } }

The ITF collected from all collider proponents the peak luminosity as a function of the center-of-mass (CM) energy. The plots below show this proponent-provided peak luminosity vs. CM energy for a single interaction point (IP). This is the peak luminosity of a fully commissioned facility. A number of proposals allow for multiple IPs and we list the proponent-provided total peak luminosity in the summary tables. All the luminosity vs CM energy plots also show on the right-hand scale the integrated luminosity for operation at peak luminosity for $10^7$ seconds or a "Snowmass" year.

All the plots show a single version per collider proposal and all the proponents were given the opportunity to select which version of their proposal should be plotted with the understanding that a more ambitious proposal would likely require more pre-project R\&D time and effort. This aspect was considered during the evaluation of the technical risk in section \ref{sec:TRL} on technical readiness and reflected in the "years of pre-project R\&D" column of the summary tables.

In this section, we also give a summary of the physics output of the various proposals for future lepton-lepton and hadron-hadron colliders, and the relationship with luminosity. A more detailed discussion can be found in a Snowmass whitepaper \cite{Liu:2022rua}. The lepton-hadron collider proposals are included in the summary table \ref{tab:ITFleptonhadron} but it is more difficult to compare their physics output with the equal-particle colliders.

\subsection{Higgs and electroweak physics colliders ($E_{\rm CM} \leq 1 $ TeV)}

\begin{figure}[h!]
    \centering
    \includegraphics[width=\textwidth]{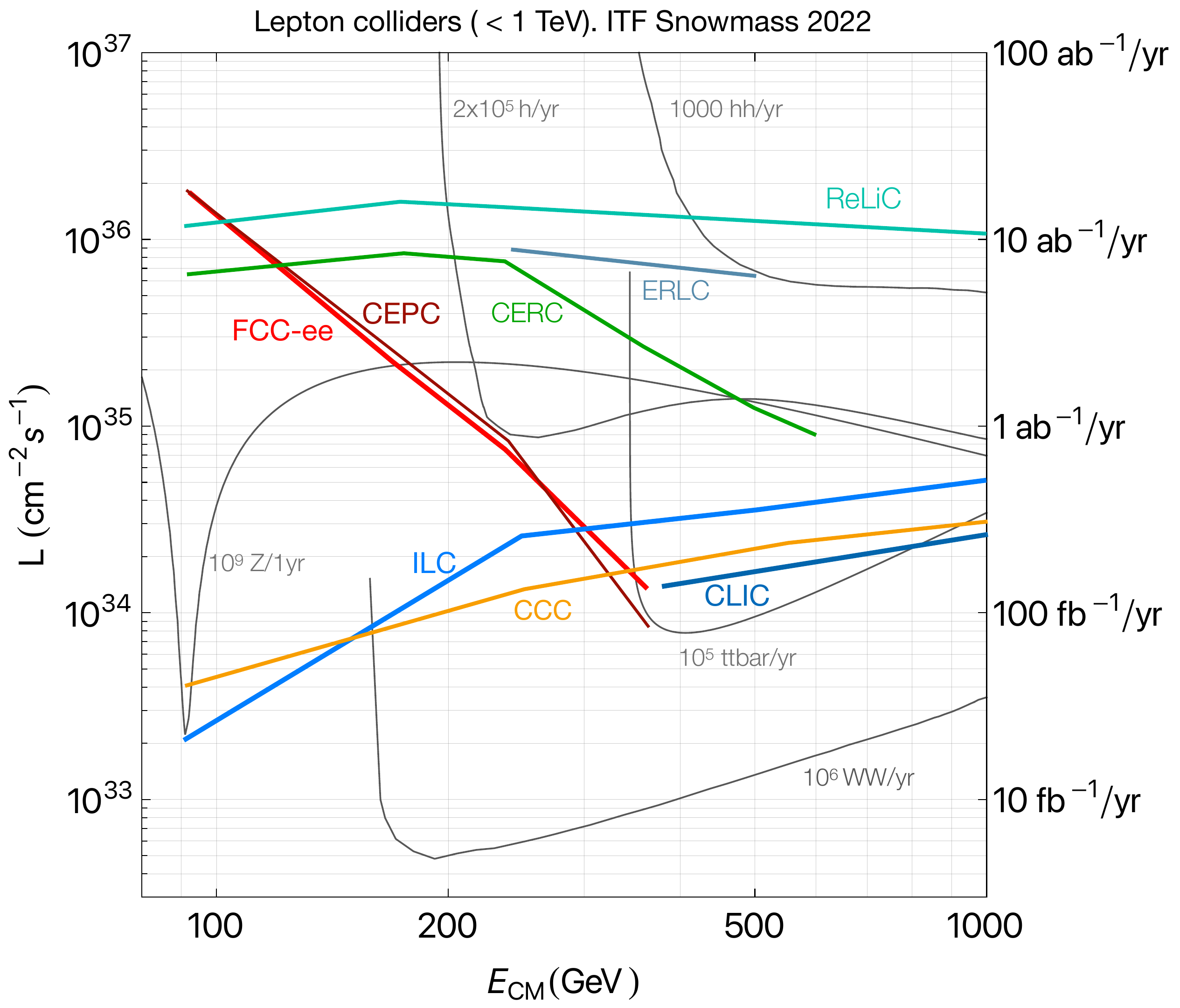}
    \caption{Peak luminosity per IP vs CM energy for the Higgs factory proposals as provided by the proponents. The right axis shows integrated luminosity for one Snowmass year (10$^7$ s). Also shown are lines corresponding to yearly production rates of important processes.}
    \label{fig:Higgscollidersplot}
\end{figure}

There have been many proposals of lepton colliders, with energies from 90 GeV - 1 TeV, including CEPC \cite{CEPC-SPPCStudyGroup:2015csa}\cite{CEPC},  CERC \cite{Litvinenko:CERC}, CLIC \cite{Linssen:2012hp}\cite{Brunner:CLIC}, ERLC \cite{ERL1}, FCC-ee \cite{FCC:2018byv}\cite{Agapov:FCC-ee}, ILC with upgraded luminosity \cite{Baer:2013cma}\cite{Aryshev:ILC}, ReLiC \cite{Litvinenko:ReLiC}, XCC\cite{Barklow:XCC}. Here, we describe the main physics output at such colliders and the corresponding luminosity. 

The main results are summarized in Fig.~\ref{fig:Higgscollidersplot}. We briefly summarize the main content in the following (see \cite{Liu:2022rua} for details). While the proposals all contain a set of specific running scenarios (energy and luminosity), keeping in mind possible updates and alternative plans in the future, we present our result for the full range of energies. 

One of the main goals of low energy lepton
colliders is to function as a Higgs factory and measure the Higgs couplings with unprecedented precision. HL-LHC can measure some of the Higgs couplings to an accuracy of a few percent.
Hence, a meaningful target for a Higgs factory would be to reach  per mil level.
Without taking into account detailed studies of specific channels, conservatively, $10^6$ Higgs boson would at least be
needed even to have a chance of measuring Higgs coupling (such as HZZ) to such a precision. Hence, we used $2 \times 10^5 {\rm \  Higgs}/{\rm yr}$ (corresponding to $10^6 {\rm \ Higgs }/{\rm 5 \ yrs}$). The dominant production mode depends on the center of mass energy. The total yield, including all production modes, is shown in Fig.~\ref{fig:Higgscollidersplot}. The required luminosity for different yields of Higgs bosons can be scaled in a straightforward way.  Another important measurement is the Higgs self-
coupling. Many studies have shown that a TeV lepton collider could extract Higgs self-coupling at around 10\% level. To set a target for the $HH$ process, we show the required luminosity for $5\times 10^3$ $HH$ within 5~years. We included the $ZHH$ associated production, VBF $HH$ production, and $t\bar t HH$ productions. 

Circular $e^+ e^-$ colliders offer the possibilities of a high statistics $Z$ factory. LEP-I produced about $10^7$ $Z$ bosons. To be significantly better, a new $Z$-factory would need to produce at least $10^9$ $Z$ bosons. We can also have a large number of $Z$s while running at energies above the Z pole via the so called radiative return process. In Fig.~\ref{fig:Higgscollidersplot}, we show the luminosity requirement to produce $10^9$ (nearly) on-shell $Z$ boson around $Z$-pole and through the ``radiative return" process. As can be seen from the figure, instead of Giga-Z, many of the proposals are aiming at producing $10^{12}$ Zs. In addition to better electroweak precision measurements, this will enable the Z-factory to serve as a powerful $b$ and $\tau$ factory, as well as offer the opportunity to probe a variety of interesting Z rare decay modes. 

The $W$ mass measurement is crucial for interpreting the electroweak precision observables at the $Z$-pole~\cite{Fan:2014vta,CEPCStudyGroup:2018ghi,FCC:2018evy}. $WW$ production near the threshold at lepton colliders is indispensible  for electroweak precision physics. While the measurement at hadron colliders will be limited at about $\delta m_{W} \sim 10 $ MeV, A $WW$ threshold scan could push the precision to the level of a few MeV. Many new physics proposals generate deviations in the $W$ boson couplings. Based on the studies of these physics goals, about $10^6$ $WW$s would be a good target for running close to the threshold.

Top mass is also a crucial input to the electroweak precision fit. The measurement at hadron colliders is limited to about $\delta m_t \sim 10^2$ MeV. Threshold scan with order $10^5$ $t \bar t$ pairs can push the precision of the top mass down to around 10~MeV~~\cite{Simon:2019axh}. The lepton colliders could probe the top gauge couplings and top EFT operators to a good precision through direct pair production and their angular correlations.

\newpage

\subsection{Energy frontier colliders}

\subsubsection{High energy lepton colliders ($E_{\rm CM} > 1 $ TeV)}
\begin{figure}[h!]
    \centering
    \includegraphics[width=\textwidth]{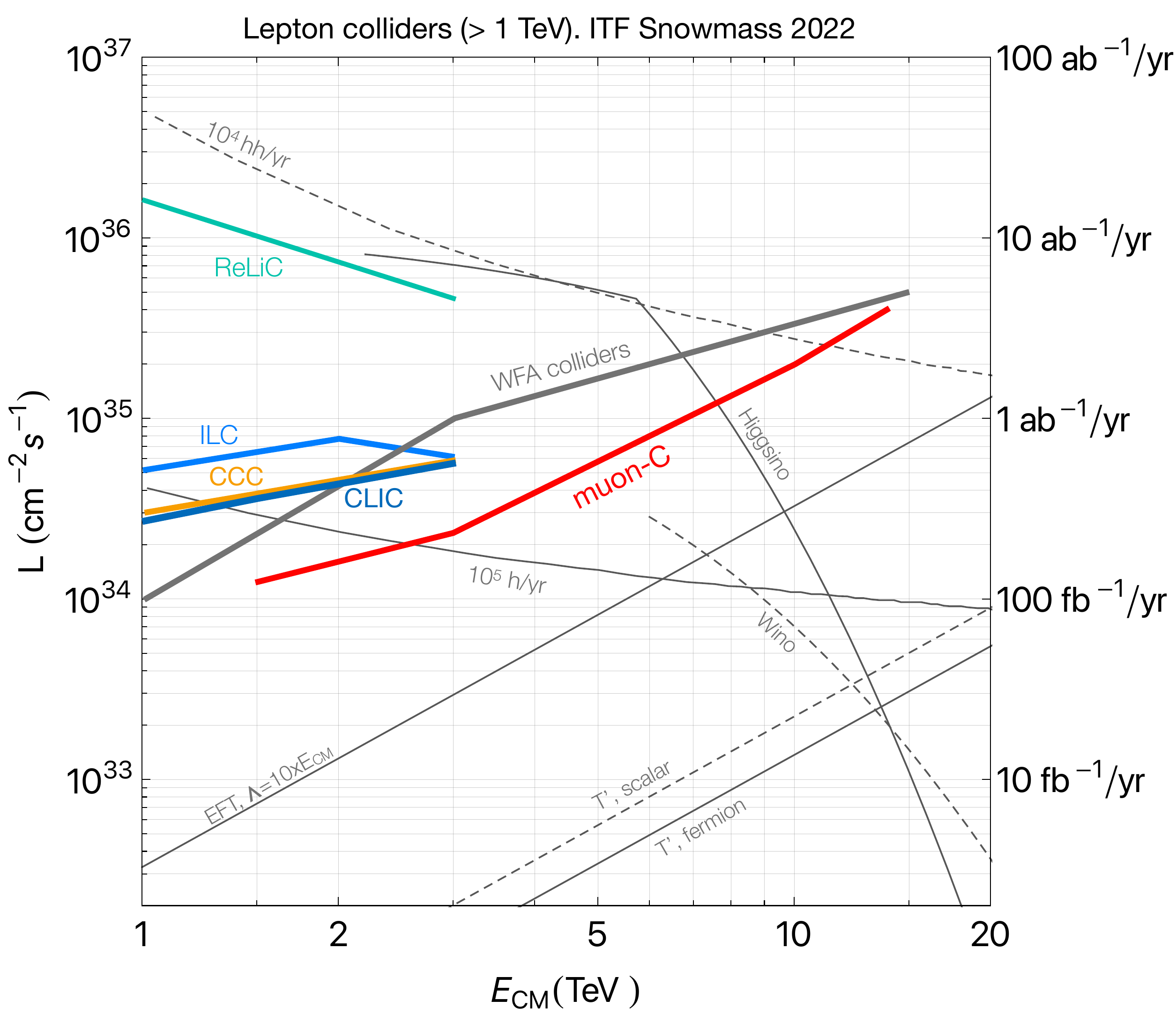}
    \caption{Peak luminosity per IP vs CM energy for the high energy lepton collider proposals as provided by the proponents. The right axis shows integrated luminosity for one Snowmass year (10$^7$ s). Also shown are lines corresponding to yearly production rates of important processes. 
    The luminosity requirement for 5$\sigma$ discovery of the benchmark DM scenarios Higgsino and Wino are also shown, see Refs.\cite{Han:2020uak,Han:2022ubw}} 
    \label{fig:HEleptoncollidersplot}
\end{figure}

In this section, we focus on high energy lepton colliders with $E_{\rm CM}$ in the range of 1 - 20 TeV. Proposals in this range include CLIC\cite{Brunner:CLIC}, CCC\cite{Dasu:CCC}, ILC\cite{Aryshev:ILC}, Muon Collider\cite{Stratakis:MC}, ReLiC\cite{Litvinenko:ReLiC}, and Wake Field Accelerators\cite{Benedetti:LWFA}\cite{PWFA}\cite{Jing:SWFA}. It was also proposed to use high power Free Electron Lasers to produce a second interaction region with high-energy and high luminosity gamma-gamma collisions at a high-energy electron-positron collider \cite{Barzi:gamma-gamma_collider}. Here, the primary goal would be searching for heavy new physics resonances. At the same time, high energy lepton colliders can contribute to the measurement of the Higgs coupling, such as Higgs precision coupling measurements, top Yukawa coupling, and Higgs self-coupling. 
Since $e^+ e^-$ and $\mu^- \mu^+$ colliders have very similar reaches in this range of energies,  we do not distinguish between them. The summary of our results are shown in Fig~\ref{fig:HEleptoncollidersplot}. In the following, we will  discuss briefly  the additional physics cases and considerations beyond those of the lower energy lepton colliders discussed in the previous section (for details, see Snowmass whitepaper \cite{Liu:2022rua}). 

The Higgs boson precision program is an essential component of a high-energy lepton collider. Similar to the low energy case, we show here the luminosity needed to produce $10^6$ Higgs particles for a 10 year running period. At the same time, higher energy is more optimal for double Higgs production, and better measurement for the Higgs self-coupling. Hence, we choose to plot a higher benchmark with $10^5$ Higgs particles for the same running period.

Addressing the hierarchy problem is a leading physics driver for future colliders. Among the new physics particles associated with the hierarchy problem, the top partner is probably the most important one due to the significant role the top quark played in the dynamics of the electroweak symmetry breaking. Pair produced through Drell Yan processes,  high energy lepton colliders will have excellent reach for top partners even very close to the kinematical threshold ($2 m_{T'} = 0.9 \times E_{\rm CM}$). one should be able to discover them.  With this in mind, we show the luminosity needed to reach a  statistics of 20 signal events, enough to discover them,  for a generic scalar or fermionic top partners  with dashed and solid lines in the figure. 
 
Testing the WIMP (Weakly Interacting Massive Particle) paradigm of dark matter is another main physics driver for future colliders. Among various possible candidates, the minimal model would be dark matter as a member of an electroweak multiplet. They can be produced copiously at high energy lepton colliders. At the same time, the signal is much more challenging to detect, and requires detailed simulation~\cite{Han:2020uak,Han:2022ubw,Capdevilla:2021fmj,Bottaro:2021snn,Bottaro:2021srh}. In Fig~\ref{fig:HEleptoncollidersplot}, we present two representative cases,  fermionic electroweak doublet (Higgsino) and triplet (Wino). The thermal relic abundance of the dark matter in the universe requires the masses to be 1.1~TeV (Higgsino) and 2.8~TeV (wino). Hence, the respective curves start at $E_{\rm CM} \sim 2 m_{\rm DM}$. We see that high energy colliders, with $E_{\rm CM} > 5 (6)$ TeV,  are required to cover the Higgsino (Wino) cases. 

Direct production of new physics particles could still be beyond the reach of the colliders. In this case, their effect can be encapsulated in higher dimensional (EFT) operators, for example, of the form ${\mathcal O}/\Lambda^2 $, where $\Lambda$ approximately corresponds to the mass scale of the new physics. High energy lepton colliders can perform good precision measurement to probe this new physics, and reach a scale above its center of mass energy. Since such effects of new physics grow with energy, we expect the reach to grow with the center of mass energy of the collider. In this figure, we show the required luminosity in order to reach a new physics scale 10 times the center of mass energy. 


\clearpage
\subsubsection{High energy hadron colliders}
\begin{figure}[h!]
    \centering
    \includegraphics[width=\textwidth]{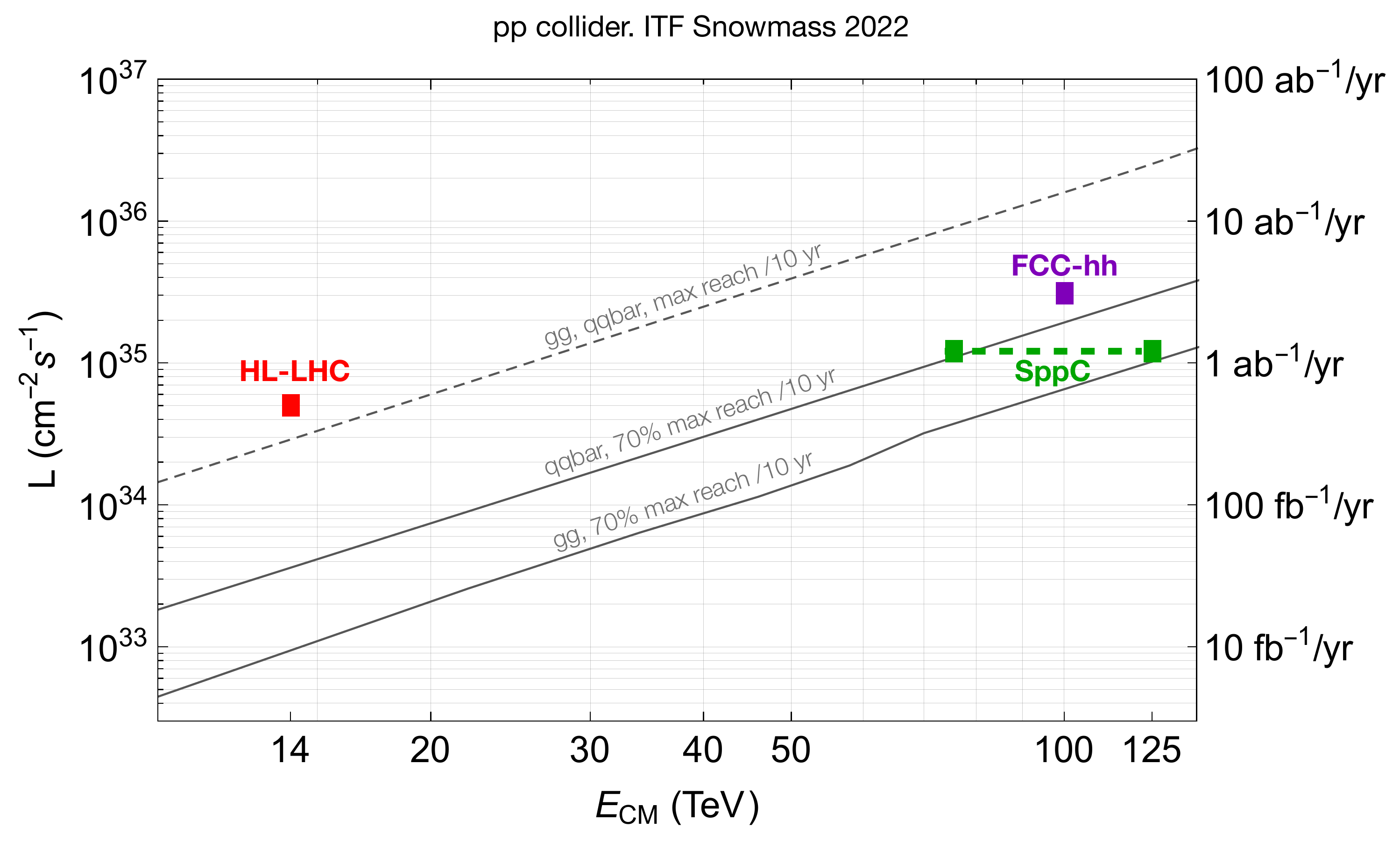}
    \caption{Peak luminosity per IP vs CM energy for the high energy hadron collider proposals as provided by the proponents. The right axis shows integrated luminosity for one Snowmass year (10$^7$ s).
    Also shown are the luminosity requirements with two possible initial states gg and $q \bar q$. The dashed curve represents the luminosity needed (assuming a 10 year run) to have linear increase of new physics mass reach with CM energy. 
    The solid lines represent the luminosity requirements for 70\% of this new physics mass reach.  }
    \label{fig:HEppcollidersplot}
\end{figure}

Two proposals of hadron colliders, submitted to ITF, are covered here: FCC-hh \cite{Benedikt:FCC-hh} and SPPC \cite{Tang:SPPC}. It is much more challenging to make projections for the hadron colliders without detailed machine and detector design, especially for searches dominated by systematics. At the same time, it is possible to make some rough estimates for searches based on the behavior of parton luminosity and statistics \cite{parton_scaling}. Our result is presented in \autoref{fig:HEppcollidersplot}. It shows the luminosity requirements for high energy future hadron colliders, with two possible initial states (gg, red; $q \bar q$ blue). 
The scaling of the reach shown here is done by statistics and using parton luminosity~\cite{parton_scaling}. We used the reach at the HL-LHC as a reference point for the extrapolation. For the gluon-gluon initial state dominated process, we assumed the reach of the mass of new physics at the HL-LHC of 3 TeV (approximately 1.5 TeV for pair production). This could be similar to the case, for example, of the stop. For $q \bar{q}$ initiated processes, we assumed a reach at the HL-LHC of 1 TeV (which would be about 500 GeV for pair production). This would be similar to the electroweak states. Changing the assumption of HL-LHC reach will give rise to some differences but not affect the qualitative feature.

In the following, we make a couple of observations (see Ref.~\cite{Liu:2022rua} for a more detailed discussion).
\begin{itemize}
    \item Given two colliders with the center of mass energies $E_1$ and $E_2$, the corresponding reach of the masses of a particular new physics particle are denoted as $M_1$ and $M_2$, respectively.
    In the ideal case,  the reach of new physics scales linearly with the CM energy, $M_2 /M_1 = E_2 / E_1$. With a mild assumption on the production rate scaling, it is straightforward to see we need ${\cal L}_2 / {\cal L}_1 = E^{2}_2 / E^2_1$, which  can be seen in the dashed curves in \autoref{fig:HEppcollidersplot}. This is similar to the scaling for the lepton collider, and it could be a large step in luminosity increase for a large increase of the CM energy.  
    \item Due to the nature of the parton luminosity as a function of parton CM energy, there is a rapid gain in reach for a relatively small amount of luminosity. Hence, to achieve a somewhat lower goal for the mass reach enhancement, for example, $70 \%$ of $E_2/E_1$, one needs a significantly smaller amount of data, shown as the solid lines in \autoref{fig:HEppcollidersplot}. 
    \item The parton luminosity for the gluon-gluon initial state falls as a function of the parton CM energy faster than the $q \bar{q}$ initial state. As a result, the luminosity needed for obtaining a significant fraction of the maximal reach in the gluon-gluon initial state-dominated processes is less than that of processes dominated by the $q\bar{q}$ processes. This is also shown in \autoref{fig:HEppcollidersplot}.
\end{itemize}
\newpage

%% file: TechnicalReadiness.tex
\section{Technical Readiness of Collider Proposals} 
\label{sec:TRL}


\subsection{General approach, TRL levels}

The ITF has developed metrics for a high-level comparison of the technical risks of key components necessary for implementing the proposed facility. Each proponent was given a spreadsheet template and asked to provide three to five critical enabling technologies (those representing the highest technical risk), then numerically evaluate them in each of five risk categories according to a prescribed scoring key. In cases where information was incomplete or missing, the ITF referred to contributed papers or applied expert judgement. The proposed projects represent a very broad range of maturity, from the CDR/TDR level to parameter lists. It is not unexpected that the less mature proposals have more high impact risks relative to those that are more mature. To provide a more equitable comparison, the Task Force expanded the list of technologies to 5 for each proposal. The five risk categories that were used in the comparison and the scoring key are discussed below.

\subsubsection{Collider component and subsystems technical risk factor based on the current Technical Readiness Level (TRL)}
\label{TRLsubsec1}

A brief description of the TRL definitions is given below (more detail  definitions used for the evaluation can be found in Appendix \ref{TRLAppendix}): 

\begin{itemize}
    \item TRL1: Basic principles observed and reported
\item TRL2: Technology concept and/or application formulated

\item TRL3: Analytical and experimental critical function and/or characteristic proof of concept.
\item TRL4: Component and/or breadboard validation in laboratory environment.
\item TRL5: Component and/or breadboard validation in relevant environment.
\item TRL6: System/subsystem model or prototype demonstration in a relevant environment. 
\item TRL7: System prototype demonstration in an operational environment.
\item TRL8: Actual system completed and qualified through test and demonstration.
\item TRL9: Actual system has proven through successful mission operations.
\end{itemize}

\begin{table}[h!]
\centering
\begin{tabular}{|p{8cm}|p{1cm}|p{2cm}|}
\hline 
\hline
{\bf Technical Risk Factor} & {\bf Score} & {\bf Color Code}
\\
TRL = 1,2 & 4 & \cellcolor{blue!100}
\\
TRL = 3,4 & 3 & \cellcolor{blue!75}\\
TRL = 5,6 & 2 & \cellcolor{blue!50}\\
TRL = 7,8 & 1 &\cellcolor{blue!25}\\
\hline
\end{tabular}
\caption{ TRL scoring chart and color codes (used below in the summary Table \ref{tab:subgroupTRL}).}
\label{TRLcolors}
\end{table}

The two Tables, \ref{tab:TRLVS1} and \ref{tab:TRLVS2} below, list key enabling technologies for the ITF collider proposals with the colors indicating the present day TRLs (lighter to darker meaning lower to higher risk - see Table \ref{TRLcolors}). Table \ref{tab:TRLVS1} lists the $e^+e^-$ and $ep$ colliders, Table \ref{tab:TRLVS2} lists the future very high energy $pp$, muon and advanced $e^+e^-$ colliders. Note, that in these tables, a facility may have more than five technologies but only the five most critical ones were used in generating the overall risk score and ranking defined later and summarized in Table \ref{tab:subgroupTRL}. 

\pagebreak
\begin{table}[htbp]
  \centering
  \caption{Technical risk registry of accelerator components and systems for future $e^+e^-$ and $ep$ colliders: lighter colors indicate progressively higher TRLs (less risk), white is for either not significant or not applicable.} 
  
  \hspace{0.1em}

    \begin{tabular}{|l|c|cccccc|ccccc|c|}
    \hline
          & {\rotatebox{90}{FCCee/CEPC}} & {\rotatebox{90}{ILC}} & {\rotatebox{90}{HE ILC}} & {\rotatebox{90}{CCC}} & {\rotatebox{90}{HE CCC}} & {\rotatebox{90}{CLIC}} & {\rotatebox{90}{HE CLIC}} & {\rotatebox{90}{CERC}} & {\rotatebox{90}{ReLiC}} & {\rotatebox{90}{HE ReLiC}} & 
          {\rotatebox{90}{ERLC}} & {\rotatebox{90}{XCC}} & {\rotatebox{90}{LHeC/FCCeh}}
          \\
    \hline
    \hline
    RF Systems & \cellcolor{blue!50}   & \cellcolor{blue!25}    & \cellcolor{blue!100}  &  \cellcolor{blue!50}    & \cellcolor{blue!75}  &   \cellcolor{blue!50}  & \cellcolor{blue!50}  & \cellcolor{blue!50}  & 
    \cellcolor{blue!75} & \cellcolor{blue!75} & \cellcolor{blue!75} & \cellcolor{blue!25} & \cellcolor{blue!25} \\
    Cryomodules & \cellcolor{blue!25}   & \cellcolor{blue!25}    & \cellcolor{blue!25}  &  \cellcolor{blue!75}    & \cellcolor{blue!75}  &   \cellcolor{blue!0}  & \cellcolor{blue!0}  & \cellcolor{blue!25}  & 
    \cellcolor{blue!25} & \cellcolor{blue!25} & \cellcolor{blue!50} & \cellcolor{blue!25} & \cellcolor{blue!25} \\
    HOM detuning/damp & \cellcolor{blue!0}   & \cellcolor{blue!25}    & \cellcolor{blue!25}  &  \cellcolor{blue!75}    & \cellcolor{blue!75}  &   \cellcolor{blue!25}  & \cellcolor{blue!25}  & \cellcolor{blue!50}  & 
    \cellcolor{blue!50} & \cellcolor{blue!75} & \cellcolor{blue!75} & \cellcolor{blue!25} & \cellcolor{blue!75} \\
    High energy ERL & \cellcolor{blue!0}   & \cellcolor{blue!0}    & \cellcolor{blue!0}  &  \cellcolor{blue!0}    & \cellcolor{blue!0}  &   \cellcolor{blue!0}  & \cellcolor{blue!0}  & \cellcolor{blue!75}  & 
    \cellcolor{blue!0} & \cellcolor{blue!0} & \cellcolor{blue!0} & \cellcolor{blue!0} & \cellcolor{blue!75} \\
    Positron source & \cellcolor{blue!50}   & \cellcolor{blue!50}    & \cellcolor{blue!50}  &  \cellcolor{blue!25}    & \cellcolor{blue!25}  &   \cellcolor{blue!25}  & \cellcolor{blue!25}  & \cellcolor{blue!25}  & 
    \cellcolor{blue!25} & \cellcolor{blue!25} & \cellcolor{blue!25} & \cellcolor{blue!0} & \cellcolor{blue!0} \\
    Arc\&booster magnets & \cellcolor{blue!50}   & \cellcolor{blue!0}    & \cellcolor{blue!0}  &  \cellcolor{blue!0}    & \cellcolor{blue!0}  &   \cellcolor{blue!0}  & \cellcolor{blue!0}  & \cellcolor{blue!50}  & 
    \cellcolor{blue!0} & \cellcolor{blue!0} & \cellcolor{blue!0} & \cellcolor{blue!0} & \cellcolor{blue!25} \\
    Inj./extr. kickers & \cellcolor{blue!0}   & \cellcolor{blue!25}    & \cellcolor{blue!25}  &  \cellcolor{blue!0}    & \cellcolor{blue!0}  &   \cellcolor{blue!0}  & \cellcolor{blue!0}  & \cellcolor{blue!0}  & 
    \cellcolor{blue!75} & \cellcolor{blue!75} & \cellcolor{blue!0} & \cellcolor{blue!0} & \cellcolor{blue!0} \\
   Two-beam acceleration & \cellcolor{blue!0}   & \cellcolor{blue!0}    & \cellcolor{blue!0}  &  \cellcolor{blue!0}    & \cellcolor{blue!0}  &   \cellcolor{blue!50}  & \cellcolor{blue!75}  & \cellcolor{blue!0}  & 
    \cellcolor{blue!0} & \cellcolor{blue!0} & \cellcolor{blue!0} & \cellcolor{blue!0} & \cellcolor{blue!0} \\
    Damping rings & \cellcolor{blue!0}   & \cellcolor{blue!25}    & \cellcolor{blue!25}  &  \cellcolor{blue!25}    & \cellcolor{blue!25}  &   \cellcolor{blue!25}  & \cellcolor{blue!25}  & \cellcolor{blue!0}  & 
    \cellcolor{blue!75} & \cellcolor{blue!75} & \cellcolor{blue!0} & \cellcolor{blue!00} & \cellcolor{blue!0} \\
    Emitt. preservation & \cellcolor{blue!0}  & \cellcolor{blue!25}    & \cellcolor{blue!50}  &  \cellcolor{blue!50}    & \cellcolor{blue!75}  &   \cellcolor{blue!50}  & \cellcolor{blue!75}  & \cellcolor{blue!75}  & 
    \cellcolor{blue!50} & \cellcolor{blue!75} & \cellcolor{blue!50} & \cellcolor{blue!50} & \cellcolor{blue!0} \\
    IP spot size/stability & \cellcolor{blue!0}  & \cellcolor{blue!25}    & \cellcolor{blue!50}  &  \cellcolor{blue!50}    & \cellcolor{blue!75}  &   \cellcolor{blue!50}  & \cellcolor{blue!75}  & \cellcolor{blue!50}  & 
    \cellcolor{blue!25} & \cellcolor{blue!50} & \cellcolor{blue!25} & \cellcolor{blue!50} & \cellcolor{blue!25} \\
    High power XFEL & \cellcolor{blue!0}   & \cellcolor{blue!0}    & \cellcolor{blue!0}  &  \cellcolor{blue!0}    & \cellcolor{blue!0}  &   \cellcolor{blue!0}  & \cellcolor{blue!00}  & \cellcolor{blue!0}  & 
    \cellcolor{blue!0} & \cellcolor{blue!0} & \cellcolor{blue!0} & \cellcolor{blue!75} & \cellcolor{blue!0} \\
    $e^-$ bunch compression & \cellcolor{blue!0}   & \cellcolor{blue!0}    & \cellcolor{blue!0}  &  \cellcolor{blue!0}    & \cellcolor{blue!0}  &   \cellcolor{blue!0}  & \cellcolor{blue!00}  & \cellcolor{blue!0}  & 
    \cellcolor{blue!0} & \cellcolor{blue!0} & \cellcolor{blue!0} & \cellcolor{blue!75} & \cellcolor{blue!0} \\
    High brightness $e^-$ gun & \cellcolor{blue!0}   & \cellcolor{blue!0}    & \cellcolor{blue!0}  &  \cellcolor{blue!0}    & \cellcolor{blue!0}  &   \cellcolor{blue!0}  & \cellcolor{blue!00}  & \cellcolor{blue!0}  & 
    \cellcolor{blue!0} & \cellcolor{blue!0} & \cellcolor{blue!0} & \cellcolor{blue!75} & \cellcolor{blue!50} \\
    IR SR and asymm.quads & \cellcolor{blue!0}   & \cellcolor{blue!0}    & \cellcolor{blue!0}  &  \cellcolor{blue!0}    & \cellcolor{blue!0}  &   \cellcolor{blue!0}  & \cellcolor{blue!00}  & \cellcolor{blue!0}  & 
    \cellcolor{blue!0} & \cellcolor{blue!0} & \cellcolor{blue!0} & \cellcolor{blue!0} & \cellcolor{blue!50} \\
        \hline
    \end{tabular}%
  \label{tab:TRLVS1}%
\end{table}%

\pagebreak

\begin{table}[htbp]
  \centering
  \caption{Technical risk registry of accelerator components and systems for future very high energy $pp$, muon and advanced $e^+e^-$ colliders: lighter colors indicate progressively higher TRLs (less risk), white is for either not significant or not applicable.} 
  
  \hspace{0 pt}
  
    \begin{tabular}{|l|ccc|ccc|ccc|}
    \hline
          & {\rotatebox{90}{FCChh}} & {\rotatebox{90}{SPPC}} & {\rotatebox{90}{Coll.Sea}} & {\rotatebox{90}{MC-0.125}} & {\rotatebox{90}{MC-3-6}} & {\rotatebox{90}{MC-10-14}} & {\rotatebox{90}{LWFA-LC}} & {\rotatebox{90}{PWFA-LC}} & {\rotatebox{90}{SWFA-LC}} 
          \\
    \hline
    \hline
    RF Systems & \cellcolor{blue!0}   & \cellcolor{blue!0}    & \cellcolor{blue!0}  &  \cellcolor{blue!25}    & \cellcolor{blue!25}  &   \cellcolor{blue!25}  & \cellcolor{blue!00}  & \cellcolor{blue!75}  & 
    \cellcolor{blue!100}  \\
    High field magnets & \cellcolor{blue!100}   & \cellcolor{blue!75}    & \cellcolor{blue!25}  &  \cellcolor{blue!50}    & \cellcolor{blue!75}  &   \cellcolor{blue!75}  & \cellcolor{blue!0}  & \cellcolor{blue!0}  & 
    \cellcolor{blue!0} \\
    Fast booster magnets/PSs& \cellcolor{blue!00}   & \cellcolor{blue!00}    & \cellcolor{blue!0}  &  \cellcolor{blue!25}    & \cellcolor{blue!50}  &   \cellcolor{blue!75}  & \cellcolor{blue!0}  & \cellcolor{blue!50}  & 
    \cellcolor{blue!0} \\
    High power lasers & \cellcolor{blue!0}   & \cellcolor{blue!0}    & \cellcolor{blue!0}  &  \cellcolor{blue!0}    & \cellcolor{blue!0}  &   \cellcolor{blue!0}  & \cellcolor{blue!100}  & \cellcolor{blue!0}  & 
     \\
    Integration and control & \cellcolor{blue!00}   & \cellcolor{blue!0}    & \cellcolor{blue!100}  &  \cellcolor{blue!00}    & \cellcolor{blue!0}  &   \cellcolor{blue!0}  & \cellcolor{blue!50}  & \cellcolor{blue!50}  & 
    \cellcolor{blue!75} \\
    Positron source & \cellcolor{blue!0}   & \cellcolor{blue!0}    & \cellcolor{blue!0}  &  \cellcolor{blue!0}    & \cellcolor{blue!0}  &   \cellcolor{blue!0}  & \cellcolor{blue!75}  & \cellcolor{blue!75}  & 
    \cellcolor{blue!75} \\
    6D $\mu$-cooling elements & \cellcolor{blue!0}   & \cellcolor{blue!0}    & \cellcolor{blue!0}  &  \cellcolor{blue!50}    & \cellcolor{blue!75}  &   \cellcolor{blue!75}  & \cellcolor{blue!0}  & \cellcolor{blue!0}  & 
    \cellcolor{blue!0}  \\
    Inj./extr. kickers & \cellcolor{blue!75}   & \cellcolor{blue!75}    & \cellcolor{blue!75}  &  \cellcolor{blue!0}    & \cellcolor{blue!0}  &   \cellcolor{blue!0}  & \cellcolor{blue!0}  & \cellcolor{blue!0}  & 
    \cellcolor{blue!0} \\ 
    Two-beam acceleration & \cellcolor{blue!0}   & \cellcolor{blue!0}    & \cellcolor{blue!0}  &  \cellcolor{blue!0}    & \cellcolor{blue!0}  &   \cellcolor{blue!0}  & \cellcolor{blue!0}  & \cellcolor{blue!75}  & 
    \cellcolor{blue!75} \\
    $e^+$ plasma acceleration & \cellcolor{blue!0}   & \cellcolor{blue!0}    & \cellcolor{blue!}  &  \cellcolor{blue!0}    & \cellcolor{blue!0}  &   \cellcolor{blue!0}  & \cellcolor{blue!100}  & \cellcolor{blue!100}  & 
    \cellcolor{blue!0} \\
    Emitt. preservation & \cellcolor{blue!0}  & \cellcolor{blue!0}    & \cellcolor{blue!25}  &  \cellcolor{blue!25}    & \cellcolor{blue!25}  &   \cellcolor{blue!25}  & \cellcolor{blue!100}  & \cellcolor{blue!100}  & 
    \cellcolor{blue!100} \\
    FF/IP spot size/stability & \cellcolor{blue!0}  & \cellcolor{blue!0}    & \cellcolor{blue!0}  &  \cellcolor{blue!25}    & \cellcolor{blue!25}  &   \cellcolor{blue!75}  & \cellcolor{blue!100}  & \cellcolor{blue!100}  & 
    \cellcolor{blue!100} \\
    High energy ERL & \cellcolor{blue!0}   & \cellcolor{blue!0}    & \cellcolor{blue!0}  &  \cellcolor{blue!0}    & \cellcolor{blue!0}  &   \cellcolor{blue!0}  & \cellcolor{blue!0}  & \cellcolor{blue!75}  & 
    \cellcolor{blue!0} \\
    Inj./extr. kickers & \cellcolor{blue!0}   & \cellcolor{blue!25}    & \cellcolor{blue!25}  &  \cellcolor{blue!0}    & \cellcolor{blue!0}  &   \cellcolor{blue!0}  & \cellcolor{blue!0}  & \cellcolor{blue!0}  & 
    \cellcolor{blue!75} \\
    High power target & \cellcolor{blue!0}   & \cellcolor{blue!0}    & \cellcolor{blue!0}  &  \cellcolor{blue!75}    & \cellcolor{blue!75}  &   \cellcolor{blue!50}  & \cellcolor{blue!00}  & \cellcolor{blue!0}  & 
    \cellcolor{blue!0}  \\
    Proton Driver & \cellcolor{blue!0}   & \cellcolor{blue!0}    & \cellcolor{blue!0}  &  \cellcolor{blue!25}    & \cellcolor{blue!25}  &   \cellcolor{blue!25}  & \cellcolor{blue!00}  & \cellcolor{blue!0}  & 
    \cellcolor{blue!0}  \\
    Beam screen & \cellcolor{blue!50}   & \cellcolor{blue!50}    & \cellcolor{blue!50}  &  \cellcolor{blue!0}    & \cellcolor{blue!0}  &   \cellcolor{blue!0}  & \cellcolor{blue!00}  & \cellcolor{blue!0}  & 
    \cellcolor{blue!0}  \\
    Collimation system & \cellcolor{blue!75}   & \cellcolor{blue!75}    & \cellcolor{blue!75}  &  \cellcolor{blue!25}    & \cellcolor{blue!25}  &   \cellcolor{blue!50}  & \cellcolor{blue!50}  & \cellcolor{blue!50}  & 
    \cellcolor{blue!50}  \\
    Power eff.\& consumption & \cellcolor{blue!75}   & \cellcolor{blue!75}    & \cellcolor{blue!100}  &  \cellcolor{blue!25}    & \cellcolor{blue!25}  &   \cellcolor{blue!50}  & \cellcolor{blue!100}  & \cellcolor{blue!75}  & 
    \cellcolor{blue!75}  \\
        \hline
    \end{tabular}%
  \label{tab:TRLVS2}%
\end{table}%


\subsubsection{Technology validation requirement}

This metric was used to indicate the level of effort required to validate the technology. For some technologies, validation can be established by a single component, while others require a full-scale demonstration. See Table \ref{tab:TRLSG1}

\begin{table}[h!]
\centering
\begin{tabular}{|p{8cm}|p{1cm}|p{2cm}|}
\hline
{\bf Technology Validation Required} & {\bf Score} & {\bf Color Code} \\
Full-scale - requires comprehensive demonstration & 3 & \cellcolor{blue!75} \\
Partial with scaling - partial demonstration sufficient & 2 &  \cellcolor{blue!50}\\
Separate - component validation & 1 & \cellcolor{blue!25}\\
\hline

\end{tabular}
\caption{ Technology validation scoring chart and color codes (used below in summary Table \ref{tab:subgroupTRL}).}
\label{tab:TRLSG1}%
\end{table}


\pagebreak

\subsubsection{Cost reduction impact}

In several cases a proposed technology is a significant cost driver, for example superconducting RF and magnets. The evaluations were made on the current status of the technology and indicates the potential impact of cost reduction.The scoring key and associated color code is shown in Table \ref{tab:TRLSG2}

\begin{table}[h!]
\centering
\begin{tabular}{|p{8cm}|p{1cm}|p{2cm}|}
\hline 
{\bf Cost Reduction Impact} & {\bf Score} & {\bf Color Code} \\
Critical - a "no-go" without significant cost reduction & 3 & \cellcolor{blue!75} \\
Significant Impact & 2 & \cellcolor{blue!50}\\
Desirable & 1 & \cellcolor{blue!25}\\
\hline
\end{tabular}
\caption{Technology cost reduction scoring chart and color codes (used below in the summary Table \ref{tab:subgroupTRL}).}
\label{tab:TRLSG2}%
\end{table}

%
%

\subsubsection{Evaluation of performance achievability.}

This metric indicates the ITF judgement on the extent of the effort needed to close the technology gap for a component or subsystem to demonstrate performance achievability. It is correlated with the Technical Risk Level and indicates the risk associated with increasing the TRL.The scoring key and associated color code is shown in Table \ref{tab:TRLSG3}

\begin{table}[h!]
\centering
\begin{tabular}{|p{11cm}|p{1cm}|p{2cm}|}
\hline 
{\bf Performance Achievability} & {\bf Score} & {\bf Color Code} \\
Significant - needs explicit demo of beyond state-of-the-art & 3 & \cellcolor{blue!75}\\
Moderate - Feasible to achieve 2 - 3X state-of-the-art & 2 & \cellcolor{blue!50}\\
Feasible - at state-of-the-art & 1 & \cellcolor{blue!25} \\
\hline
\end{tabular}
\caption{ Technical component and subsystems' performance achievability scoring chart and color codes (used below in the summary Table \ref{tab:subgroupTRL}).}
\label{tab:TRLSG3}%
\end{table}

\subsubsection{Technically limited timescale}

This metric is an estimate of the timescale required to reduce the TRL of a colliders' technical components and subsystems to seven to eight (corresponding to the ITF Risk Level of one to two, see Sec.\ref{TRLsubsec1} above). It does not include time for industrialization which could overlap with later stages of the R\&D activity. These scores, derived from Table \ref{tab:subgroupTRL} were combined with the other four risk scores to obtain an overall score, but are not shown in the Technical Risk summary tables.. Results of a detailed analysis of R\&D timescales are given the the cost and schedule Section \ref{sec:Cost}. The numeric scoring key is shown in Table \ref{tab:TRLSG4} below.

\begin{table}[h!]
\centering
\begin{tabular}{|p{8cm}|p{1cm}|}
\hline 
{\bf R\&D Timescale} & {\bf Score} \\
> 20 years & 4 \\
15 - 20 years & 3 \\
10 - 15 years & 2 \\
5 - 10 years & 1 \\
0 - 5 years & 0.5  \\
\hline
\end{tabular}
\caption{ R\&D time frame scoring chart.}
\label{tab:TRLSG4}%
\end{table}

\pagebreak

\subsubsection{ITF technical risk evaluation process}

For each technology, the risk scores were squared and averaged across the five risk categories. Squaring was done to increase the weight of the higher scores. The average of the squares from each category was then summed to obtain an overall risk score. The scores were then grouped into four broad ranges or tiers, from lower to higher risk. The highest scores reflect the general lack of maturity of the proposed technologies and the large uncertainties in risk mitigation. Generally, the proposals with higher scores also have a larger number of high-risk technologies. The interpretation is that more and longer-term R\&D is needed to realize the potential of the proposed collider facilities.
As an example of the process, the completed evaluation for the ILC Higgs Factory is shown in Table \ref{TRLexampleILC}. 

The results for the risk categories are summarized in Table \ref{tab:subgroupTRL}. For each facility proposal, the highest score for each risk category was used, based on the color code charts, to indicate where the dominant risks reside. The table also indicates the design status of each proposal and the risk tier based on the overall technical risk score.

\begin{table}[h!]
\centering
\begin{tabular}{|l|ccccc|c|}
\hline 
{\bf ILC Higgs Factory} &  &  &  &  &  &  
\\
{\bf Critical Enabling Technologies} & {\rotatebox{90}{\bf Risk Factor}} & {\rotatebox{90}{\bf Technology Validation }}  & {\rotatebox{90}{\bf Cost Reduction Impact}} & {\rotatebox{90}{\bf Performance Achievability}} & {\rotatebox{90}{\bf R\&D Timescale}} &  {\rotatebox{90}{\bf Average of Squares}}\\
\hline
SRF Cavities & 1 &1  & 1 & 1  & 0.5 & 0.85\\ 
Cryomodules/Assembly & 1 & 2  & 2 & 1.5  & 0.5 & 2.3
\\
Positron Source & 2 & 2  & 1 & 2  & 0.5 & 3.65\\
nm Spot Size/Stability at IP & 1 & 2  & 1 & 1  & 0.5 & 1.45\\
Damping rings inj and extr & 1 & 1  & 1 & 1  & 0.5 & 0.85\\
\hline
\end{tabular}
\caption{ ILC Higgs Factory Scoring Example}
\label{TRLexampleILC}
\end{table}


\begin{table}
\caption{Table summarizing the TRL categories, technology validation requirements, cost reduction impact and the judgement of performance achievability on technical components and subsystems  for the evaluated collider proposals.  Colors and categories are described above in Sec.\ref{sec:TRL} and go from lighter/lower/easier to darker/higher/more challenging.  The first column "Design Status" indicates current status of the design concepts: I - TDR complete, II - CDR complete, III - substantial documentation; IV - limited documentation and parameter table; V - parameter table. The last column indicates the overall risk tier category, ranging from Tier 1 (lower overall technical risk) to Tier 4 (multiple technologies that require further R\&D).}
\begin{center}
\hspace*{0.1em}
\begin{tabular}{|  l || P{1.2cm}|| P{1.6cm} | P{2cm} | P{1.6cm} | P{2cm} || P{1.1cm}| }
\hline
Proposal Name &Collider & Lowest & Technical & Cost & Performance & Overall  \\
(c.m.e. in TeV)   & Design & TRL & Validation & Reduction & Achievability & Risk \\ 
 & Status & Category & Requirement & Scope &  & Tier\\
\hline 
FCCee-0.24 & II & \cellcolor{blue!50} &  \cellcolor{blue!25}&  \cellcolor{blue!25}& \cellcolor{blue!5} & 1 \\  
\hline 
CEPC-0.24   &  II & \cellcolor{blue!50} &  \cellcolor{blue!25}&  \cellcolor{blue!25}& \cellcolor{blue!5} & 1 \\
\hline 
ILC-0.25   &   I & \cellcolor{blue!50} &  \cellcolor{blue!25}&  \cellcolor{blue!25}& \cellcolor{blue!25} & 1 \\  
\hline 
CCC-0.25   & III & \cellcolor{blue!75} &  \cellcolor{blue!25}&  \cellcolor{blue!25}& \cellcolor{blue!25} & 2 \\   
\hline 
CLIC-0.38   & II & \cellcolor{blue!50} &  \cellcolor{blue!25}&  \cellcolor{blue!25}& \cellcolor{blue!25} & 1 \\  
\hline 
CERC-0.24  &   III & \cellcolor{blue!75} &  \cellcolor{blue!25}&  \cellcolor{blue!25}& \cellcolor{blue!50} & 2 \\ 
\hline 
ReLiC-0.24  &  V & \cellcolor{blue!75} &  \cellcolor{blue!25}&  \cellcolor{blue!50}& \cellcolor{blue!25} & 2  \\
\hline
ERLC-0.24  & V & \cellcolor{blue!75} &  \cellcolor{blue!25}&  \cellcolor{blue!50}& \cellcolor{blue!50} & 2 \\
\hline 
XCC-0.125  & IV & \cellcolor{blue!75} &  \cellcolor{blue!50}&  \cellcolor{blue!25}& \cellcolor{blue!50} & 2 \\ 
\hline 
MC-0.13   & III & \cellcolor{blue!75} &  \cellcolor{blue!50}&  \cellcolor{blue!25}& \cellcolor{blue!50} & 3 \\ 
\hline
\hline
ILC-3   & IV & \cellcolor{blue!100} &  \cellcolor{blue!50}&  \cellcolor{blue!50}& \cellcolor{blue!25} & 2 \\ 
\hline 
CCC-3  &  IV & \cellcolor{blue!75} &  \cellcolor{blue!50}&  \cellcolor{blue!50}& \cellcolor{blue!25} & 2 \\
\hline 
CLIC-3  & II & \cellcolor{blue!75} &  \cellcolor{blue!25}&  \cellcolor{blue!50}& \cellcolor{blue!25} & 1 \\
\hline 
ReLiC-3  & IV & \cellcolor{blue!75} &  \cellcolor{blue!25}&  \cellcolor{blue!50}& \cellcolor{blue!50} & 3 \\
\hline 
MC-3   & III & \cellcolor{blue!75} &  \cellcolor{blue!50}&  \cellcolor{blue!25}& \cellcolor{blue!50} & 3 \\
\hline 
LWFA-LC 1-3   & IV & \cellcolor{blue!100} &  \cellcolor{blue!50}&  \cellcolor{blue!100}& \cellcolor{blue!50} & 4 \\
\hline 
PWFA-LC 1-3   & IV & \cellcolor{blue!100} &  \cellcolor{blue!50}&  \cellcolor{blue!50}& \cellcolor{blue!50} & 4 \\
\hline 
SWFA-LC 1-3   & IV & \cellcolor{blue!100} &  \cellcolor{blue!50}&  \cellcolor{blue!50}& \cellcolor{blue!50} & 4 \\
\hline
\hline
MC 10-14   & IV & \cellcolor{blue!75} &  \cellcolor{blue!50}&  \cellcolor{blue!50}& \cellcolor{blue!50} & 3 \\
\hline
LWFA-LC-15     &  V & \cellcolor{blue!100} &  \cellcolor{blue!50}&  \cellcolor{blue!100}& \cellcolor{blue!50} & 4 \\
\hline 
PWFA-LC-15   &  V & \cellcolor{blue!100} &  \cellcolor{blue!50}&  \cellcolor{blue!50}& \cellcolor{blue!50} & 4 \\  
\hline 
SWFA-LC-15    &  V & \cellcolor{blue!100} &  \cellcolor{blue!50}&  \cellcolor{blue!50}& \cellcolor{blue!50} & 4 \\
\hline
FCChh-100   & II & \cellcolor{blue!100} &  \cellcolor{blue!50}&  \cellcolor{blue!50}& \cellcolor{blue!50} & 3 \\
\hline 
SPPC-125    & III & \cellcolor{blue!75} &  \cellcolor{blue!50}&  \cellcolor{blue!25}& \cellcolor{blue!50} & 3 \\
\hline 
Coll.Sea-500    &  V & \cellcolor{blue!100} &  \cellcolor{blue!50}&  \cellcolor{blue!50}& \cellcolor{blue!50} & 4 \\
\hline
\end{tabular}
\end{center}
\label{tab:subgroupTRL}
\end{table}

\begin{table}[htbp]
  \footnotesize
  \centering
  \caption{Duration and integrated cost of the past and present, and proposed R\&D programs and facilities (the latter indicated by a shift to the right). Funding sources for the past and present programs are indicated ("OHEP" - directed R\&D in the DOE OHEP, "GARD" - General Accelerator R\&D and facilities operation program in the OHEP, "LDG/CERN" - aspirational support requested as part of the European Accelerator R\&D Roadmap \cite{Mounet:European_roadmap}). Separately listed $input$s are estimates from the proponents on the  total cost of demonstration projects and on the pre-CD2 validation (R\&D, design and industrialization); sources of support of these developments are "tbd". 
\\}
    \begin{tabular}{lccccp{4cm}}
    \hline
    \hline
    R\&D Program & Benefiting & Duration & Integrated  & Funding  & Key Topics \\
    Facility Name & Concept & (Years) & Cost (M\$)  & Source &  Rationale \\
    \hline
    \multicolumn{2}{l}{\bf Linear $e^+e^-$ colliders}&       &       &       &       \\
     NLC/NLCTA/FFTB & NLC/C$^3$ & 14   & 120   & OHEP  & NC RF gradient, final focus \\
     TESLA/TTF & ILC   & $\sim$10   & 150  & DESY/Collab & SCRF CMs and beam ops\\
     ILC in US/FAST & ILC   & 6 & 250  & OHEP  & SCRF CMs and beam ops\\
     ILC in Japan/KEK & ILC   & 10   & 100 & KEK   & SCRF CMs and beam ops\\
     ATF/AFT2 & ILC   & 15   & 100 & KEK/Intl & LC DR and final focus\\
     CLIC/CTF/CTF3 & CLIC  & 25   & 500 & CERN/Intl & 2-beam scheme and driver\\
     General RF R\&D & All LCs & 8     & 160  & GARD  & see RF Roadmap; incl facilities\\
     \hspace{0.3cm}ILC in Japan/KEK  & ILC   & 5     & 50    & KEK   & next 5 yr request \\
     \hspace{0.3cm}High-$G$ RF \& Syst. & CLIC/SRF & 5     & 150 & LDG/CERN &  
     NC/SC RF and klystrons\\
    \hspace{0.3cm}C$^3$ $input$ & C$^3$  & 8     & 200 &  tbd     & 72-120 MV/m CMs,  design\\
    \hspace{0.3cm}HELEN $input$ & HELEN &  n/a     & 200 &  tbd     & pre-TDR,   TW SRF tech\\
    \hspace{0.3cm}ILC-HE $input$ & ILC-HE & 20   & 100 &  tbd     & 10 CMs 70MV/m  $Q$=2e10\\
    \hspace{0.3cm}ILC-HighLumi $input$ & ILC-HL & 10   & 75    &  tbd  & 31.5 MV/m at $Q$=2e10 \\
    \hline
     \multicolumn{2}{l}{\bf Circular/ERL $ee/eh$ colliders} &            &       &       &      \\
     CBB & LCs & 6     & 25    & NSF   & high-brightness sources\\
     CBETA & ERLCs & 5     & 25    & NY State & multi-turn SRF ERL demo\\
     \hspace{0.3cm}ERLs/PERLE & ERLCs & 5     & 80*  & LDG/CERN & NC/SC RF, klystrons\\
    \hspace{0.3cm}FNAL$ee$ $input$ & FNALee &  n/a     & 100   &  tbd     & design and demo efforts\\
    \hspace{0.3cm}LHeC/FCC$eh$ $input$ & $eh$-coll. &  n/a     & 100   &  tbd     & demo facility, design\\
    \hspace{0.3cm}CEPC $input$ & CEPC  & 6     & 154   &  tbd     & SRF, magn. cell, plasma inj. \\
    \hspace{0.3cm}ReLiC $input$ & ReLiC & 10    & 70 &  tbd     & demo $Q$=1e10 at 20 MV/m \\
    \hspace{0.3cm}XCC $input$ & XCC   & 7     & 200 &  tbd     & demo and design efforts \\
    \hspace{0.3cm}CERC $input$ & CERC  & 8     & 70 &  tbd     & demo high-$E$ ERL at CEBAF \\
    \hline
       \multicolumn{2}{l}{\bf Muon colliders}  &       &       &       &      \\
     NFMCC & MC    & 12   & 50   & OHEP  & design study, prototyping\\
     US MAP & MC   & 7    & 60   & OHEP  & IDS study, components \\
     MICE & MC   & 12   & 60  & UK/Collab & 4D cooling cell demo\\
      \hspace{0.3cm}IMCC/pre-6D demo & MC-HE & 5     & 70  & LDG/CERN & pre-CDR work,  components\\
      \hspace{0.3cm}IMCC/6D cool. & MC-HE & 7     & 150 & CERN/Collab & 6D cooling facility and R\&D\\
    \hline
     {\it {\bf Circular $hh$ colliders}} &       &       &       &       &      \\
     LHC Magnet R\&D & LHC & 12    & 140  & CERN  & 8T NbTi LHC magnets\\
     US LARP & LHC & 15 & 170   & OHEP  & more LHC luminosity faster\\
     SC Magnets General & $pp, \mu\mu$ & 10    & 120  & GARD  & HF-magnets and materials\\
      \hspace{0.3cm}US MDP & $pp, \mu\mu$ & 5     & 40    & GARD  & see HFM Roadmap \\
      \hspace{0.3cm}HFM Program & FCChh & 7     & 170  & LDG/CERN & 16 T magnets for FCChh\\
     \hspace{0.3cm}FNAL$pp$ $input$ & FNAL$pp$ &  n/a     & 100   &  tbd     & 25T magnets demo\\
     \hspace{0.3cm}FCChh $input$ & FCChh & 20    & 500 &  tbd & large demo,  R\&D and design\\
 \hspace{0.3cm}Coll.Sea $input$ & CollSea & 16    & 400 &  tbd     & 300m magnets underwater \\
    \hline
     {\it {\bf AAC colliders}} &       &       &       &       &      \\
     SWFA/AWA & SWFA-LC & 8    & 40  & GARD  & 2-beam accel in THz structures\\
     LWFA/BELLA & LWFA-LC & 8    & 80  & GARD  & laser-plasma WFA R\&D\\
     LWFA/DESY & LWFA-LC & 10    & 30    & DESY  & laser-plasma WFA R\&D\\
     PWFA/FACET-I,II & PWFA-LC & 13    & 135  & GARD  & 2-beam PWFA, facility\\
     AWAKE & PWFA-LC & 8    & 40  & CERN/Collab  & proton-plasma PWFA, facility\\
      \hspace{0.3cm}EUPRAXIA & LWFA-LC & 10   & 570 & EUR/Collab. & high quality/eff. LWFA R\&D\\
      \hspace{0.3cm}LWFA/DESY & LWFA-LC & 10    & 80    & DESY  & laser WFA R\&D\\
     \hspace{0.3cm}SWFA $input$ & SWFA-LC & 8     & 100 &  tbd     & 0.5 \& 3GeV demo facilities \\
     \hspace{0.3cm}LWFA $input$ & LWFA-LC & 15    & 130 &  tbd     & 2nd BL, $e^+$, kBELLA project\\
     \hspace{0.3cm}PWFA $input$ & PWFA-LC & 10    & 100 &  tbd     & demo and design effort\\
    \hline
    \hline
    \end{tabular}%
  \label{tab:RandDcosts}%
\end{table}%


\subsection{Summary}

The overall technical risk for each proposal was evaluated by identifying five critical enabling technologies and generating a relative technical risk score. Some input was provided by the proponents, but in certain cases it was augmented (to bring the number of enabling technologies up to five) and/or normalized to make a consistent comparison. Where data was not provided, the ITF applied expert judgement. The scores were then grouped into four broad tiers that range from those based on mature, well-understood technologies, to those with multiple, as-yet unproven critical technologies. Generally, the scores indicate the relative need and extent of R\&D required to develop a given technology. It is important to note that there is a large spread within each group of proposals and there can be a significant difference between the low and high ends. More detail can be obtained from the risk registry Tables for the proposal components and systems. For reference, Table \ref{tab:RandDcosts} summarizes integrated cost and duration of the past and present, and proposed R\&D programs and facilities.

%% file: PowerComplexity.tex
\section{Power, Complexity and Environmental Impact of Colliders}
\label{sec:Power}


\subsection{Summary table}

\begin{table}[h!]
\caption{Table summarizing the categories of power consumption, size, complexity and required radiation mitigation for the evaluated collider proposals.  Color schemes and categories are explained in Sec. \ref{sec:power2} (power consumption), Sec. \ref{sec:size} (size), \ref{sec:complexity} (complexity) and Sec. \ref{sec:radiation} (radiation). For linear colliders, the size of the machine includes main linac and final focus, but excludes damping rings, except where otherwise noted.}
\begin{center}
\hspace*{-2em}
\begin{tabular}{|  c | P{2cm}| P{2cm} | P{2cm} | P{2cm} |}
\hline
Proposal Name & Power & Size & Complexity & Radiation \\
 & Consumption& & & Mitigation\\ 
\hline 
\hline 
FCC-ee (0.24 TeV) &  290 \cellcolor{blue!50} &  \textcolor{white}{91 km} \cellcolor{blue!100}&  I \cellcolor{blue!15}& I \cellcolor{blue!15} \\  
\hline 
CEPC (0.24 TeV) &  340 \cellcolor{blue!50} &  \textcolor{white}{100 km} \cellcolor{blue!100}&  I \cellcolor{blue!15}& I \cellcolor{blue!15} \\
\hline 
ILC (0.25 TeV) &  140 \cellcolor{blue!15}& 20.5 km \cellcolor{blue!50}& I \cellcolor{blue!15}& I \cellcolor{blue!15}\\  
\hline 
CLIC (0.38 TeV)  &  110 \cellcolor{blue!15}&  11.4 km\cellcolor{blue!50}&  II \cellcolor{blue!50}&  I \cellcolor{blue!15}\\  
\hline
CCC (0.25 TeV) &  150 \cellcolor{blue!15}& 3.7 km \cellcolor{blue!15}& I \cellcolor{blue!15}& I \cellcolor{blue!15}\\   
\hline 
CERC (0.24 TeV)&  90 \cellcolor{blue!15}&  \textcolor{white}{91 km} \cellcolor{blue!100}&  II \cellcolor{blue!50}&  I \cellcolor{blue!15}\\ 
\hline 
ReLiC (0.24 TeV)& 315 \cellcolor{blue!50}&  20 km \cellcolor{blue!50}& II \cellcolor{blue!50}& I \cellcolor{blue!15}\\
\hline 
ERLC (0.24 TeV)& 250\cellcolor{blue!50}& \textcolor{white}{30 km}\cellcolor{blue!75}& II\cellcolor{blue!50}& I\cellcolor{blue!15}\\
\hline 
XCC (0.125 TeV)& 90 \cellcolor{blue!15}&  1.4 km \cellcolor{blue!15}& II \cellcolor{blue!50}& I \cellcolor{blue!15}\\ 
\hline 
MC (0.13 TeV) & 200 \cellcolor{blue!50}& 0.3 km  \cellcolor{blue!15}& I \cellcolor{blue!15}& II \cellcolor{blue!50}\\ 
\hline
\hline
ILC (3 TeV) & $\sim$400 \cellcolor{blue!50}& \textcolor{white}{59 km} \cellcolor{blue!100}& II \cellcolor{blue!50}& II \cellcolor{blue!50}\\ 
\hline 
CLIC (3 TeV)& \textcolor{white}{$\sim$550} \cellcolor{blue!75}&  \textcolor{white}{50.2 km}\cellcolor{blue!75}& \textcolor{white}{III}  \cellcolor{blue!75}& II \cellcolor{blue!50}\\
\hline 
CCC (3 TeV)&  \textcolor{white}{$\sim$700} \cellcolor{blue!75}& \textcolor{white}{26.8 km}\cellcolor{blue!75}& II \cellcolor{blue!50}& II \cellcolor{blue!50}\\
\hline 
ReLiC (3 TeV)& \textcolor{white}{$\sim$780}\cellcolor{blue!75}& \textcolor{white}{360 km} \cellcolor{blue!100}& \textcolor{white}{III}  \cellcolor{blue!75}& I \cellcolor{blue!15}\\
\hline 
MC (3 TeV) & $\sim$230\cellcolor{blue!50}& 10-20 km \cellcolor{blue!50}& II \cellcolor{blue!50}& \textcolor{white}{III} \cellcolor{blue!75}\\
\hline 
LWFA (3 TeV) & $\sim$340\cellcolor{blue!50}&  1.3 km (linac) \cellcolor{blue!15}& II \cellcolor{blue!50}& I \cellcolor{blue!15}\\
\hline 
PWFA (3 TeV) &$\sim$230\cellcolor{blue!50}&  14 km\cellcolor{blue!50}& II \cellcolor{blue!50}& II \cellcolor{blue!50}\\
\hline 
SWFA (3 TeV) & $\sim$170\cellcolor{blue!15}& 18 km \cellcolor{blue!50}& II \cellcolor{blue!50}& II \cellcolor{blue!50}\\
\hline
\hline
MC (14 TeV) & \textcolor{white}{$\sim$300} \cellcolor{blue!75}& \textcolor{white}{27 km}\cellcolor{blue!75}&  \textcolor{white}{III}\cellcolor{blue!75}&  \textcolor{white}{III}\cellcolor{blue!75}\\
\hline
LWFA (15 TeV)   & \textcolor{white}{$\sim$1030} \cellcolor{blue!75}& 6.6 km\cellcolor{blue!15}&  \textcolor{white}{III} \cellcolor{blue!75}&  I\cellcolor{blue!15}\\
\hline 
PWFA (15 TeV) & \textcolor{white}{$\sim$620} \cellcolor{blue!75}& 14 km \cellcolor{blue!50}&  \textcolor{white}{III} \cellcolor{blue!75}&  II\cellcolor{blue!50}\\  
\hline 
SWFA (15 TeV)  & $\sim$450 \cellcolor{blue!50}& \textcolor{white}{90 km}\cellcolor{blue!100}& \textcolor{white}{III} \cellcolor{blue!75}&  II\cellcolor{blue!50}\\
\hline
FCC-hh (100 TeV) &  \textcolor{white}{$\sim$560}\cellcolor{blue!75}& \textcolor{white}{91 km}\cellcolor{blue!100}& II \cellcolor{blue!50}& \textcolor{white}{III}\cellcolor{blue!75}\\
\hline 
SPPC (125 TeV)  & $\sim$400\cellcolor{blue!50}& \textcolor{white}{100 km}\cellcolor{blue!100}& II \cellcolor{blue!50}&  \textcolor{white}{III}\cellcolor{blue!75}\\
\hline 
\hline

\end{tabular}
\end{center}
\label{tab:WGpowersizecompenvimpactsummary}
\end{table}

\subsection{Power consumption}
\label{sec:power2}
Estimates of power consumption for collider proposals are summarized in Table \ref{tab:WGpowersizecompenvimpactsummary} and refer to the total site power required by the collider complex for operation. Numbers provided by the proponents were grouped into three categories. The lowest category is light blue (1) and indicates a power consumption below 200 MW. The next category is blue (2), for a power consumption between 200 and 500 MW. The highest category is dark blue (3) and indicates a consumption larger than 500 MW. For reference, CERN's annual electric energy consumption is about 1.3 TWh (2015), with a peak power of about 230 MW at the times of the entire accelerator complex operational with the LHC machine alone requiring some 120 MW. 

\begin{figure}[h]
    \centering
    \includegraphics[width=\textwidth]{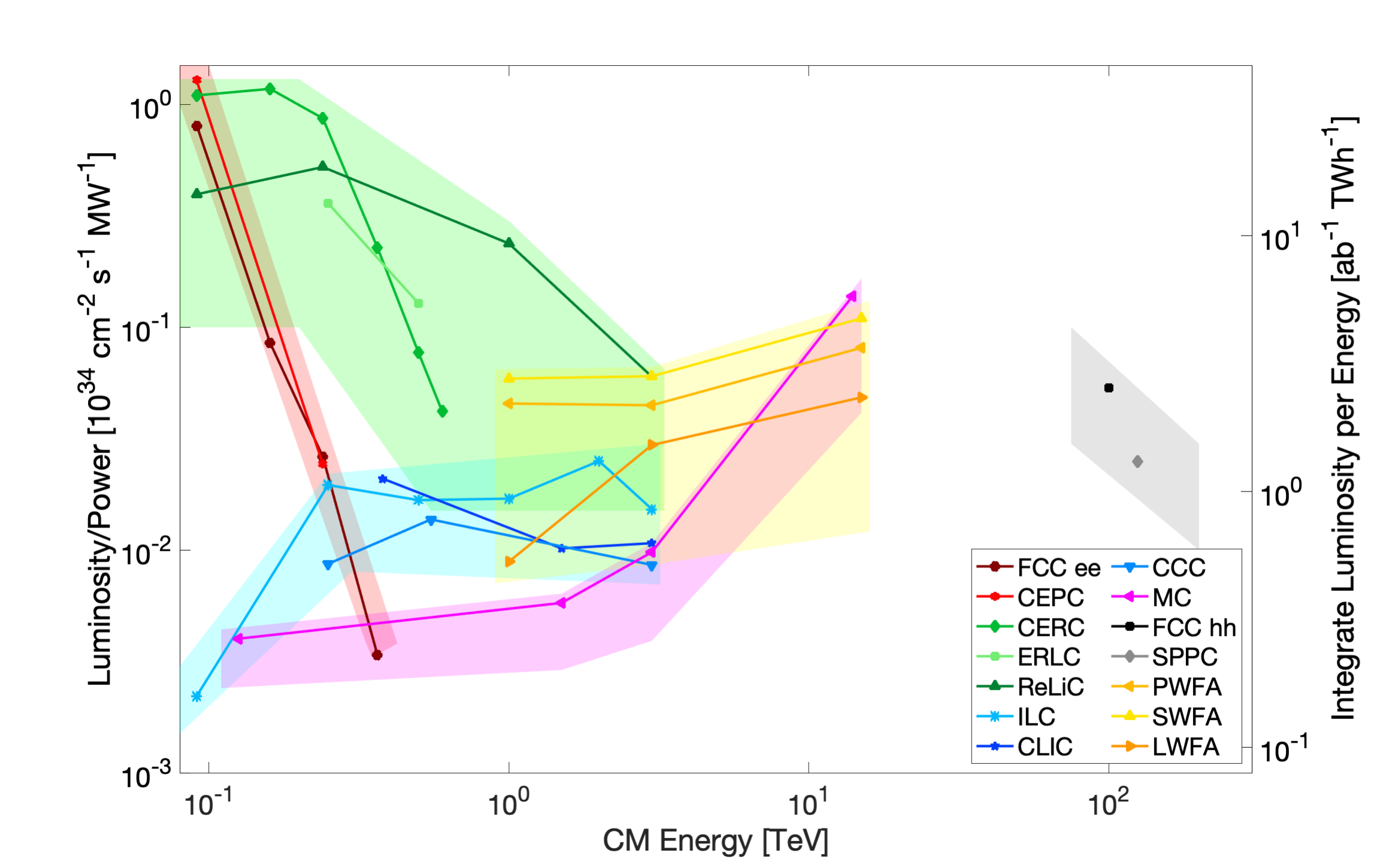}
    \caption{Figure-of-merit Peak Luminosity (per IP) per Input Power and Integrated Luminosity per TWh. Integrated luminosity assumes $10^7$ seconds per year. The luminosity is per IP. Data points are provided to the ITF by proponents of the respective machines. The bands around the data points reflect approximate power consumption uncertainty for the different collider concepts.}
    \label{fig:lumi_pow}
\end{figure}

One of the figures-of-merit for a collider is the luminosity-per-site power. Figure~\ref{fig:lumi_pow} shows the luminosity-per-site power for each of the machines plotted in Figs.~\ref{fig:Higgscollidersplot}-\ref{fig:HEppcollidersplot}.

\subsection{Facility size}
\label{sec:size}
An overview of collider sizes (as provided by proponents) is shown in column 3 of Tab.~\ref{tab:WGpowersizecompenvimpactsummary}. {\it Collider Size} refers to either the length of a linear collider (main linac plus final focus) or the circumference of a circular collider main ring, without the injector complex. The ITF defined four size categories (shown in Tab.~\ref{tab:WGpowersizecompenvimpactsummary}): light blue (1) for colliders that are designed to be shorter then 10 km, medium blue (2) for colliders between 10-20 km, blue (3) for colliders between 20-50 km and dark blue (4) for machines with a length or circumference larger than 50 km.

The length of HEP linear colliders is typically dominated by the distance required for particle acceleration and is proportional to final beam energy (approximately the product of $2\times$ the final beam energy and the accelerating gradient). Using acceleration technologies with higher accelerating gradients allows to decrease acceleration length and is responsible for the different lengths of similar energy linear colliders. For example, superconducting radio-frequency cavities accelerate with a gradient of $\sim$ 30 MV/m (ILC), CLIC is based on the two-beam acceleration scheme with copper cavities and accelerates with $\sim$ 100 MV/m, while plasma-based accelerators can provide peak gradients of $10^{3}-10^{5}$ MV/m (LWFA, PWFA). Adding to the length required for acceleration is the length required for the beam delivery system (final focusing), which also increases with increasing with beam energy.\\ 

\vspace{0.2cm}
\textbf{Overview of linear collider sizes:}
\begin{itemize}
    \item $<10$ km, Category 1: CCC (0.25 TeV), XCC (0.125 TeV), LWFA (3 TeV), LWFA (15 TeV)
    \item $10-20$ km, Category 2: ReLiC (0.24 TeV), ILC (0.25 TeV), CLIC (0.38 TeV), PWFA (3 TeV), SWFA (3 TeV), PWFA (15 TeV)
    \item $20-50$ km, Category 3: ERLC (0.24 TeV), CCC (3 TeV), CLIC (3 TeV)
    \item $>50$ km, Category 4: ILC (3 TeV), ReLiC (3 TeV), SWFA (15 TeV)
\end{itemize}
\vspace{0.2cm}

The size (circumference) of circular colliders is defined by final beam energy, the strength of the dipole magnets in the accelerator ring as well as the rest mass of the accelerated particles (e.g. electrons, muons or protons). The beam energy for a given circumference accelerator is limited by synchrotron radiation losses. At the same beam energy, heavier particles (e.g. muons or protons compared to electrons) radiate less allowing for comparatively higher final energies in the same collider ring.\\

\vspace{0.2cm}
\textbf{Overview of circular collider sizes:}
\begin{itemize}
    \item $<10$ km, Category 1: MC (0.13 TeV)
    \item $10-20$ km, Category 2: MC (3 TeV),
    \item $20-50$ km, Category 3: MC (14 TeV)
    \item $>50$ km, Category 4: FCC-ee (0.24 TeV), CEPC (0.24 TeV), CERC (0.24 TeV), ReLiC (3 TeV), FCC-hh (100 TeV), SPPC (125 TeV), C-sea (500 TeV)
\end{itemize}

\vspace{0.2cm}

\textbf{Additional comments on collider size:}
\begin{itemize}
    \item The FCC-ee and CEPC (0.24 TeV) Higgs factory designs reuse the FCC-hh and SPPC tunnels, respectively. The lengths have been partially chosen for the hadron colliders.
    \item The length of the 3 and 15 TeV PWFA colliders is 14 km as they are proposed upgrades to the ILC (0.25 TeV).
\end{itemize}

\subsection{Complexity of accelerators}
\label{sec:complexity}
It is generally accepted that modern accelerators are very sophisticated systems and that, e.g., {\it "...the LHC is the most complex scientific instrument of our time"} \cite{shiltsev2020particle}. Sometimes distinguished are a) complexity to design and build many dissimilar systems containing many subsystems and elements of various sizes and levels of operational functionality, and b) complexity to reach the beam energy ({\it “make it work reliably")}  and 
to reach the design performance, e.g., luminosity of colliders \cite{shiltsev2011complexity}.  In general, the issue of complexity of systems (and accelerators in particular) is far from being comprehensively resolved. The problem is that complexity is something that we immediately recognize when we see it, but it is very hard to define it quantitatively when it comes to questions like {\it How hard is it to describe?; How hard is it to create?
What is its degree of organization?} Ref.\cite{lloyd2001measures} lists four dozens different possible of complexity measures that can be roughly divided into two categories: i) computational/descriptive complexities; and ii)  effective/physical "structural" complexities. The former can be represented by, e.g., the Kolmogorov complexity that is the length of the shortest description (in a given language) of the object of interest \cite{kolmogorov1963tables}.  Such an approach results in overestimated complexity of random systems not very suitable for complex physical phenomena. Among the second type of complexity definitions is the approach of Ref.\cite{bagrov2020multiscale} focused on hierarchies and patterns that generally treats complexity the measure of dissimilarity at various scales.  

Within the paradigm of the structural complexity, accelerators indeed, can be characterised by dissimilarity of various components and subsystems  (magnets, RF cavities, plasma cells, beam cooling systems, beam drivers, injectors, final focus systems, etc) each having it's own hierarchy. For example, the LHC collider can be seen as  "1 ring, consisting of 	$O$(10) sectors, $O$(100) cells with $O$(1000) main SC magnets in addition to $O$(10$^4$) auxiliary magnets and $O$(10$^5$) control channels". In such approach, the collider complexity measure {\bf C} can be estimated as a sum over key subsystems:
\begin{equation}
{\bf C}  = \sum_{subsystems} (C_i+\Delta_i),
\label{Complexity}    
\end{equation}
where individual subsystem complexity is $C_i\simeq \log_{10}$(Number of elements in $i$-th subsystem). For example, a 3 TeV linear collider has about factor of 10 more RF or plasma accelerating cells than the same type EW/Higgs factory and would have about "1 unit" higher {\bf C}. Extra factor $\Delta_i$ is supposed to account for 
additional complications due to the nature of basic elements: for example, 1-2 T normal-conducting magnets are "off-the-shelf" items nowadays and might be considered as less complex ($\Delta_i \sim -1$) compared to 8 T superconducting NbTi magnets, while 16 T Nb$_3$Sn magnets are more complex $\Delta_i \simeq +1$). When accounting various subsystems one should not forget that not only colliders themselves can be complex, but also their injectors or/and boosters. The most notable examples are those for the FCCee/CEPC, Muon Colliders and FCChh/SPPC. Subsystems related to particle production and, if necessary, cooling may significantly contribute, too, such as positron sources or damping rings for all linear $e^+e^-$ colliders, proton drivers and muon ionization cooling systems for muon colliders, and high power drive systems for CLIC, BPLC and SWLC (low energy high-power electron beams), and for LPLC (high peak and average power lasers). Finally, rather complex are multi-km long final focus systems employed in all $e^+e^-$ linear colliders. 

Such analysis indicates that all future collider projects under the ITF consideration are significantly more complex ($\Delta {\bf C} \geq 1-2$) than the LHC or the Tevatron. Table \ref{tab:WGpowersizecompenvimpactsummary} summarizes the collider project complexities in color-coded schemes with about 2 units in ${\bf C}$ increase between light blue, blue and dark blue colors. Obviously, the complexity of colliders changes in time as corresponding R\&D programs and technologies progress, and experience accumulates; for example, construction of an LHC-type machine does not look as difficult now as 20 years ago. 

\subsubsection{Machine commissioning}

Due to their complexity colliders often require significant time to reach their design or ultimate luminosity - see Table \ref{tabLumiTime}. Analysis of Ref.\cite{shiltsev2011complexity} indicates that the luminosity improvements usually come in incremental steps and as long as those occur regularly, the peak luminosity progresses exponentially : 
\begin{equation}
{\cal L} (t)  = {\cal L}(0) \cdot \exp{\Large( t/ {\cal C} \Large)}. 
\label{CPT}    
\end{equation}
The coefficient $\cal C$ has dimension of [time], it might serve as an indirect indicator of facilities' complexities, and  for the leading colliders of the past varies between 2 and 4 years. It is hard to project Eq.(\ref{CPT}) predictions onto the future colliders, but given the insights into the structural complexity of future facilities it is reasonable to expect to expect event larger coefficient  $\cal C$ for them. Given challenging luminosity goals for all of the future colliders projects, the expected long commissioning periods should be taken into account in planing for the integrated luminosity and operational schedules. 

\begin{table}[htbp]
  \centering
  \caption{Time required to reach design peak luminosity for several recent lepton and hadron particle colliders. The last column indicates maximum achieved luminosity w.r.t. to the design luminosity. (* colliders still in high-luminosity operation; ** - RHIC operation in $pp$ collider mode was intermittent with heavy ions collisions runs.) }
    \begin{tabular}{|l|c|c|}
    \hline
    \hline
    Collider & Time to Reach Design Peak $L$ & Record $L$ / Design $L$ \\
    \hline
    LEP-I & 5 years & $\times$ 2 \\
    SLC   & Not achieved (9 years) & $\times$ 0.5 \\
    LEP-II & 0.3 years & $\times$ 3 \\
    PEP-II & 1.5 years & $\times$ 4 \\
    KEK-B & 3.5 years & $\times$ 2 \\
    BEPC-II & 7.5 years & $\times$ 1.0 \\
    DAFNE & Not achieved (9 years) & $\times$ 0.9 \\
    Super-KEK-B & Not yet achieved (4 years) & $\times$ 0.05 * \\
    \hline
    TEV-Ib & 1.5 years & $\times$ 1.5 \\
    HERA-I & 8 years & $\times$ 1 \\
    RHIC-$pp$ & 10 years ** & $\times$ 1.2 * \\
    TEV-II & 5 years & $\times$ 2.1 \\
    HERA-II & 5 years & $\times$ 1 \\
    LHC   & 6 years & $\times$ 2.1 * \\
    \hline
    \hline
    \end{tabular}%
  \label{tabLumiTime}%
\end{table}%

\subsection{Radiation mitigation}
\label{sec:radiation}

Particle colliders produce ionizing radiation and activate material. The hazard level and therefore required mitigation techniques depend, among other factors, on the particle energy, the particle type and species and the total amount of energy that is stored in the beams. The ITF defined three categories of radiation mitigation required (see Tab.\ref{tab:WGpowersizecompenvimpactsummary}): light blue (1) where required radiation safety measures comparable to the ones used in current facilities are sufficient; medium blue (2) where moderately higher or more complex radiation mitigation techniques are required (e.g. more activated material produced) and dark blue (3) if much higher and significantly more advanced and sophisticated efforts and considerations will be required to mitigate radiation risks (e.g., actively moving the beam to reduce average radiation levels).

\vspace{0.2cm}
\textbf{Overview of radiation categories:}
\begin{itemize}
    \item Category I: FCC-ee (0.24 TeV), CEPC (0.24 TeV), ILC (0.25 TeV), CCC (0.25 TeV), CLIC (0.38 TeV), CERC (0.25 TeV), ReLiC (0.25 TeV), ERLC (0.25 TeV), XCC (0.125 TeV), ReLiC (3 TeV), LWFA (3 TeV), LWFA (15 TeV)
    \item Category II: MC (0.13 TeV), ILC (3 TeV), CCC (3 TeV), PWFA (3 TeV), SWFA (3 TeV)
    \item Category III:  MC (3 TeV), MC (14 TeV), FCC-hh (100 TeV), SPPC (125 TeV)
\end{itemize}

Higgs factories (except the muon collider), the 3 TeV ReLiC design and the LWFA wakefield linear collider designs (3 and 15 TeV) are in the light blue category. Radiation levels and the amount of activated material produced are comparatively low and can be safely mitigated with standard shielding. The amount of activated material produced will be comparatively low, because of the particle type (electrons and positrons) and either moderate beam powers (ReLiC, LWFA) or moderate beam energies (Higgs factories).

The medium category is populated by the muon collider Higgs factory, the 3 TeV ILC, CCC, CLIC, PWFA and SWFA designs as well as the 15 TeV PWFA and SWFA collider designs. These electron-positron collider use high beam energies (3 or 15 TeV) together with high beam powers. 

The highest safety mitigation efforts are required for the 3 and 14 TeV muon colliders, as well as the 100 TeV FCC-hh and  and 125 TeV SPPC designs. The very high energy hadron colliders (FCC-hh and SPPC) will produce large amount of activated material, both along the accelerator as well as in the final beam dump. Muon colliders pose special, new challenges. For example, they require new, active neutrino radiation mitigation techniques because of the muon decay. Even if the collider is underground, neutrino radiation can exceed environmental safety limits on the earth surface without mitigation.

%% file: Cost.tex
\section{Collider Facilities Costs and Time to Construct}
\label{sec:Cost}
High-energy colliders are large projects funded by public money, developed over many years and constructed via major laboratory and industrial contracts both in advanced technology and in more conventional domains such as civil engineering and infrastructure, for which they often constitute one-off markets. Assessing their costs, as well as the risks and uncertainties is therefore an essential part of project preparation and a justified requirement by the funding agencies. Below we present previous analyses of larger accelerator projects, discuss in detail the cost models developed by this Task Force, and summarize main results and conclusions regarding the cost estimates for the majority of the future collider projects submitted to the ITF as well as their construction timelines and the R\&D needed to either improve performance or to make the accelerator more affordable.

\subsection{Introduction to cost estimates}

Over almost a century of existence, high-energy particle accelerators have undergone sustained development of
their performance, as exemplified in the "Livingston"-type diagrams - see the most up-to-date one in Figs. 2 and 3 in Ref.\cite{shiltsev2021modern}. The corresponding increases in size and cost, however, proceeded at slower pace thanks to implementation of novel technologies and application of industrial construction methods. Still, recent large HEP accelerator projects such as LHC have costs amounting to several years of funding of the particle physics discipline, therefore, drawing significantly on public research budgets in their construction years.

It was clearly understood already a decade ago that assessing risk in costing such projects is therefore an important issue, whether these risks are later mitigated by suitable R\&D, project de-scoping, stretching of construction schedules or reallocation of additional resources - see Ref.\cite{lebrun2012riskassessment}. The assessment was rendered more difficult by the fact that these projects are usually one-off or single-time activity, without a market outside the project proper enabling the establishment of real market prices. It was also noted then that high-technology accelerator components account only for a fraction of the total project cost (TPC) while much of the budget goes into civil engineering, infrastructure and services for which market prices are usually available. 

The very first attempts to project the past and present knowledge onto future colliders \cite{lebrun2013costing, shiltsev2014costmodel} have indicated that: i) the cost of larger facilities does scale more slowly than linearly with any parameter such as, e.g., size, energy or power consumption; and ii) there are significant variations in the cost estimating methodology worldwide. As for the latter, the cost estimates of some construction projects
 included the industrial contracts for major items like civil engineering, the accelerator elements
and corresponding labor requirements (such approach is often referred as the “European accounting”)
while other estimates included full accounting of all associated expenses (the “US accounting”). For
example, all scientific facilities supported by the US DOE Office of Science are required to prepare and report estimates of “the total project cost” (TPC) which includes not only the cost of the technical components, materials, contracts, services, civil construction and conventional systems, and associated labor, but also costs of the required in-project R\&D, development of the engineering design, project management, escalation due to inflation, installation, threshold commissioning, contingency, overhead funds, project-specific facility site development, sometimes — detectors, etc. The difference between the TPC and “European accounting” could often be as big as factor of 2.0 \cite{shiltsev2014costmodel}. Note that such a big difference is typical not only for accelerators, but for other large scientific projects as well, and, e.g., the 2018 estimates of the ITER construction costs are 41BEuro and 65B\$, correspondingly \cite{kramer2022further}. 

The first phenomenological three-parameter cost model Ref.\cite{shiltsev2014costmodel} was attempted on the basis of publicly available cost estimates for 17 large accelerators of the past, present and those then (as of 2014) in the planning stage, including: SSC in Texas; VLHC in Illinois; NLC at SLAC; TESLA at DESY; ILC; Main Injector and Project-X at FNAL; CERN’s LHC, SPL and CLIC; the Beta-Beam and Neutrino Factory projects; RHIC at BNL;  SNS at ORNL; 
ESS at Lund; XFEL at DESY; and FAIR at GSI. All the costs were reduced to the TPC methodology (“the US accounting”) and broken up into three major parts corresponding to “civil construction”, “accelerator components”, and “site power infrastructure” in such a manner that they total the derived TPC ranges. 
The model utilized just three parameters — the length of the tunnels $L$, the
center-of-mass or beam energy $E$, and the total required site power $P$  — and found that over almost
3 orders of magnitude of $L$, 4.5 orders of magnitude of $E$ and more than 2 orders of magnitude
of $P$ the following cost model works with $\sim$30\% accuracy: 
\begin{equation}
TPC_{\alpha \beta \gamma} \approx \alpha \cdot (Length)^{p_1}
 + \beta \cdot (Energy)^{p_2} + \gamma \cdot (Power)^{p_3}
\label{ModelOld}    
\end{equation}
where the exponents are $p_1 \approx 0.55$, $p_2 \approx 0.46$, $p_3 \approx 1.0$, and coefficients $\alpha \approx 1.1$B\$ if $L$ is in the units of 10 km, $\gamma \approx 1.7$B\$ if $P$ is in the units of 100 MW, and accelerator technology dependent coefficient $\beta_{\rm MAG} \approx 1.2$B\$ for SC magnets if the c.m.e. $E$ is in the units of 1 TeV. For RF based facilities, such as linear colliders, $p_2 \approx 0.53$ and $\beta_{\rm RF} \approx 9.1$B\$. 
The above three-parameter $\alpha \beta \gamma$-model could be further simplified for equal exponents $p_1=p_2=p_3=1/2$, i.e., the square root cost scaling, without loss of $\sim$ 30\% accuracy and had been applied to several proposed collider facilities to obtain the TPC ranges or the cost of their parts which are expected to be built on the base of the currently known accelerator technologies. It was remarked that besides the feasibility of the cost, very important are the feasibility of the performance and availability of expertise for large machine construction projects, and that significant investments into the R\&D on the novel advanced accelerator techniques or on the cost reduction of the existing technologies are required before one can evaluate opportunities for financially feasible, next generation energy frontier accelerators.

\subsection{ITF comparison approach}
The Snowmass'21 ITF approach is significantly more sophisticated and detailed than a few parameter  $\alpha \beta \gamma$-model. First of all, it does not only use communicated cost estimates given by the projects' proponents but is also based on independent cost estimates of various machine components, labor and accounting factors. Thus, a multi-parameter (30) model was  initiated to contrast with the $\alpha \beta \gamma$-model.\\ 

\subsubsection{Description of the "30 Parameter Cost Model"}

A “30-parameter model” was developed by the ITF sub-committee on cost and schedule and applied to all colliders and their various respective energies. This model was used, along with several 3 parameter models, to reconstruct, extrapolate, and compare the submitted costs of the various collider proponents. 

Some of the submitted projected costs by the collider proponents were the results of detailed studies leading to a CDR and were very complete (e.g. ILC). Some proponent project cost projections represent very early calculations with uneven inputs. Some proponent estimates had missing sub-systems (several examples of missing sections are $e^+$ source, beam injector, damping rings, beam transport, IR BDS, IR FF). Some proponent cost estimates were for one center-of-mass energy only and had to be extrapolated to other $E_{\rm CM}$. Several projects submitted no cost estimates.

The “30-parameter model” in conjunction with the 3 parameter models were developed to alleviate many of these problems and to put all the collider costs, covering all energies, on a level cost-comparison field. The general assumptions of the 30-parameter model will be discussed first then each of the parameters.

The general assumptions for the 30-parameter model are:\\
A)	Each collider was divided into the “main collider” and the “injector+power-drivers+particle-sources”. \\
a.	The main collider encompasses all components and tunnels needed to get the beam from the particle sources, accelerated, collided, and then transported to the beam dump after the collision point (or to be stored or recirculated). \\
b.	The “injector+beam-drivers+particle-sources” include the costs for $e^-$ or $p$ sources, $e^+$ production, $\mu$ production, RF systems to accelerate the injected beams, damping rings, injector klystrons, modulators, high-power beams sources for acceleration, drive-beam transport, drive-beam dumps ($\sim$300 kW), drive-laser power sources, laser-transport, drive-laser dumps ($\sim$300 kW), and associated tunnels, shafts, and power conversion.\\
B)	All colliders are assumed to have green field sites including all new $hh$ colliders. The exceptions are $eh$ colliders where the $e^-$ accelerator is often added to an already-built $hh$ collider. A few colliders located near an existing laboratory site will reused some accelerator equipment (e.g., FNAL or CERN).\\
C)	The cost calculations for the same exact type of components for the main collider and, separately, for the injector+power-drivers+particle-sources will use the same scaling coefficients.\\
D)	The cost calculation for all colliders use the same coefficients for similar technical items (e.g., magnets, klystrons, lasers, cryomodules, or cryo-plants) independent of geographical region.\\
E)	For the same collider but with a higher energy, the main collider costs often increases, but sometimes the injector costs do not (e.g., ILC) (unless of course more or fewer particles are needed, more laser drivers are needed, more beam drivers are needed, or the repetition rate changes).\\
F)	Since the era of airplane manufacturing in the 1940s, it is well known that the cost of building technical components become cheaper if more components are made. The 30-parameter model uses a number scaling for each accelerator parameter depending on the initial costs and the number of units to be made. The scaled cost $c(n)$ for the nth constructed unit is $c(n)=c(1)\cdot/n^b$ where $c(1)$ is the cost of the first unit and $b$ is the “cost reduction coefficient” typically in the range 0.2 to 0.4 depending several factors including staffing costs per unit, technical complexity, and needed raw materials. To build, say, $n$ units, one integrates this equation from 1 to n obtaining the {\it “total cost  of $n$ units”} = $c(1)\cdot n^{(1-b)}/(1-b)$. The expected cost per unit shrinks faster with a larger $b$ factor. The cost reduction coefficients for accelerator components are taken from industry for off-the-shelf equipment or from LHC, XFEL, LCLS-II, PIP-II, or SwissFEL experience for specific items such as superconducting magnets, CuRF or SCRF. Of note, the equation for "Total cost of $n$ units" is an approximation and is very accurate for $n$ greater than about 10 but can be up to about 10 percent high for smaller $n$. \\
G)	For positrons the ILC $e^+$ source (guns, RF and damping ring) making about 6 kHz of positron bunches is well studied and documented. If a proposed collider did not specify its chosen positron source, then the technical scope and cost of the ILC positron source is used, scaled to the new collider’s repetition rate and particles per bunch.\\
H)	The resulting cost coefficients and models developed for this exercise were used to estimate the costs of six existing but recent accelerators or under construction: XFEL, LHC, Swiss-FEL, NSLS-II, LCLS-II+HE, and PIP-II covering a broad range of technologies and energies. The resulting cost estimates were compared to the known costs. The estimates and the reals costs were within $\pm$20\% setting a scale for the expected projection uncertainties.\\
I)	The beam collisions parameters and generation rates for the various colliders are listed in Appendices, Sec. \ref{sec:Appendices}. The cost technical risks and cost production risks are discussed in Section 5.2.3.\\
J)	The length of new accelerators for a collider often do not match the length of new tunnels for that collider as multiple accelerators, rings, or transport lines share the same tunnels.\\

\subsubsection{Parameters for cost model} 
\label{30parametermodel}

\begin{table}[htbp]
  \centering
  
    \begin{tabular}{|l|l|l|}
        \hline
System & Item & Unit \\
    \hline
Civil Infrastructure &  Length of new tunnels & km \\
Civil Infrastructure &  Length of reused tunnels & km \\
Power Infrastructure & Total wall plug accelerator power & MW \\
Vacuum Systems & Length of new accelerators & km \\
Vacuum Systems & Length of reused accelerators & km \\
Vacuum Systems & High power vacuum chamber length & km \\
Vacuum Systems & Low power vacuum chamber length & km \\
Physics Infrastructure & Number of new interaction regions & \# \\
Physics Infrastructure & Number of new particle sources & \# \\
Physics Infrastructure & Number of high power beam dumps & \# \\
Magnets & SC dipole field & T \\
Magnets & SC dipole length & km \\
Magnets & SC quadrupole field & T/m \\
Magnets & SC quadrupole length & km \\
Magnets & NC dipole field & T \\
Magnets & NC dipole length & km \\
Magnets & NC quadrupole field & T/m \\
Magnets & NC quadrupole length & km \\
RF & Cu RF voltage & GeV \\
RF & Cu RF length & km \\
RF & SC RF voltage & GeV \\
RF & SC RF length & km \\
Cryo & Number of SC RF cryomodules & \# \\
Cryo & Number of SC RF cryoplants & \# \\
Cryo & Number of SC RF cryomodules & \# \\
Cryo & Number of cryoplants for pulsed magnets and SCRF & \# \\
Plasma & Number of plasma acceleration cells & \# \\
Plasma & Number of short pulse drive-lasers (50 kHz) & \# \\
Plasma & Number of laser support buildings & \# \\
Design & CD2-3, R\&D, commissioning (25\% multiplier) & n/a \\
Controls & Diagnostics, cables, etc (30\% multiplier) & n/a \\
    \hline

    \end{tabular}%
\caption{Table of parameters for cost model}
    
  \label{ModelParams}%
\end{table}%

Table~\ref{ModelParams} shows the inputs for the cost model. Since many of the project input spread sheets were not fully completed or had uncertainties, the ITF needed to collect additional input from proponents, research published papers, and use comparable “component” costs. Of note, it was decided not to include contingency and escalation in the the 30-parameter model cost estimates.  The projected cost figures are in 2021 USD.

\subsubsection{Costing the collider projects}

As was discussed above, the projected costs of the various proposed colliders (often with several energies per collider) were estimated using the three "3 parameter" models (differentiated by different exponential scaling with component numbers) and by the "30 parameter model" (also with exponential scaling with component numbers). All collider costs are for a “green field” assuming no previous infrastructure, with the exceptions of the CERN related proposals.

Discussed for each collider and for various designated energies, there are "cost drivers" with present day prices, "cost risks" (driven by unproven technologies or manufacturing that needs improving), and, finally, possible "cost reduction with future R\&D". The last indicates that future R\&D may reduce manufacturing costs as well as proving the viability of new technologies. All of the individual colliders with their parameters are discussed in more detail in the Appendices. The collider descriptions are listed in the order of appearance in the executive summary Tables \ref{tab:ITFHiggs}-\ref{tab:ITFsiteFillers}.

Finally, the cost estimates by the proponents are included, if available, as submitted to the ITF study in the work sheets. The proponents in the ITF input form listed four costs (in 2021 USD), namely: 1) new accelerator systems, 2) new accelerator infrastructure, 3) new civil construction, and 4) explicit person-years of personnel. Here, the ITF simply added the first three costs then added the person-years at 0.2 MUSD/FTE-yr to make the proponent’s accelerator cost. Note, that the cost estimates submitted by the proponents came in various, and sometimes non-uniform, accounting that is often quite different from that of the ITF, and, therefore, are given here for reference only.

Each collider described below lists a design level from I to V. These design levels are described in Table \ref{tab:subgroupTRL}.\\

{\it Lepton colliders }\\

FCCee is a Level II (CDR report) proposed circular $e^+e^-$ collider (91 km) with about 8800 stored bunches per beam (1.4A at the $Z$). Costs were estimated by the ITF at CM energies of 0.25 and 0.37 TeV. The cost drivers are the tunnel, storage ring magnets, and full-energy top-up injector at 0.25 TeV then adding SCRF systems and cryo-plants at 0.37 TeV. R\&D needed to reduce technical cost risk is SCRF cryomodule HOMs. R\&D items to reduce production cost risks are higher efficiency klystrons and higher-gradient CW cryomodules at higher energy $E$. The proponents submitted FCCee estimated costs from 12.0 BUSD at 0.25 TeV and 13.3 BUSD at 0.37 TeV, both without personnel costs.

CEPC is a Level II (CDR report) proposed circular $e^+e^-$ collider (100 km) with about 12 thousand stored bunches per beam (0.8A at the $Z$). Costs were estimated by the ITF at c.m.e. of 0.25 and 0.37 TeV. The cost drivers are the tunnel, storage ring magnets, and full-energy top-up injector at 0.25 TeV then adding SCRF systems and cryo-plants at 0.37 TeV. R\&D needed to reduce technical cost risk is SCRF cryomodule HOMs. R\&D items to reduce production cost risks are higher efficiency klystrons and higher-gradient CW cryomodules at higher $E$. The proponents submitted a CEPC estimated cost of 4.6 BUSD at 0.25 TeV c.m.e. 

ILC is a Level I (TDR report) proposed linear $e^+e^-$ collider (20 to 67 km for 0.25 to 3 TeV c.m.e.) with about 6550 to 5000 bunch collisions per second. Costs were estimated by the ITF at c.m. energies of 0.25, 0.37, 0.5, 1, 2, and 3 TeV. The cost drivers are cryomodule production and the e+ and e- generation complex at 0.25 TeV plus the cryomodules with increased-gradient up to 3 TeV. R\&D items needed to reduce technical cost risks are SCRF cryomodule higher-gradients and efficient polarized e+ generation. The R\&D item to reduce production cost risk is reducing cryomodule assembly costs for large quantities. The proponents submitted an ILC estimated cost from 7.7 to 8.3 BUSD for 0.25 TeV.

CLIC is a Level II (CDR report) proposed linear $e^+e^-$ collider (11 to 50 km) up to about 16 kHz bunch collisions per second (0.38-3 TeV). Costs were estimated by the ITF at CM energies of 0.38 and 3 TeV. The cost drivers are X-band (12 GHz) Cu RF two-beam accelerators, nano-beam IR configuration, and 15.6 kHz $e^+$ production. R\&D items needed to reduce technical and cost risks are efficient Cu X-band RF two-beam accelerating and decelerating structures and the nano-beam IR. The R\&D item to reduce production cost risk is producing high electron drive beam currents and 16 kHz positron production. The proponents submitted an estimated cost of 12.4 BUSD to reach 3 TeV without personnel costs.

CCC is a Level III (substantial documentation) proposed linear $e^+e^-$ collider (8 to 38 km) up to about 9 to 16 kHz bunch collisions per second (0.25-3 TeV). Costs were estimated by the ITF at CM energies of 0.25, 0.55, 2, and 3 TeV. The cost drivers are C-band (5.7 GHz) Cu RF, nitrogen cooled cryomodules, and 9 to 16 kHz $e^+$ production. R\&D items needed to reduce technical and cost risks are efficient Cu c-band RF structures, LN$_2$ cryomodules, and positron production design. The R\&D item to reduce production cost risk is producing low cost cryomodules. The proponents submitted estimated costs of 4 BUSD for a  0.25 TeV collider and an additional 6 BUSD to reach 3 TeV, without personnel costs.

CERC is a Level III (substantial documentation) proposed “circular-ERL” $e^+e^-$ collider (100 km tunnel) colliding about 9 to 99 thousands bunches/second (0.09 to 0.6 TeV). Beams are accelerated to full energy in 4 turns, collided, and decelerated in four turns recovering the energy and particles which are then damped and topped-up, i.e. four turn ERLs. Costs were estimated by the ITF at CM energies of 0.25, 0.37, and 0.6 TeV. The cost drivers are the tunnel, NC ring magnets (16 turns), SCRF of up to 75 GeV per turn, and damping for up to 99 k positron bunches per second. R\&D needed to reduce technical risk is reducing the number of transport rings. R\&D items needed to reduce production cost risks are low cost ring magnets and low cost SCRF cryomodules. The proponents submitted a  CERC estimated costs from 11.5 BUSD at 0.25 TeV to 13.5 BUSD at 0.6 TeV. 

ReLiC is a Level V (unreviewed parameter table) proposed “ERL” SCRF linear $e^+e^-$ collider colliding about 3.7 to 12.6 MHz bunches/second (0.25 to 3 TeV). Beams are accelerated to full energy in opposing SC linacs, collided, and decelerated in the opposite side SCRF linacs recovering the energy and particles which are then damped and topped-up. Costs were estimated by the ITF at CM energies of 0.25 and 3 TeV. The cost drivers are the tunnel, SCRF cryomodules, and up to 12.6 MHz bunch damping. R\&D needed to reduce technical risk is increasing the gradient in the SCRF cryomodules. R\&D items needed to reduce production cost risks are low cost SCRF cryomodules and 12.6 MHz bunch damping design with good particle acceptance. The proponents submitted a ReLiC estimated costs of 22.5 BUSD at 0.25 TeV and 112.5 BUSD at 3 TeV.  

ERLC is a Level V (unreviewed parameter table) proposed “ERL” SCRF linear $e^+e^-$ collider colliding about 400 MHz bunches/second, moderated 2 seconds on and 4 seconds off (0.25 to 1 TeV). Beams are accelerated to full energy in opposing duel-bore SC linacs, collided, and decelerated in the opposite side duel-bore SCRF linacs, recovering the energy and particles which are then damped and topped-up.  The beams are accelerated in one of the two bores and decelerated in the other bore followed by the RF power being transferred from the decelerating bore to the accelerating bore to be reused with very little additional RF power needed. Costs were estimated by the ITF at CM energies of 0.25 and 3 TeV. The cost drivers are the tunnel, SCRF double-bore RF structures, newly designed cryomodules, and up to 400 MHz bunch damping. R\&D needed to reduce technical risk is a design for a duel-bore high-gradient SC cryomodule. R\&D items needed to reduce production cost risks are low cost new SCRF cryomodules and a 400 MHz bunch damping design with good particle acceptance. The proponents did not submit a cost estimate. 

XCC is a Level IV (limited documentation) proposed linear $e^-e^-$ or $\gamma \gamma$ collider (10 km) up to about 120 Hz “bunch” collisions per second (0.140 TeV). Costs were estimated by the ITF at CM energies of 0.140 TeV. The cost drivers are two 70 GeV $e^-$ Cu linacs, two FELs, two IR $e^-$ to $\gamma$ conversions and the $\gamma \gamma$ final focus IR including beam dumps. R\&D items needed to reduce technical and cost risks are IR $e^-$ to $\gamma$  conversion and the $\gamma \gamma$  final focus IR with beam dumps. The R\&D items to reduce production cost risks are producing inexpensive Cu $e^-$ linacs and two FELs. The Snowmass XCC white paper contains a cost estimate for the XCC of 2.3 BUSD, without personnel costs.

MC (3-14 TeV) is a Level III (substantial documentation)
proposed $\mu^+\mu^-$ collider, with an option for the highest c.m.e. option to be located in the 27 km CERN LEP tunnel. Costs were estimated by the ITF at CM energies of 0.13, 3 and 10-14 TeV. The cost drivers are the $\mu^+$ and $\mu^-$ sources using a proton driver, the emittance cooling channel, fast muon acceleration in a rapid cycling synchrotron RCS or equivalent, and a muon collider ring. Some CERN infrastructure can be reused. R\&D items needed to reduce technical cost risks are the cooling channel, RF acceleration, high-field SC dipoles, and $\sim$2T NC rapid-cycling dipoles with suitable pulsed power supplies. R\&D items to reduce the production cost risks are efficient proton drivers, cost effective RCS magnet power supplies, and radiation shielding. The proponents did not submit a cost estimate.

LWFA-LC is a Level IV (limited documentation) proposed linear $e^+e^-$ collider ($\sim$1 to $\sim10$ km) with about 50 kHz bunch collisions per second (1 to 15 TeV). Costs were estimated by the ITF at CM energies of 1, 3, and 15 TeV. The cost drivers are laser-power drivers, laser-plasma cells, and the 50 kHz $e^+$ bunch generation complex. The laser-power driver costs will dominate to 15 TeV. R\&D items needed to reduce technical cost risks are efficient fiber lasers [with high-pulse-energy (5 J), high-average power (300 kW),  and short-pulse ($\sim$40 fs)], 50 kHz $e^+$ bunch production, 50 kHz  plasma-cells, and also the final focus system at high energy. The laser driver above has been demonstrated at low repetition rates, but with low efficiency. New laser technologies show promise for delivering the required average power and high efficiency. Plasma cells have an upcoming demonstration of a few kHz. The R\&D item to reduce production cost risk is developing the high-power laser technology at extremely low costs compared to present solid-state systems. The proponents submitted a cost estimate only for the LWFA-LC laser drivers.

PWFA-LC is a Level IV (limited documentation) proposed linear $e^+e^-$ collider (5 to 20 km) up to about 10 kHz bunch collisions per second (1 TeV). Costs were estimated by the ITF at CM energies of 1, 3, and 15 TeV. The cost drivers are beam-power drivers, beam-plasma cells, the 10 kHz $e^+$ bunch generation complex, and also the final focus system at high energy. The beam-power driver costs will dominate to 15 TeV. R\&D items needed to reduce technical cost risks are efficient $e^-$ beam drivers, 10 kHz $e^+$ bunch production, 10 kHz  beam-plasma-cells, transport optics between cells, and the final focus system at high energy. The first three of these have been demonstrated at 30 to 120 Hz but not at a higher rate. The R\&D item to reduce production cost risk is producing $e^-$ bunch-drivers at a low cost. The proponents did not submit a cost estimate for PWFA-LC.

Structure-WFA-LC is a Level IV (limited documentation) proposed linear $e^+e^-$ collider (5 to 80 km) up to about 1 kHz bunch collisions per second (1-15 TeV). Costs were estimated by the ITF at CM energies of 1, 3, and 15 TeV. The cost drivers are $e^-$ drive-bunch generation, Cu wake structures, 1 kHz $e^+$ bunch generation complex, pulsed kicker magnets, and also the final focus system at high energy. R\&D items needed to reduce technical cost risks are efficient e- drive-bunch generation and Cu structure wake cells. The R\&D item to reduce production cost risk is producing e- bunch-drivers at a low cost. The proponents did not submit a cost estimate for Structure-WFA-LC.\\

{\it Energy Frontier Hadron Colliders\\}

FCChh is a (considered here a green field) Level II (CDR completed)
proposed $hh$ collider (91 km) to be located near CERN and Geneva with about 10,000 stored bunches per beam. Costs were estimated by the ITF at a CM energy of 100 TeV. The cost drivers are the tunnel length, 15 T SC dipole magnets, and the cryogenic plant. Some CERN infrastructure can be reused. R\&D needed to reduce technical cost risk is high-field SC dipoles. R\&D item to reduce the production cost risk is the SC cold-masses. The proponents submitted an FCChh estimated costs of 23.9 BUSD for 100 TeV including the tunnel but without personnel costs.

SPPC is a (green field) Level III (substantial documentation) proposed $hh$ collider (100 km) with about 10,000 stored bunches per beam. Costs were estimated by the ITF at a CM energy of 125 TeV. The cost drivers are the tunnel length, high-field SC dipole magnets, and the cryogenic plant. R\&D needed to reduce technical cost risk is medium-high-field SC dipoles ($\sim$20 T). R\&D item to reduce the production cost risk is the SC cold-masses. The proponents did not submit a cost estimate.

Collider-in-the-Sea (C-Sea) is a Level V (unreviewed parameter table) proposed hh collider (2100 km) to be located underwater in the Gulf of Mexico with about 200 thousands stored bunches per beam. Costs were estimated by the ITF at a CM energy of 500 TeV. The cost drivers are the underwater “tunnel”, superferric SC magnets, the cryogenic plant, and a new p/h injector. R\&D items needed to reduce technical cost risk are low-field superferric SC dipoles and the $p/h$ source(s). R\&D items to reduce the production cost risk are the SC cold-masses and underwater water-tight aligned magnet “girders”. The proponents submitted a Collider-in-the-Sea estimated cost of 35.4 BUSD for 500 TeV including the tunnel.\\

{\it TeV lepton-hadron colliders\\}

LHeC is a Level II (CDR completed)
proposal for $e-h$ collider using the LHC (27 km) and a new ERL electron accelerator (10 km), producing collisions with 50 GeV on 7 TeV. The project assumes a working LHC and other CERN infrastructure. Costs were estimated by the ITF at collision energies of 0.05 TeV $\times$ 7 TeV. The cost drivers are the 50 GeV $e^-$ ERL and the modification of one interaction point.  R\&D item needed to reduce technical cost risk is the ERL SCRF system. R\&D items to reduce the production cost risk are the ERL turn-around magnet systems and SCRF cryomodule production. The proponents submitted a LHeC estimated cost of about 1.4 BUSD, without personnel costs.

FCCeh is a Level II (CDR completed)
proposal for $e-h$ collider using the FCChh (91 km) and a new ERL $e^-$ accelerator (11 km), producing collisions with 60 GeV on 50 TeV. The project assumes a working FCChh and other CERN infrastructure. Costs were estimated by the ITF at collision energies of 0.06 TeV $\times$  50TeV. The cost drivers are the 60 GeV $e^-$ ERL and the modification of one interaction point.  R\&D item needed to reduce technical cost risk is the ERL SCRF system. R\&D items to reduce the production cost risk are the ERL turn-around magnet systems and SCRF cryomodule production. The proponents submitted a LHeC estimated cost of about 1.5 BUSD, without personnel costs.

SPPCep is a Level III (substantial documentation) proposed add-on circular $e-h$ collider (100 km) using SPPC ($h$) and CEPC ($e^-$ or $e^+$) with about 250 stored bunches per beam. The project assumes that the CEPC collider ring (not detectors) is kept installed and operational when SPPC is installed. Costs were estimated by the ITF at collision energies of 0.12 TeV $\times$ 62.5 TeV. The cost driver is the IR reconfiguration involving several km of tunnel. The related cost risks are low. The proponents did not submit a cost estimate for SPPC-eh.\\

{\it FNAL collider proposals\\}

FNAL-HELEN (Higgs-Energy LEptoN) is a Level IV (limited documentation) proposed linear $e^+e^-$ collider (7-12 km) with about 6600 bunch collisions per second. Costs were estimated by the ITF at a CM energy of 0.25 TeV. The cost drivers are high-gradient travelling wave (TW) SCRF cryomodule production and the $e^+$ and $e^-$ generation complex. R\&D items needed to reduce technical cost risks are “demonstrated” TW SCRF cryomodules at high-gradients and efficient $e^+$ generation. The R\&D item to reduce production cost risk is reducing cryomodule assembly costs for large quantities. The proponents submitted a FNAL-HELEN cost estimate to be 87.5\% of the ILC-250 construction cost.

FNAL-ee is a Level IV (limited documentation) proposed circular $e^+e^-$ collider (16 km) with 2 stored bunches per beam. Costs were estimated by the ITF at CM energies of 0.25 TeV. The cost drivers are the tunnel, storage ring magnets, full-energy top-up injector at 0.25 TeV, and SCRF systems and cryo-plants to support 0.25 TeV. R\&D items needed to reduce technical cost risks are SCRF cryomodule HOMs and medium-low-field ring magnets. R\&D items to reduce production cost risks are higher efficiency klystrons and higher-gradient CW cryomodules at this smaller radius ring. The proponents submitted a FNAL-ee estimated cost of 6 BUSD.

FNAL MC is a Level IV (limited documentation) proposed $\mu^+ \mu^-$ collider (up to 16 km) to be located on the FNAL site. Costs were estimated by the ITF at CM energies of 0.25, 3, and 6 TeV. The cost drivers are the  $\mu^+$ and  $\mu^-$  sources using a proton driver, the emittance cooling channel, fast muon acceleration in a rapid cycling synchrotron RCS or equivalent, and the muon collider ring. Some FNAL infrastructure can be reused. R\&D items needed to reduce technical cost risks are the cooling channel, fast RF acceleration, high-field SC and $\sim$2T NC rapid-cycling dipoles. R\&D items to reduce the production cost risks are efficient $p$ drivers, cost effective RCS magnet power supplies, and radiation shielding. The proponents did not submit a cost estimate.

FNAL-hh is a Level V (unreviewed parameter table) proposed $hh$ circular collider (16 km) to be located on the FNAL site about 2000 stored bunches per beam. Costs were estimated by the ITF at a CM energy of 24 TeV. The cost drivers are the very high field SC dipole magnets and the cryogenic plant. Some FNAL infrastructure can be reused. R\&D needed to reduce technical cost risk is high-field ($\sim$25 T) SC dipoles. R\&D item to reduce the production cost risk is the SC high-field-magnet cold-masses. The proponents did not submit a cost estimate.

\subsubsection{Cost of labor}
In order to treat the collider projects on a comparable basis, the estimates of construction cost for the collider projects follow the “{\it Value + Explicit Labour}” methodology, developed for large international projects such as the ILC and CLIC. “{\it Value}” is defined as the lowest reasonable estimate of the procurement cost of a component or system with the required specification and in the appropriate quantity and schedule, based on production costs in a major industrial nation. Here it is expressed in current USD.
\begin{figure}
\begin{center}
\includegraphics[width=0.75\textwidth]{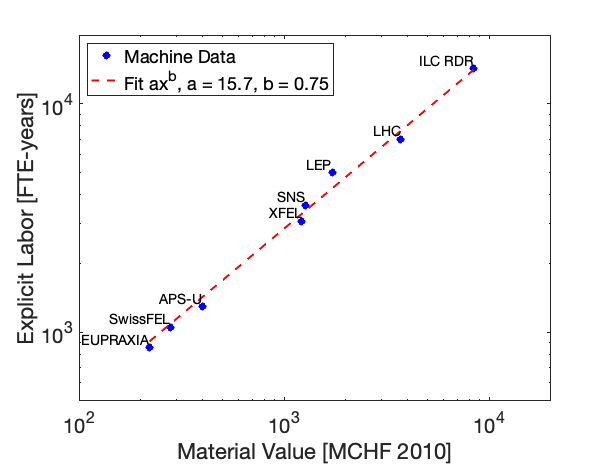}
\caption{Explicit labor for several large accelerator projects vs. project value. }
\label{figLabor}
\end{center}
\end{figure}
“{\it Explicit Labour}” may be provided by the collaborating laboratories and institutions, or may be purchased from industrial firms. This is to be distinguished from a company’s “implicit” labour associated with the industrial production of components and contained (implicitly) within the purchase price. The implicit labour is included in the value part of the estimate. Explicit Labour is expressed in Full-Time-Equivalent-years (FTE-years).
Note that the cost estimates do not include either contingency or provision for escalation over the construction period. To obtain full costs, the $Explicit \, Labour$ requirements are converted from FTE-years into USD, on the basis of an average cost of about 200k\ USD/FTE-year (averaged over several laboratories' staffing costs).

In case the project proposals do not quote $Explicit \, Labour$ for construction, we have estimated it from the Value, based on a correlation developed for existing colliders (LEP, LHC, SNS, European-XFEL, SwissFEL) and some proposals containing detailed estimates (ILC, APS-U, Eupraxia). See Fig. 5. The correlation, established over one and a half orders of magnitude, shows some economy of scale for large projects and suggests 
\begin{equation}
Explicit \, Labor  = 15.7 \cdot (Value)^{0.75},
\label{ModelLabor}    
\end{equation}
with $Explicit \, Labour$ in FTE-years and $Value$ in MCHF of 2010. \\

One should note that for very large projects, the $Explicit \, Labor$ requirement may exceed the current regional or global pool of available highly qualified accelerator workforce. As the result, the project timeline will need to be extended either to assure effective spending of the project funds with available experts or to create (attract or train) the team adequate to the task (see also below in Sec.\ref{ProjectTime}). \\

For very large projects where the projected costs are dominated by a few large number of identical components, then the exponent in the explicit labor calculation may be in the range 0.5 to 0.7 where oversight and testing that will be needed will be more efficient.

\subsubsection{Cost of magnets: SC and NC}

Several future colliders rely on extensive use of high-field superconducting magnets ranging from 3.2 T NbTi dipoles for the "Collider-in-the-Sea" project, to state-of-the art Nb$_3$Sn magnets with 12-15 T  field for all Muon Collider variants and 16 T for the FCChh, and 20 T IBS (iron-based superconductor) magnets for the SPPC.  In addition, some projects need "beyond-state-of-the-art" HTS magnets: few of them with fields exceeding 20 T for the muon collider target and final ionization cooling solenoids, many kilometers of such with 24-25 T fields for the 16 km circumference Fermilab site-filler proton-proton collider proposal. 

The basis of the superconducting magnet cost analysis are known costs of 
the NbTi magnets built for HERA (5.3 T), RHIC (3.5 T) and LHC(8.3 T). The latter was 1 MCHF per 15-m long double aperture dipole in 2000, with a share of approximately 1/3 for the conductor, 1/3 for the structure and 1/3 for the assembly \cite{rossi2015manufacturing}. "Aspirational" cost model of the high-field Nb$_3$Sn magnets for FCChh \cite{tommasini2017considerations} results in the cost target of 2 MCHF per dipole, and combined with the NbTi magnet data, the following equation can be derived for the total collider magnet cost: 
\begin{equation}
Cost_{\rm MAG}  \approx C_{\rm MAG} \cdot (Total \, Magnet \, Length) \cdot B^{0.8},  
\label{ModelMAG}    
\end{equation}
where the $Total \, Magnet \, Length$ doubles for two-aperture-magnets,  $C_{\rm MAG}$ is the cost coefficient and $B$ is the maximum field. Note that the continuity of cost with performance expressed by this formula does not translate the present discontinuities at the change of technology: It can only be considered valid for "aspirational" costs, once all R\&D has been successfully performed to attain performance and cost reduction. For example, in the current state of the Nb$_3$Sn magnet technology, represented by 11 T dipoles and 12 T (pole field) interaction region quadrupoles developed for the HL-LHC project, the cost per km is about twice the aspirational value Eq.(\ref{ModelMAG}) \cite{apollinari16T}. The present-day cost estimates for the much higher field HTS magnets are at least another factor of two higher. 
    
Normal conducting magnets are "off-the-shelf" items easily available from industry nowadays and expected to be routinely used in essentially all future colliders. While for most of the future projects they are "secondary" and employed in injectors, transport beamlines or such, few projects anticipate large scale use of them. The FCCee and CEPC projects need fast electron/positron booster rings in the same 91-100 km tunnels with the collider rings - and all of those are based on the NC magnets.  For example, about 2900 twin aperture dipoles: 2800 twin aperture quads, and 4700 single aperture sextupoles are required in the FCCee, totaling some 80 km of the NC magnet length. These magnets, however, operate well below the saturation of iron and call for original, cost-saving designs. Despite other alternatives potentially available, muon collider design at present assumes RCSs for very fast acceleration of muons to the top energies, that requires some 12-16 km of pulsed $\sim$2 T NC magnets for a 10 TeV machine. A potentially more economical option of using HTS-based pulsed magnets for fast acceleration of muons \cite{piekarz2022record} has been taken into account only as a possible aspirational target (i.e., not as a baseline assumption). 

Main items to account in the NC magnet cost  include production specific tooling, cost of materials (steel sheets/laminations and copper conductor), 
yoke manufacturing, and coil manufacturing. Usually, the cost of manufacturing is greater than that of laminations and conductor. The ratio of the latter two (costs of “yoke” and “coils”) is usually about 1:1 for high field magnets ($B \sim 1$ T or above), and while the "yoke" part scales with the aperture width, the "coil" part is $\propto B \times$(gap). Typically, the cost depends on length, aperture, maximum magnetic field and type of magnets - dipoles, quadrupoles, sextupoles, correctors, etc. Our model is built on the base of known or well-estimated costs for large- and medium-size circular accelerators, including 105 GeV LEP ring (some 20 km of magnets, 100 mm gap, 0.13T maximum dipole field), 10-18 GeV EIC-Electron Storage Ring ($\sim$2.1 km, 36 mm bore,  0.25-0.45 T), 4 GeV SuperKEKB-LER (0.58 km, 110 mm, 0.3 T) and 3 GeV NSLS-II-Booster ring (0.14 km, 24 mm, 1.1 T) and looks functionally similar to Eq.(\ref{ModelMAG}): $Cost_{\rm NC \, MAG}  \approx C_{\rm NC \, MAG} \cdot (Total \, Magnet \, Length)^{0.78}$, where coefficient $C_{\rm NC \, MAG}$ is taken to be twice as high for high-field magnets (like in the Muon Colliders) than for relatively low field magnets (as in the FCCee and CEPC projects). 

\subsubsection{Cost of RF: SC and NC}

The RF costs are related to the technology used (NC or SC), total acceleration needed (GeV), the repetition rate, and the time duration of the RF (pulsed or CW). \\

Most of the NC RF used for the proposed future colliders or their drivers  uses either "S (3GHz) or C (6GHz)" band and pulse rates up to 200 Hz. These "relatively standard" Cu linacs have costs that are well known (e.g. SACLA and SwissFEL), and are used in the 30-parameter model (parameters \#19-20). \\

The superconducting RF systems needed for future colliders have more variations. A SC linac costs more if the linac is CW because more RF power is needed, the cryogenic heat load is more, and the acceleration gradients may be more expensive to maintain. The Task Force looked at XFEL and ILC for actual costs related to a pulsed SC linac,  high gradient pulsed cryo-modules, RF klystron drivers, and the needed cryo-plants and cryogenic transport. The Task Force looked at LCLS-II-HE for actual costs for a CW linac, high gradient CW cryo-modules, with solid-state-amplifiers RF drivers, and CW related cryo-plants and cryogenic transport. \\

Estimating the cost of SCRF cryo-modules has been an ongoing discussion in the accelerator field. XFEL produced 101 cryo-modules. LCLS-II-HE has produced about 35 cryo-modules with 25 more in construction. The ILC has made many pre-production cryo-modules. If approved, the ILC (at 0.25 TeV) and its upgrades (up to 3 TeV) will make between 1000 and 6000 cryo-modules. All these units are about 12 m long. These produced cryomodules cost about 4 MUSD per cryomodule for about 50 units, 2 MUSD per cryomodule for about 100 units, and is projected to cost about 1 MUSD (2021) per cryomodule for 6000 units.

\subsubsection{Cost of civil construction}
Large collider projects require extensive civil works underground (accelerator tunnels, klystron and power converter galleries, stub tunnels for equipment, access and service shafts) as well as at ground level (buildings on technical sites). Shared infrastructure, for example with an academic campus, such as access roads, usually not charged to the project construction budget, is not considered here.

Although the cost of underground works, highly dependent upon geotechnical aspects and tunnelling techniques, may show large regional variations, we use a simplified approach in which the cost of all civil works is proportional to the total length of new tunnels to be built, with a coefficient of proportionality including all underground and ground level construction. The cost data for construction of the existing LEP/LHC tunnel (built in the 1980s) were considered to be too old. The coefficients used here were derived from the detailed costs of civil works for projects such as NSLS-II, XFEL, and SwissFEL and engineering studies for ILC, CLIC, and FCC.

\subsubsection{Cost of diagnostics, vacuum, and power supplies}
Large future colliders will have many standard sub-systems which are similar. All colliders will need diagnostics like beam position monitors, beam size monitors, charge monitors, bunch length monitors, beam loss monitors, and beam feedbacks. In general, these systems are lumped into the 30th parameter (controls, diagnostics, cables, survey, installation) of the 30-parameter cost model and scaled by length. However, almost every future collider will need specialized diagnostics depending on the beam parameters and usually around the interaction region. These special diagnostics are included in the interaction region costs. 

In the 30-parameter cost model there are two vacuum chamber cost coefficients depending on whether the chambers have heavy power absorption requirements or low. The vacuum chambers of some high current lepton storage rings and transport lines need to absorb many 10s of watts of power per cm requiring (water) cooling continuously along its length. Whereas, some vacuum chambers need little or no cooling per meter of length. In the 30-parameter model the highly powered chambers are costed at 5 times more per meter than the low power chambers. No cost was allocated to superconducting sections of the accelerator, since that cost is assumed to be included in the magnet or SRF system cost.

Power supplies, in general, are related to magnet fields and lengths and are included in the cost for the magnets. However, there can be specialty magnets in a collider, for example fast ramping NC magnets (muon colliders), fast kicker magnets (ILC damping rings), or fast diversion kickers in a SWLC that can be quite expensive. These specialty magnets are included in the power and magnet considerations. \\

\subsubsection{Cost of advanced accelerators: plasma, beams, lasers, structures}

Advanced colliders, such as those based on wakefield acceleration (LWFA-LC, PWFA-LC, and SWFA-LC) employ technologies which are not yet fully developed (see Sec.\ref{sec:TRL}) and making reliable cost estimates is more difficult. The ITF approach, therefore, was to come up with a range of estimates where the highest values come from operating test facilities and the lowest one reflect anticipated  advances and cost goals which proponents formulate on the basis of the current trends in corresponding novel technologies. It has to be noted, that for the more conventional accelerator sub-systems such as beam sources, damping rings, vacuum systems, final focus systems (except at CM energy >10TeV), beam transport lines, diagnostics and controls, the cost can be estimated for ACC colliders as reliably as for more traditional or well studied colliders (take CLIC as an example) and, generally, are a significant part of the cost of advanced accelerator colliders. \\

For the costs of beam-driven plasma wake accelerators for a TeV level lepton collider (PWFA-LC), the main concerning issues are the drive electron beam, the plasma cells, the transport optics of the colliding beam between plasma cells, and the energy transfer from drive beam to colliding beam. The electron drive bunches can already be produced at present using a conventional copper RF linac and a storage ring stacking system. A few high-current ring issues need to be verified. A CW SC linac (with well-known cost like LCLS-II) could also produce the needed electron bunches, but a demonstration test would be desirable. The plasma cells have already been built for 10 Hz operation. These cells to not have solid materials near the drive beam bunches (several cm length). This rate must be extended to >10 kHz to make a high luminosity collider. Flowing gas transversely in the meter-long cells will be the likely solution and recent tests at FLASHforward have demonstrated plasma recovery times compatible with high bunch rates. Beam transport optics between plasma cells are demanding to provide accurate matching and avoid emittance growth, but are in principle conventional with well-known costs. Beam tests at FACET-II and FLASHforward may demonstrate some of the anticipated options. Simulations have shown that a drive beam to plasma stored energy can have an efficiency of 70 percent and that the plasma energy to colliding bunch energy can also be 70 percent. Thus, the overall efficiency from drive beam to colliding beam can be about 50 percent if fully optimized. This efficiency can be tested at FACET-II. \\

For the costs of beam-driven structure wake accelerators for a TeV level lepton collider (SWFA-LC), the main concerning issues are the drive electron beam, the RF generating and transfer cells, and the energy transfer from drive beam to colliding beam. The drive beam is generated by a 1.3 GHz SW linac with a photo-cathode gun producing 32 bunch rains at 50 nC per bunch. These batches of 32 bunches make a short RF pulse in a low cost dielectric accelerator structure. This RF pulse is transferred to a main accelerating cell to accelerate the colliding bunches. This is similar to CLIC but at a higher RF frequency and with different RF cells. For the SWFA-LC, one drive linac feeds fifty 3 GeV modules, thus, making 150 GeV total acceleration. The drive beam to collider beam energy efficiency is in the range of 20 to 50 percent.Only about 1000 positron bunches are needed per second which is only 20 percent of an ILC positron source. R\&D is need to demonstrate the SW linac, the RF pulse generation, and transfer efficiencies. The cost can be relatively well estimated for most of the components.\\

For the costs of laser-driven wake accelerators for a TeV-level lepton collider (LWFA-LC), the main issues of concern are the drive lasers, the plasma cells, 50~kHz positron bunch production, and the energy transfer from drive laser to colliding beam. Present Ti:Sapphire amplifier technology can deliver 40 J of energy in 40 fs at 800 nm laser pulse at 1 Hz (for example, the BELLA facility at LBNL, with a cost of $O$(10 M\$) including auxiliary systems ca. 2007). The next generation of the BELLA experimental facility, "kBELLA" calls for J-class, kHz repetition rate  Ti:Sa laser based on present technology or new fiber-based laser technology that is under development. The recently published TDR of the EuPRAXIA \cite{assmann2020eupraxia} laser-driven site includes three Ti:Sa lasers (5J/20 fs, 15J/20 fs, 50J/50 fs, all 20 Hz, upgradable to 100 Hz), with the total cost -- established by the EuPRAXIA consortium members and European laser industry -- of 73.5 MEuro that can be broken down into 41.5 MEuro (materials) plus labor costs. Note that the laser drivers for an LWFA-LC will not be based on Ti:Sa laser technology owing to the fundamental quantum defect of Ti:Sa that restricts efficiency. \\

\begin{figure}
\begin{center}
\includegraphics[width=0.5\textwidth]{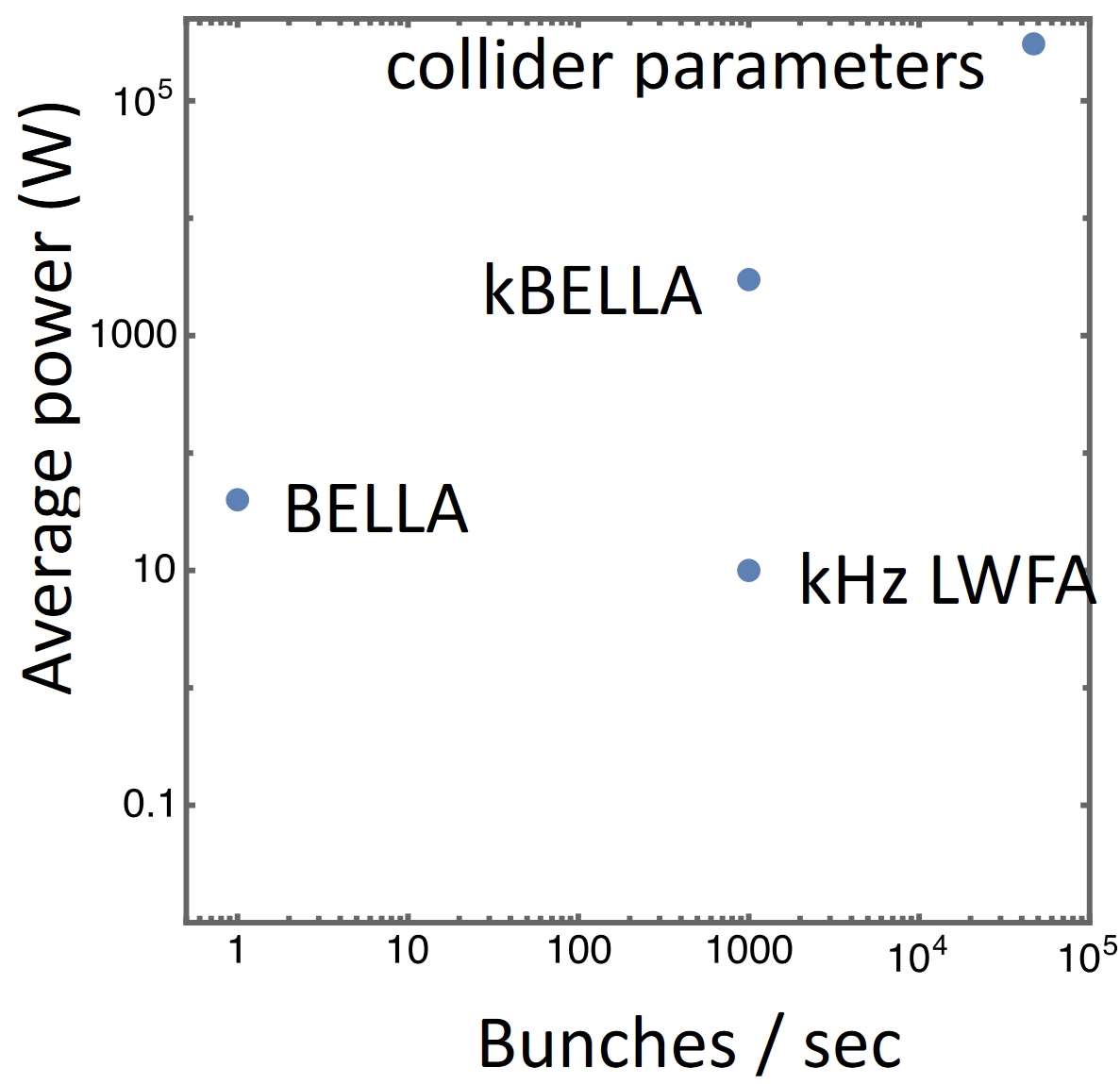}
\caption{Laser power and repetition rate for present (BELLA) and future (kBELLA) LWFA-LC facilities and those required for 1 TeV collider (from Ref.\cite{hoganAgora5}). }
\label{figLWFA}
\end{center}
\end{figure}

As indicated in Fig.\ref{figLWFA}, the present day laser systems for LWFA-LC are several orders of magnitude away from the collider requirements on the average power and repetition rate, not to mention needs to substantially improve stability, robustness, durability, reliability, cost and lifetime required. An additional difficulty in estimating the LWFA-LC collider cost is that the Ti:Sa lasers, presently used for R\&D, will certainly be substituted with another, currently emerging laser technology, such as either coherently combined Yb fiber lasers, big aperture Tm:YLF lasers, or Yb:YAG lasers. Therefore, as for other novel technologies, the ITF cost estimates for the LWFA-LCs present both the present-day costs of the Ti:Sa lasers used for R\&D prorated as $sqrt$(rep.rate, power, number of lasers), and the cost estimate using fiber laser technology of $O$(1B\$ to 6B\$) for all laser drivers of the 1TeV collider -- provided by the experts \cite{esareyLASERS} -- and scaled as $E_{cme}^{0.8}$ for higher energies. High repetition rate, feedback-stabilized laser development is under way, e.g. at the KALDERA project at DESY and BELLA/kBELLA at LBNL. \\

The laser plasma acceleration cells are under development to allow 50 kHz long term operation.  Operation of plasma cells will soon be demonstrated at about a kHz.  The 50 kHz positron bunch production rate is about 10 times that of the ILC and work is underway to develop a mechanism to produce this unprecedented rate.

\subsubsection{Simple "three-parameters" cost models}
For the purpose of comparison, we have decided to further update the simple "three-parameters" $\alpha \beta \gamma$-model. For that we have calculated the costs of all the projects under the ITF analysis by taking into account some 21\% inflation since 2014, i.e.: 
\begin{equation}
Cost \, Model_{2} = 1.21 \cdot TPC_{\alpha \beta \gamma},  
\label{ModelVS1}    
\end{equation}
where $TPC_{\alpha \beta \gamma}$ is given above in Eq.(\ref{ModelOld}). 

Yet another, but similar approach was to take into account not only the pre-2014 cost estimates but also more recent ones, such as those for HL-LHC, HE-LHC, several center-of-mass energy options of FCCee, FCChh including one with 6T SC magnets, 
LHeC, CLIC alternative with klystrons, new estimates for the ILC at 1 TeV and beyond, CEPC, NICA, XFEL, SwissFEL, 
FRIB, LCLS-II/HE, PIP-II, APS-U, Super-KEKB, and EIC. The analysis of those four dozens of inputs, similar to the one in Ref.\cite{shiltsev2014costmodel} resulted in the following "square-root-scaling" model:
\begin{equation}
Cost \, Model_{3} =  \Big[ 0.6 \sqrt{L \over {\rm 10\, km}} +
1.0 \sqrt{E_{\rm MAG} \over {\rm TeV}} + 4.1 \sqrt{E_{\rm RF} \over {\rm TeV}}  \Big] \times
F_{rest} F_{labor} F_{mngmt}F_{R\&D}F_{escal}F_{contg}. 
\label{ModelVS2}    
\end{equation}
Above, the units are B\$, $L$ is the length of required tunnels, $E_{\rm RF}$ and $E_{\rm MAG}$ are center-of-mass energies for projects based on RF acceleration (linacs) or magnets (rings), or, sometimes, both (like in the ERL-based $e^+e^-$ colliders in circular tunnels).  The multipliers account for the cost of the rest of the machine (e.g. injectors, power converters, diagnostics, safety, etc) $F_{rest} \approx 1.66$, the cost of labor $F_{labor} \approx 1.36$ that corresponds to about 1.8 FTEs per M\$ of construction cost, the cost of management $F_{mgmnt} \approx 1.10$, the cost of project pre-construction R\&D $F_{R\&D} \approx 1.075$,  escalation due to inflation over $\sim$10 years of construction $F_{escal} \approx 1.15$, and contingency $F_{contg} \approx 1.35$. The latter two factors are dropped to come up with the TPC without the escalation and contingency. 

\begin{figure}
\begin{center}
\includegraphics[width=0.8\textwidth]{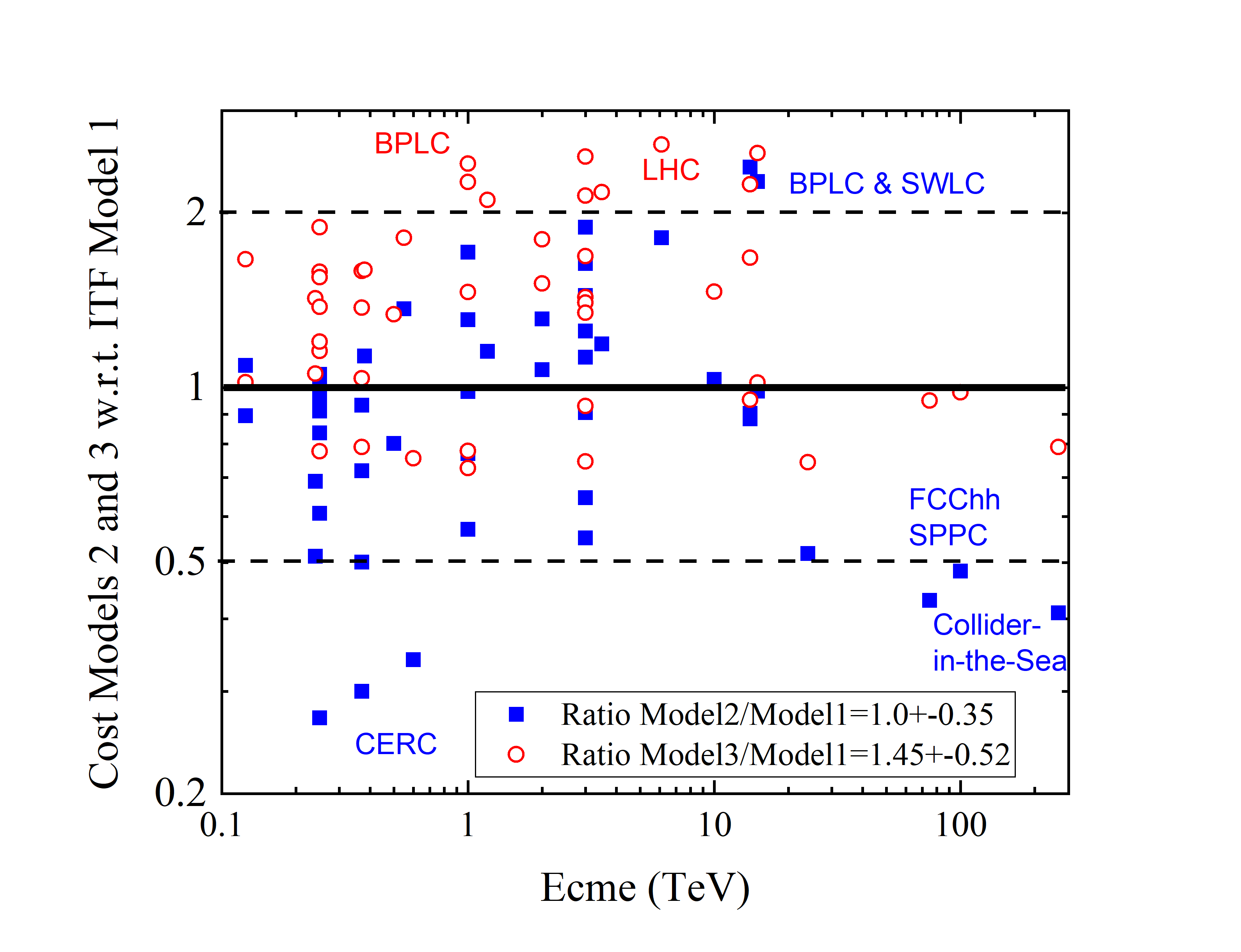}
\caption{Ratio of the  estimates of the three-parameters Models 2 and 3 to the 30-parameter cost Model 1 for future collider projects vs $E_{CM}$. All the values are for TPC without escalation and contingency, i.e., Model 2 estimates Eq.\ref{ModelVS1} are divided by $F_{contg} \approx 1.3$, and Model 3 values correspond to $F_{escal}=F_{contg}=1.0$ in Eq.\ref{ModelVS2}.  }
\label{figModels}
\end{center}
\end{figure}

Direct comparisons of the two "simple three-parameter" models with the "30-parameter" model ($Cost \, Model_{1}$) are shown in Fig.\ref{figModels}. One can see that while the $Cost \, Model_{2}$ is about the same as the 30-parameter cost model  within some 35\%, the $Cost \, Model_{3}$ is systematically higher by some 45\%. Notably, the few-parameters accounts well for only well established technologies (magnets, RF, civil construction, etc) and can substantially deviate from the ITF main model for the proposals based on novel technologies, such as CERC, "Collider-in-the-Sea", PWFA-LC and SWLC. 

\subsection{Summary on costs}

The projected costs of the various proposed colliders (with several energies per collider) were estimated with two "three-parameter" models (each with different exponential scaling factors) and by the "30 parameter model". A spreadsheet was made of these various cost estimates. The ITF Task Force then reviewed all the costs, the ones given by the proponents comparing them with those of the task force cost models. Using the combined judgement of the Task Force, a resulting range of costs for each collider and energy was determined. 

\begin{figure}
\begin{center}
\includegraphics[width=0.95\textwidth]{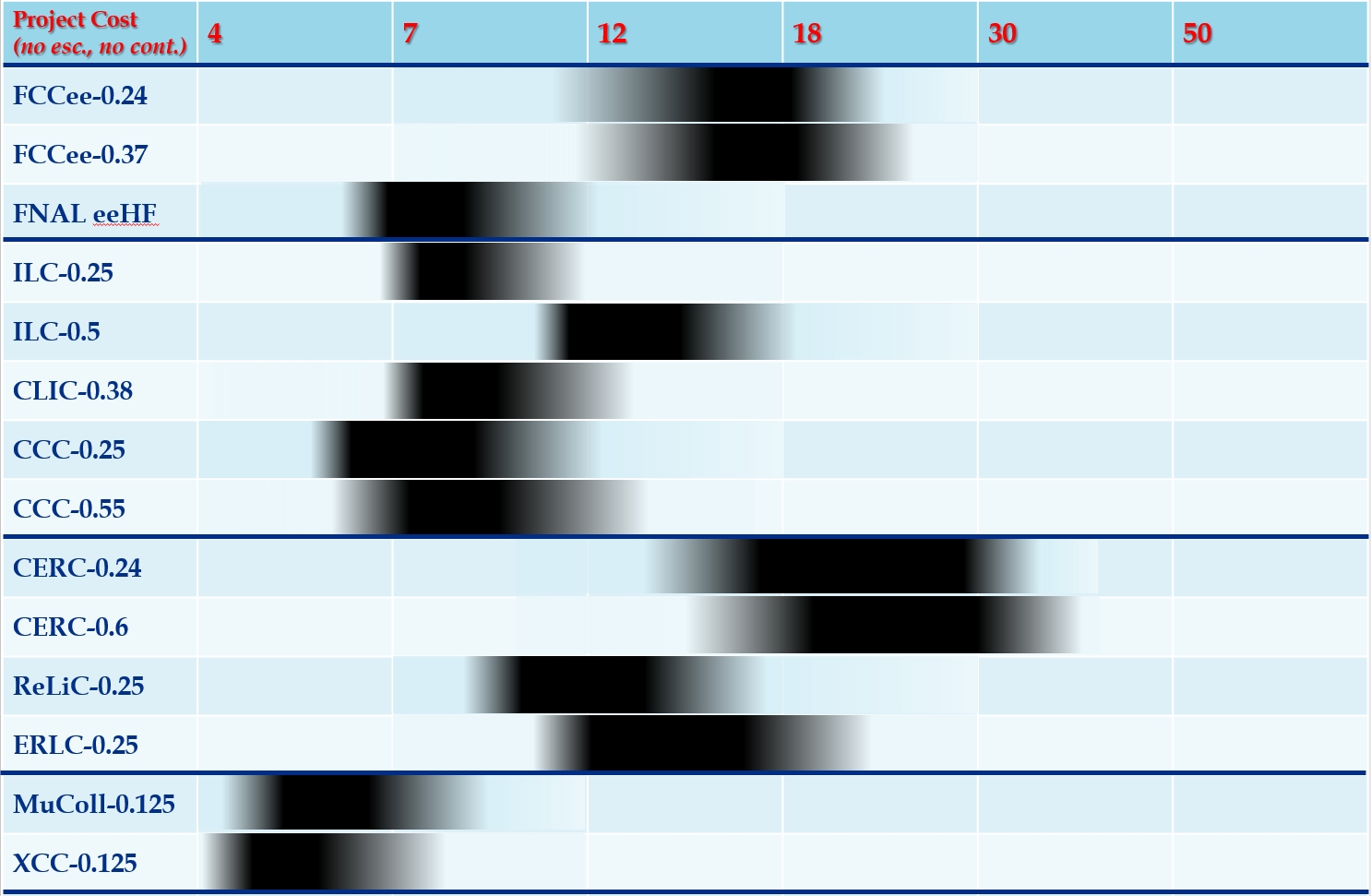}
\caption{The ITF cost model for the EW/Higgs factory proposals. Horizontal scale is approximately logarithmic for the project total cost in 2021 B\$ without contingency and escalation. Black horizontal bars with smeared ends indicate the cost estimate range for each machine.}
\label{figITFcost1}
\end{center}
\end{figure}

\begin{figure}
\begin{center}
\includegraphics[width=0.95\textwidth]{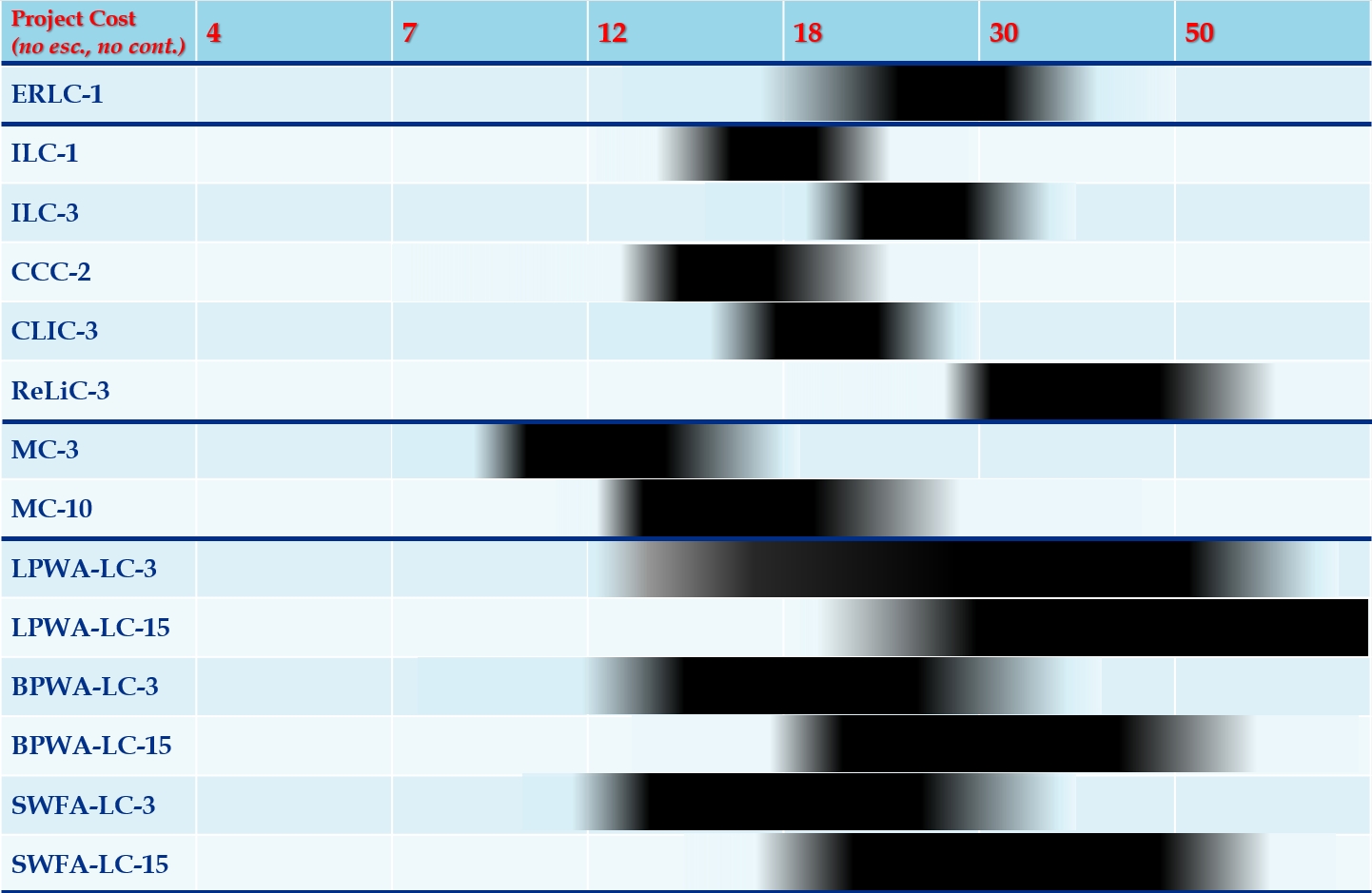}
\caption{The ITF cost model for the multi-TeV lepton collider proposals. Horizontal scale is approximately logarithmic for the project total cost in 2021 B\$ without contingency and escalation. Black horizontal bars with smeared ends indicate the cost estimate range for each machine.}
\label{figITFcost2}
\end{center}
\end{figure}

\begin{figure}
\begin{center}
\includegraphics[width=0.95\textwidth]{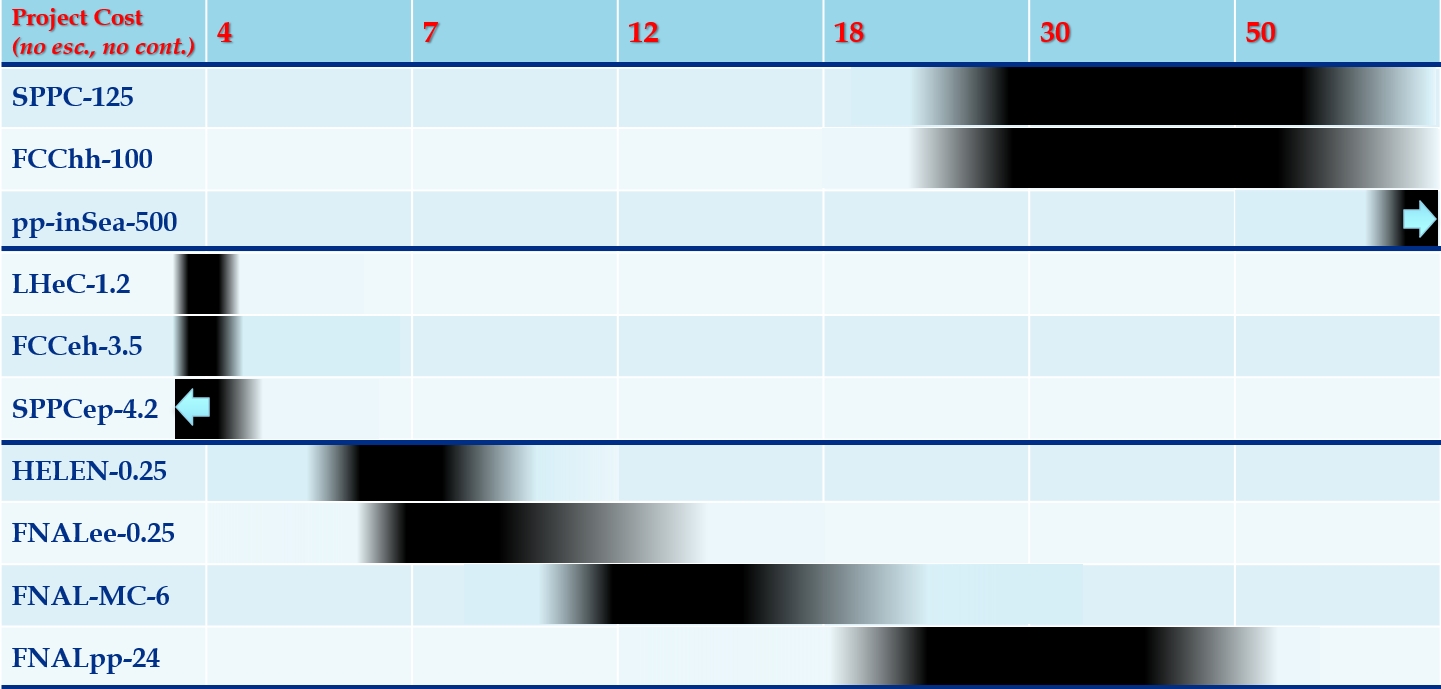}
\caption{The ITF cost model for the energy frontier hadron collider, electron-proton colliders (incremental cost from hadron collider only)  and for the proposed Fermilab site-filler colliders. Horizontal scale is approximately logarithmic for the project total cost in 2021 B\$ without contingency and escalation. Black horizontal bars with smeared ends are the cost estimate range for each machine. Right-arrow for the 500 TeV "Collider-in-the-Sea" indicates higher than 80B\$ cost.  Left-arrow for the electron-proton "SPPC-CEPC" collider concept indicates smaller than 4B\$ cost.}
\label{figITFcost3}
\end{center}
\end{figure}

 Figures \ref{figITFcost1}, \ref{figITFcost2}, and \ref{figITFcost3} summarize the ITF cost analysis for all future colliders, respectively for the EW/Higgs factory proposals, for multi-TeV lepton colliders, and for the energy frontier hadron and $ep$ colliders alongside with Fermilab site-filler proposals. The cost range of each machine was mostly defined by the 30-parameter model (\# 1) while taking predictions of simple models \#2 and \#3 into some account. The cost estimate range for each collider is indicated by a horizontal bar with smeared ends. The horizontal scale is approximately logarithmic for the project total cost without contingency and escalation (see  Sec.\ref{30parametermodel} above) with the marks approximately a factor of 1.6 from each other. The length of each bar reflects a combination of the cost model model uncertainties, differences between different models, spread of the cost parameters for not yet fully developed technologies ("aspirational" values usually correspond to lower cost bar ends, while "nowadays" estimates determine at the upper ends). Naturally, the ranges (bar lengths) of well developed projects, like ILC, CLIC, FCCee, CEPC, etc are smaller (shorter bars) than those based on less developed concepts and technologies. The extent of the smeared ("fuzzy") ends of the bars attempts to illustrate the probability of the lower cost estimates (usually smaller) and the upper cost range (usually larger).  

In somewhat reduced form, these cost estimates are also presented in the Executive Summary of this ITF Report - see Sec.\ref{sec:ES}. There, the summary tables \ref{tab:ITFHiggs},\ref{tab:ITFmultiTeV},\ref{tab:ITFfrontier},\ref{tab:ITFleptonhadron}, and \ref{tab:ITFsiteFillers} present the ITF estimates of the project costs in 2021 B\$ - without contingency and escalation, as described in Sec.\ref{30parametermodel} above, indicating one or multiple of the ranges <4B\$, 4-7B\$,  7-12B\$,  12-18B\$, 18-30B\$, 30-50B\$, and 50-80B\$.

\subsection{Construction timeline analysis and summary}
\label{ProjectTime}

\begin{figure}
\begin{center}
\includegraphics[width=0.85\textwidth]{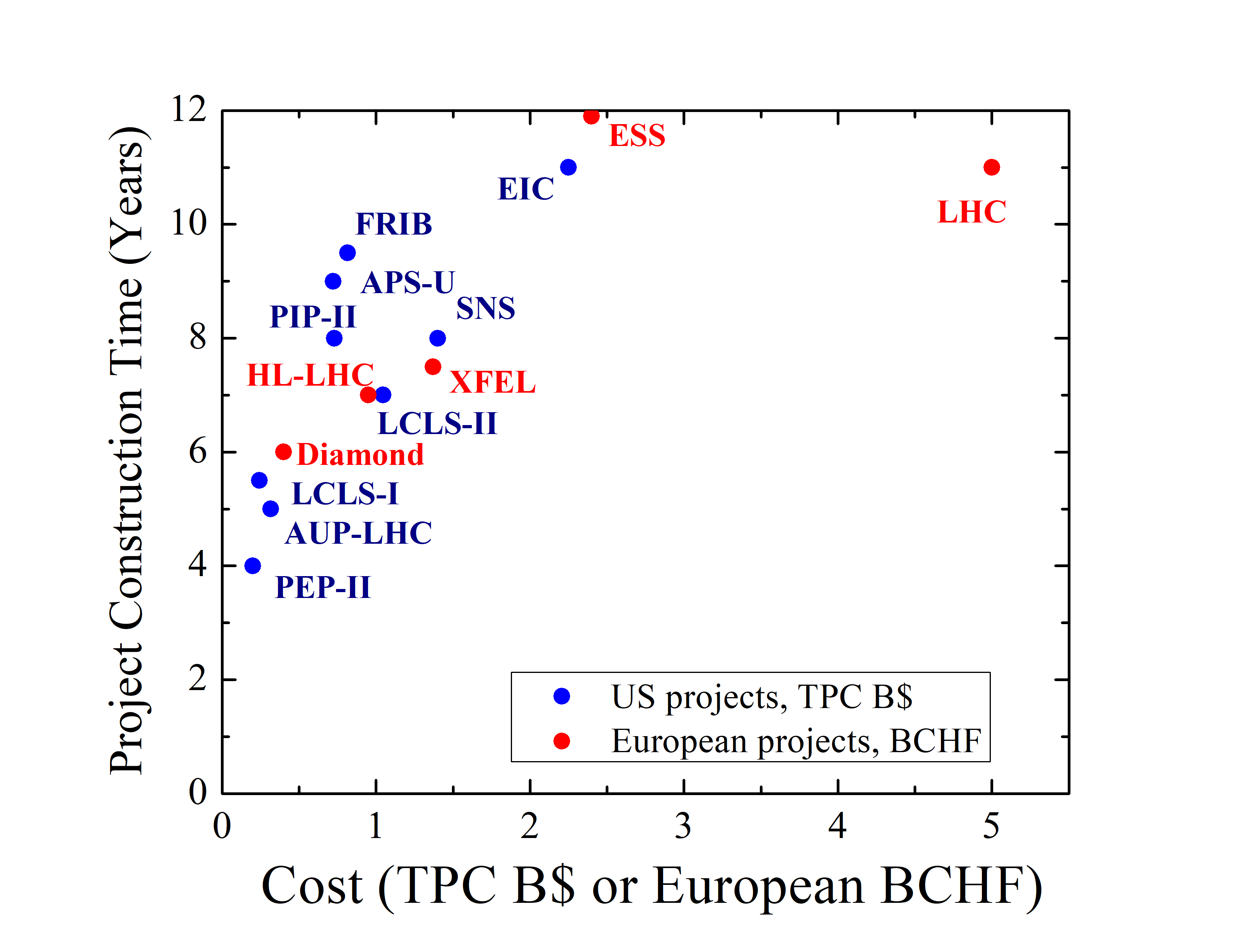}
\caption{Construction time for recent large accelerator projects in the US and Europe.}
\label{figITFConstrTime}
\end{center}
\end{figure}

Construction time, historically, has been one of most important aspects when it comes to comparative evaluation of future projects. Corresponding experience of recent large accelerator projects in the US and Europe -- see Fig.\ref{figITFConstrTime} -- tells us that larger projects usually take longer to construct. Of course, there many nuances to take into account starting with the time needed to establish the project (of any size), often-limited annual spending rate, availability of technical experts, limited pace of civil construction and fabrication of main components by existing industries, not-unusual financial and political hiccups, etc. 

The ITF approach was to address several main stages of the collider project timelines separately, namely, for each proposal estimate: \\

A) The time needed to carry out basic design and pre-project laboratory technical R\&D to the level of making a CDR available. \\

B) The time needed to design and engineer to the TDR level and the duration of industrialization with first articles needed to come up with a comprehensive TDR and any needed followup.\\

C) Duration of the construction itself, that corresponds to CD-4 in the US DOE project management system. \\

D) "Time to first physics" (years from present day to first data), which is based on combining the three previous factors. \\

The ITF judgments on the collider project duration had several assumptions: \\ 
1) 	We start with the durations submitted by the projects' proponents on "Preproject R\&D", "Design, Collaboration, and Industrialization", and "Construction time".	\\							
2)		All time durations start from now.	For example, "R\&D duration" starts now, not including past work.							\\
3)		Prototypes and demonstatrions are included in "Preproject R\&D".\\	
4) 		All projects have technically limited schedules.	\\
5)		All durations indicated are for a {\it green field} (stand alone) projects not built on prior   projects. (e.g. ILC (1 TeV) has no ILC (250), FCChh has no FCCee), etc. \\
6)		"Time to First Physics" is not just an addition of the three periods as  the stages of the projects can be partially done in parallel. \\

The results of the Task Force's timescale evaluations are shown in Table \ref{tab:constrtime} where all colliders including variants at several energies are included. The last four columns in Table \ref{tab:constrtime} summarize the ITF judgments on future collider timelines. Two of these columns - time to CDR and time to first physics -  are also reproduced in the Executive Summary of this ITF Report. The first three columns present these timescales
as submitted to the ITF by the project proponents. Not all the proponents submitted their respective timescales. These are marked as "tbd".

\begin{table}[htbp]
  \centering
  \caption{Summary of the ITF judgment on collider projects' R\&D duration, design and industrialization, construction, and combined time to first physics. The first three columns present these timescales as submitted to the ITF by the project proponents. The first group of rows are  Higgs and electroweak physics colliders, the second group are energy-frontier lepton colliders and the third group includes hh and eh colliders.}
    \begin{tabular}{l|ccc|cccc}
\hline
          &   Subm'd    & Subm'd      & Subm'd       & ITF   & ITF   & ITF   & ITF \\
    Collider &R\&D & Design  & Project & Judgement & Judgement & Judgement & Judgement \\
    Name  & Durat'n & to TDR & Constrn. & Duration & Design \& & Project  & Combined \\
    - c.m.e.  & to CDR & Durat'n & Time & Preproject & Industr'n & Constrn. & "Time to \\
     (TeV)     & (yrs) & (yrs) & (yrs) & R\&D  & Duration & Duration & the First \\
          &    &  &     & to CDR & to TDR & post CD4 & Physics" \\
\hline
    ILC-0.25 & 0     & 4     & 9     & 0-2 yrs & 3-5 yrs & 7-10 yrs & < 12 yrs \\
    ILC (6x lumi) & 10    & 5     & 10    & 3-5 yrs & 3-5 yrs & 7-10 yrs & 13-18 yrs \\
    CLIC-0.38 & 0     & 6     & 6     & 0-2 yrs & 3-5 yrs & 7-10 yrs & 13-18 yrs \\
    FCCee-0.36 & 0     & 6     & 8     & 0-2 yrs & 3-5 yrs & 7-10 yrs & 13-18 yrs \\
    CEPC-0.24  & 6     & 6     & 8     & 0-2 yrs & 3-5 yrs & 7-10 yrs & 13-18 yrs \\
    CCC-0.25   & 2-3   & 4-5   & 6-7   & 3-5 yrs & 3-5 yrs & 7-10 yrs & 13-18 yrs \\
    FNALee-0.24 & tbd   & tbd   & tbd   & 3-5 yrs & 3-5 yrs & 7-10 yrs & 13-18 yrs \\
    CERC-0.6  & 3     & 5     & 10    & 5-10 yrs & 3-5 yrs & 7-10 yrs & 19-24 yrs \\
    HELEN-0.25 & tbd   & tbd   & tbd   & 5-10 yrs & 5-10 yrs & 7-10 yrs & 19-24 yrs \\
    ReLiC-0.25 & 3     & 5     & 10    & 5-10 yrs & 5-10 yrs & 10-15 yrs & > 25 yrs \\
    ERLC-0.25  & 8     & 5     & 10    & 5-10 yrs & 5-10 yrs & 10-15 yrs & > 25 yrs \\
    MC-0.125 & 11    & 4     & tbd   & > 10 yrs & 5-10 yrs & 7-10 yrs & 19-24 yrs \\
    XCC-0.125   & 2-3   & 3-4   & 3-5   & 5-10 yrs & 3-5 yrs & 7-10 yrs & 19-24 yrs \\
    SWLC-0.25 & 8     & 5     & 10    & 5-10 yrs & 3-5 yrs & 7-10 yrs & 19-24 yrs \\
\hline
    ILC-1 & 10    & 5     & 5-10  & 5-10 yrs & 3-5 yrs & 10-15 yrs & 13-18 yrs \\
    ILC-2 & 10    & 5     & 5-10  & > 10 yrs & 3-5 yrs & 10-15 yrs & 19-24 yrs \\
    ILC-3 & 20    & 5     & 10    & > 10 yrs & 3-5 yrs & 10-15 yrs & 19-24 yrs \\
    CLIC-3  & 0     & 6     & 6     & 3-5 yrs & 3-5 yrs & 10-15 yrs & 19-24 yrs \\
    CCC-2   & 2-3   & 4-5   & 6-7   & 3-5 yrs & 3-5 yrs & 10-15 yrs & 19-24 yrs \\
    ReLiC-2 & 3     & 5     & 10    & 5-10 yrs & 5-10 yrs & 10-15 yrs & > 25 yrs \\
    MC-1.5 & 11    & 4     & tbd   & > 10 yrs & 5-10 yrs & 7-10 yrs & 19-24 yrs \\
    MC-3 & 11    & 4     & tbd   & > 10 yrs & 5-10 yrs & 7-10 yrs & 19-24 yrs \\
    MC-10 & 11    & 4     & tbd   & > 10 yrs & 5-10 yrs & 10-15 yrs & > 25 yrs \\
    MC-14 & 11    & 4     & tbd   & > 10 yrs & 5-10 yrs & 10-15 yrs & > 25 yrs \\
    PWFA-LC-1 & 15    & tbd   & tbd   & > 10 yrs & 5-10 yrs & 7-10 yrs & 19-24 yrs \\
    PWFA-LC-15 & 15    & tbd   & tbd   & > 10 yrs & 5-10 yrs & 10-15 yrs & > 25 yrs \\
    LWFA-LC-3 & 15    & tbd   & tbd   & > 10 yrs & > 10 yrs & 10-15 yrs & > 25 yrs \\
    LWFA-LC-15 & 15    & tbd   & tbd   & > 10 yrs & > 10 yrs & > 16 yrs & > 25 yrs \\
    SWFA-LC-1 & tbd   & tbd   & tbd   & > 10 yrs & 5-10 yrs & 7-10 yrs & 19-24 yrs \\
    SWFA-LC-15 & tbd   & tbd   & tbd   & > 10 yrs & 5-10 yrs & 10-15 yrs & > 25 yrs \\
\hline
    FCChh-100 & 2   & 20    & 15    & > 10 yrs & 5-10 yrs & 10-15 yrs & > 25 yrs \\
    SPPC-75  & 15    & 6     & 8     & > 10 yrs & 5-10 yrs & 10-15 yrs & > 25 yrs \\

    Coll.-Sea-500 & 10    & 6     & 6     & > 10 yrs & 5-10 yrs & > 16 yrs & > 25 yrs \\
    CEPC-SPPC & tbd   & tbd   & tbd   & 3-5 yrs & 3-5 yrs & < 6 yrs & > 25 yrs \\
    LHeC  & 0   & 5   & 5   & 0-2 yrs & 3-5 yrs & < 6 yrs & 13-18 yrs \\
    FCC-eh & 0   & 5   & 5   & 0-2 yrs & 3-5 yrs & < 6 yrs & > 25 yrs \\
\hline
    \end{tabular}%
\label{tab:constrtime}%
\end{table}%

%% file: Summary.tex
\section{Summary}
\label{sec:ES}

As part of the Snowmass'21 US HEP community strategic planning exercise, the Implementation Task Force (ITF) has compiled information for proposals for future colliders and evaluated these proposals with regard to technical risks, cost, schedule and environmental impact with the goal to address the key question: {\it "What kind of large-scale global accelerator project(s) can we envision undertaking to advance the frontiers in particle and accelerator physics, and to answer the fundamental questions of our field?"}. The ITF grouped future collider proposals into four categories that address similar physics:
\begin{itemize}
\item Higgs factory colliders with a typical CM energy of 250 GeV
\item High energy lepton colliders with up to 3 TeV CM energy
\item Lepton and hadron colliders with 10 TeV or higher parton CM energy
\item Lepton-hadron colliders
\end{itemize}
A separate group consists of versions of the proposals from these categories that could be located at Fermilab. Some collider concepts have entries in several categories. 

For each of the groups of proposals summary tables \ref{tab:ITFHiggs},\ref{tab:ITFmultiTeV},\ref{tab:ITFfrontier},\ref{tab:ITFleptonhadron}, and \ref{tab:ITFsiteFillers} list the main parameters along with four columns with a summary value for technical risk (years of pre-project R\&D needed), technically limited schedule (years until first physics), project costs (2021 B\$ without contingency and escalation), and environmental impact (most important impact is the estimated operating electric power consumption). The significant uncertainty in these values was addressed by giving a range where appropriate. The basis for these values are given in Sections \ref{sec:TRL}, \ref{sec:Power}, and \ref{sec:Cost} and more detailed information about the proposals are given in the Appendices \ref{sec:Appendices}.

The years of required pre-project R\&D is just one aspect of the technical risk, but it provides a relevant and comparable measure of the maturity of a proposal and an estimate of how much R\&D time is required before a proposal could be considered for a project start (CD0 in the US system). Pre-project R\&D includes both feasibility R\&D, R\&D to bring critical technologies to a technical readiness level (TRL) of 4-5, as well as necessary R\&D to reduce cost and electric power consumption. The extent of the cost and power consumption reduction R\&D is not well defined and it was assumed that it can be accomplished in parallel with the other pre-project R\&D. Nevertheless this R\&D is likely most important for the realization of any of these proposals. 

Note that by using the proponent-provided luminosity values ITF chose not to evaluate the risk of not achieving this level of performance directly as this would be beyond the manageable scope of the task force. However, performance risk was included in the evaluation of the technical readiness in Section \ref{sec:TRL}.

The time to first physics in a technically limited schedule is most useful to compare the scientific relevance of the proposals. It includes the pre-project R\&D, design, construction and commissioning of the facility.

The total project cost follows the US project accounting system {\it but without escalation and contingency.} Various models were used by ITF to estimate this cost and proponents were asked to also provide a cost estimate according to the guidance listed above. The ITF cost estimates use known costs of existing installations and reasonably expected costs for novel equipment. For future technologies, this cost estimate is likely very conservative and one should expect that the cost for these items can be greatly reduced, maybe by significant factors, through pre-project cost reduction R\&D.

Finally the electric power consumption is for a fully operational facility including power consumption of all necessary utilities. It is also important to consider luminosity per facility power consumption, a figure-of-merit that is shown in Figure~\ref{fig:lumi_pow}. A high figure-of-merit could achieve a certain amount of integrated luminosity faster and at a lower total energy consumption. However, for the summary tables we decided to list the electric power consumption since such large and expensive scientific facilities are expected to be in operation for at least two decades independent of whether a certain integrated luminosity goal was achieved in a shorter time.

We used the information of power consumption from the proponents if they provided it, otherwise we made a rough estimate ourselves as described in Section \ref{sec:Power}. For the future collider proposals that have not yet undergone detailed engineering studies the power consumption numbers should be considered “R\&D goals” with a considerable possibility that the actual power consumption is higher than listed in the tables. This is reflected in the luminosity per MW plot (Fig.\ref{fig:lumi_pow}) with a shaded area that extends below the proponent-provided values. Nevertheless, for all proposals pre-project R\&D focused on improving energy efficiency throughout the facility and on developing more energy efficient accelerator concepts, such as energy recovery technologies, has the potential to reduce the electric power consumption below the values listed in the tables.

Any of the future collider projects constitute one of, if not, the largest science facility in particle physics \cite{shiltsev2021modern}. The cost, the required resources and, maybe most importantly, the environmental impact in the form of large energy consumption will approach or exceed the limit of affordability. We urge to give high priority to the R\&D topics aimed at the reduction of the cost and the energy consumption of future collider projects.


\begin{table}[b!]
\begin{center}
\hspace*{-2em}
\begin{tabular}{| l | c | c | c | c | c | c |}
\hline
\hline
 Proposal Name & CM energy & Lum./IP & Years of & Years to  & Construction & Est. operating \\ 
 & nom. (range) & @ nom. CME & pre-project & first & cost range & electric power \\  
  & [TeV] & [10$^{34}$ cm$^{-2}$s$^{-1}$] & R\&D & physics & [2021 B\$] & [MW] \\  
\hline
FCC-ee$^{1,2}$ & 0.24 & 7.7 (28.9) & 0-2 & 13-18 & 12-18 & 290 \\  
 & (0.09-0.37) & & & & & \\
\hline 
CEPC$^{1,2}$ & 0.24 & 8.3 (16.6) & 0-2 & 13-18 & 12-18 & 340 \\  
 & (0.09-0.37) & & & & & \\
\hline 
ILC$^{3}$ - Higgs & 0.25 & 2.7  & 0-2 & <12 & 7-12 & 140 \\  
factory & (0.09-1) & & & & & \\
\hline 
 CLIC$^{3}$ - Higgs & 0.38 & 2.3  & 0-2 & 13-18 & 7-12 & 110 \\  
factory & (0.09-1) & & & & & \\
 \hline 
 CCC$^{3}$ (Cool & 0.25 & 1.3  & 3-5 & 13-18 & 7-12 & 150 \\  
Copper Collider) & (0.25-0.55) & & & & & \\
\hline 
CERC$^{3}$ (Circular& 0.24 & 78  & 5-10 & 19-24 & 12-30 & 90 \\  
ERL Collider) & (0.09-0.6) & & & & & \\
 \hline 
ReLiC$^{1,3}$ (Recycling & 0.24 & 165 (330)  & 5-10 & >25 & 7-18 & 315 \\  
Linear Collider) & (0.25-1) & & & & & \\
 \hline 
ERLC$^{3}$ (ERL & 0.24 & 90  & 5-10 & >25 & 12-18 & 250 \\  
linear collider) & (0.25-0.5) & & & & & \\
 \hline 
XCC (FEL-based & 0.125 & 0.1  & 5-10 & 19-24 & 4-7 & 90 \\  
$\gamma \gamma$ collider) & (0.125-0.14) & & & & & \\
 \hline 
Muon Collider & 0.13 & 0.01  & >10 & 19-24 & 4-7 & 200 \\  
Higgs Factory$^{3}$ & & & & & & \\
\hline 
\hline 
\end{tabular}
\end{center}
\caption{Main parameters of the submitted Higgs factory proposals. The cost range is for the single listed energy. The superscripts next to the name of the proposal in the first column indicate
(1) Facility is optimized for 2 IPs. Total peak luminosity for multiple IPs is given in parenthesis; 
(2) Energy calibration possible to 100 keV accuracy for $M_Z$ and 300 keV for $M_W$; 
(3) Collisions with longitudinally polarized lepton beams have substantially higher effective cross sections for certain processes} 
\label{tab:ITFHiggs}
\end{table}

\begin{table}
\begin{center}
\hspace*{-2em}
\begin{tabular}{| l | c | c | c | c | c | c |}
\hline
\hline
 Proposal Name & CM energy & Lum./IP & Years of & Years to  & Construction & Est. operating \\ 
 & nom. (range) & @ nom. CME & pre-project & first & cost range & electric power \\  
  & [TeV] & [10$^{34}$ cm$^{-2}$s$^{-1}$] & R\&D & physics & [2021 B\$] & [MW] \\   
\hline
High Energy ILC & 3 & 6.1 & 5-10 & 19-24 & 18-30 & $\sim$400 \\  
 & (1-3) & & & & & \\
\hline 
 High Energy CLIC & 3 & 5.9  & 3-5 & 19-24 & 18-30 & $\sim$550 \\  
 & (1.5-3) & & & & & \\
\hline 
 High Energy CCC & 3 & 6.0 & 3-5 & 19-24 & 12-18 & $\sim$700 \\  
 & (1-3) & & & & & \\
\hline 
High Energy ReLiC & 3 & 47 (94)  & 5-10 & >25 & 30-50 & $\sim$780 \\  
 & (1-3) & & & & & \\
\hline 
Muon Collider & 3 & 2.3 (4.6)  & >10 & 19-24 & 7-12 & $\sim$230 \\  
 & (1.5-14) & & & & & \\
\hline 
LWFA - LC & 3 & 10  & >10 & >25 & 12-80 & $\sim$340 \\  
(Laser-driven) & (1-15) & & & & & \\
 \hline 
PWFA - LC & 3 & 10  & >10 & 19-24 & 12-30 & $\sim$230 \\  
(Beam-driven) & (1-15) & & & & & \\
 \hline 
Structure WFA - LC & 3 & 10  & 5-10 & >25 & 12-30 & $\sim$170 \\  
(Beam-driven) & (1-15) & & & & & \\
\hline 
\hline 
\end{tabular}
\end{center}
\caption{Main parameters of the lepton collider proposals with CM energy higher than 1 TeV. Total peak luminosity for multiple IPs is given in parenthesis. The cost range is for the single listed energy. Collisions with longitudinally polarized lepton beams have substantially higher effective cross sections for certain processes.} 
\label{tab:ITFmultiTeV}
\end{table}

\begin{table}
\begin{center}
\hspace*{-2em}
\begin{tabular}{| l | c | c | c | c | c | c |}
\hline
\hline
Proposal Name & CM energy & Lum./IP & Years of & Years to  & Construction & Est. operating \\ 
 & nom. (range) & @ nom. CME & pre-project & first & cost range & electric power \\  
  & [TeV] & [10$^{34}$ cm$^{-2}$s$^{-1}$] & R\&D & physics & [2021 B\$] & [MW] \\  
\hline
Muon Collider & 10 & 20 (40)  & >10 & >25 & 12-18 & $\sim$300 \\  
 & (1.5-14) & & & & & \\
 \hline 
LWFA - LC & 15 & 50  & >10 & >25 & 18-80 & $\sim$1030 \\  
(Laser-driven) & (1-15) & & & & & \\
 \hline 
PWFA - LC & 15 & 50  & >10 & >25 & 18-50 & $\sim$620  \\  
(Beam-driven) & (1-15) & & & & & \\
 \hline 
Structure WFA & 15 & 50  & >10 & >25 & 18-50 & $\sim$450  \\  
(Beam-driven) & (1-15) & & & & & \\
\hline
FCC-hh & 100 & 30 (60) & >10 & >25 & 30-50 & $\sim$560 \\  
 &  & & & & & \\
\hline 
SPPC& 125 & 13 (26)  & >10 & >25 & 30-80 & $\sim$400 \\  
 & (75-125) & & & & & \\
\hline 
\hline 
\end{tabular}
\end{center}
\caption{Main parameters of the colliders with 10 TeV or higher parton CM energy. Total peak luminosity for multiple IPs is given in parenthesis. The cost range is for the single listed energy. Collisions with longitudinally polarized lepton beams have substantially higher effective cross sections for certain processes. The relevant energies for the hadron colliders are the parton CM energy, which can be substantially less than hadron CM energy quoted in the table.
} 
\label{tab:ITFfrontier}
\end{table}

\begin{table}
\begin{center}
\hspace*{-2em}
\begin{tabular}{| l | c | c | c | c | c | c |}
\hline
\hline
Proposal Name & CM energy & Lum./IP & Years of & Years to  & Construction & Est. operating \\ 
 & nom. (range) & @ nom. CME & pre-project & first & cost range & electric power \\  
  & [TeV] & [10$^{34}$ cm$^{-2}$s$^{-1}$] & R\&D & physics & [2021 B\$] & [MW] \\  
\hline
LHeC & 1.2 & 1 & 0-2 ? & 13-18 & <4 & $\sim$140 \\  
 &  & & & & & \\
\hline 
FCC-eh & 3.5 & 1  & 0-2 ? & >25 & <4 & $\sim$140  \\  
 &  & & & & & \\
\hline 
 CEPC-SPPC-ep & 5.5 & 0.37  & 3-5 & >25 & <4 & $\sim$300 \\  
 &  & & & & & \\
\hline 
\hline 
\end{tabular}
\end{center}
\caption{Main parameters of the lepton-hadron collider proposals. For lepton-hadron colliders only, the parameters (years of pre-project R\&D, years to first physics, construction cost and operating electric power) show the increment needed for the conversion of the hadron-hadron collider to a lepton-hadron collider.} 
\label{tab:ITFleptonhadron}
\end{table}

\begin{table}
\begin{center}
\hspace*{-2em}
\begin{tabular}{| l | c | c | c | c | c | c |}
\hline
\hline
Proposal Name & CM energy & Lum./IP & Years of & Years to  & Construction & Est. operating \\ 
 & nom. (range) & @ nom. CME & pre-project & first & cost range & electric power \\  
  & [TeV] & [10$^{34}$ cm$^{-2}$s$^{-1}$] & R\&D & physics & [2021 B\$] & [MW] \\  
 \hline
High Energy LeptoN & 0.25 & 1.4  & 5-10 & 13-18 & 7-12 & $\sim$110 \\  
(HELEN) $e^+e^-$ collider & (0.09-1) & & & & &  \\
 \hline 
$e^+e^-$ Circular Higgs & 0.24 & 1.2  & 3-5 & 13-18 & 7-12 & $\sim$200 \\  
Factory at FNAL & (0.09-0.24) & & & & & \\
 \hline 
 Muon Collider & 10 & 20 (40)  & >10 & 19-24 & 12-18 & $\sim$300 \\  
at FNAL & (6-10) & & & & & \\
 \hline 
$pp$ Collider & 24 & 3.5 (7.0)  & >10 & >25 & 18-30 & $\sim$400 \\  
at FNAL &  & & & & & \\
\hline 
\hline 
\end{tabular}
\end{center}
\caption{Main parameters of the collider proposals located at FNAL.Total peak luminosity for multiple IPs is given in parenthesis. The cost range is for the single listed energy. There is also a recent proposal for a CCC version that can be located at FNAL \cite{Bath:Future_Colliders}. Other recently developed collider proposals, such as CERC, ReLiC, or wake field accelerators, could also be evaluated for being located at FNAL.} 
\label{tab:ITFsiteFillers}
\end{table}

%% file: Appendices.tex
\section{Appendices}
\label{sec:Appendices}

Below are brief descriptions of the main types of the collider proposals evaluated by the ITF: circular $e^+e^-$ colliders, RF-based linear $e^+e^-$ colliders, ERL-based $e^+e^-$ Colliders,  muon colliders, hadron colliders, and advanced concept $e^+e^-$ colliders. More detail descriptions and comprehensive list of references are available in Ref.\cite{shiltsev2021modern}. 

\subsection{Circular $e^+e^-$ colliders}
\label{sec:eecirc}

Circular $e^+e^-$ collider proposals include FCCee (Future Circular Collider $e^+e^-$ machine), CEPC (Chinese Electron Positron Collider) and Fermilab site filler. Their high-level technical parameters are given in Table \ref{TeeCirc}. Note, that several similar collider proposed had been discussed in the past, like, e.g., the TLEP Higgs factory in the LHC tunnel \cite{blondel2012high}. 

\subsubsection{FCCee ($Z$, $W$, $ZH$, $t\bar t$) collider proposal}

 Overview: The proposed circular FCCee is a well-studied $e^+e^-$ collider to be located surrounding CERN and Geneva \cite{blondel2012high}. The double-ring collider would operate at four CM energies albeit needing very different beam parameters ranging from the $Z$ (91 GeV c.e.e.) to $t \bar t$ (365 GeV c.m.e.). The present optimized main tunnel length is 91.2 km. Ampere level bunched beams maintained by SC RF cavities would be circulated in the two rings, one per beam, and made to collide in up to four interaction regions. The projected luminosity per IP ranges from 1.8$\cdot10^{36}$ cm$^{-2}$s$^{-1}$ at the $Z$ to 1.25$\cdot10^{34}$ cm$^{-2}$s$^{-1}$ at the $t\bar t$ within the limit of 50 MW of synchrotron radiation power loss per beam. A full energy injector located in the same tunnel would top-up the beam currents in the two colliding rings. The injector would reuse significant parts of the present CERN infrastructure. A CDR has been written in 2018 (with 2 IPs) and recently updated to a 4-IP lattice. Significant design efforts and R\&D have been completed including lattice, magnets, IR, site, and staging. The crucial future technical R\&D will concentrate on the 7.7 GeV SC RF cavity systems including HOM(high order modes) damping with ampere level bunched beams and, also, highly efficient RF klystrons. The magnet systems have very low fields to minimize the synchrotron radiation power. Considerable attention is given to the interaction region for clean experimental conditions, and to the center-of-mass energy calibration, especially at Z and W energies with resonant depolarization.

Main advantages: Circular $e^+e^-$ colliders overall have a successful 50 year history including LEP at CERN. Multi-ampere beams have been demonstrated at PEP-II and KEKB. The SuperKEKB e+e– collider in Tsukuba, now in operation, will demonstrate in the next few years nearly all the required accelerator physics techniques for FCCee, as will the future electron ring for the EIC at Brookhaven.

Main challenges: The peak luminosity within given synchrotron radiation power limit drops at higher beam energies approximately as $1/E^3$. Crab waist collision scheme with a large crossing angle, high bunch charges and mm-level vertical beam beta functions needs solid verification. SC RF cavities with multi- ampere beams with strong HOM damping require reliable demonstrations. Studies are ongoing for cost and power reduction. Pre-project cost and schedule: Technically, the project is nearly ready to proceed. R\&D and prototyping is
ongoing. However, the project needs to wait for the LHC-HL operational program to be completed leading to a start date for FCCee of around 2042.

Project construction time and cost: The project cost for $Z$ (91 GeV c.m.e.) to $ZH$ (240 GeV c.m.e.) operation is projected by the proponents to be about 10.5 BCHF (in 2021) using European accounting. Add 1.1 BCHF for the RF needed to go to $t \bar t$. The feasibility study of FCC has been approved by council and launched, technical study addressing all aspects but the financial study concentrated on the tunnel and the first collider (FCC-ee).

Prototypes: LEP1/2 and B-factories are considered as prototypes of FCCee. The level of achievements depend on the aspect: Beam energy: $\sim1/4$ for $t\overline t$, stored beam energy: $\sim0.008$, luminosity: $\sim1/50$.

\subsubsection{CEPC ($Z$, $W$, $ZH$) collider proposal}

Overview: The proposed circular CEPC is a well-advanced $e^+e^-$ collider to be located at one of several potential sites in China. The collider would operate at four CM energies although needing very different beam parameters ranging from the $Z$ to $t \bar t$. The present optimized main tunnel length is 100.0 km. Ampere level bunched beams maintained by SC RF cavities would be circulated in the two rings, one per beam, and made to collide in up to two interaction regions. The projected luminosity ranges from 1.15$\cdot10^{36}$ cm$^{-2}$s$^{-1}$ at the $Z$ to 0.5$\cdot10^{34}$ cm$^{-2}$s$^{-1}$ at the $t\bar t$. A full energy injector located in the same tunnel would top-up the beam currents in the two colliding rings. The full injector would be a new accelerator. A CDR has been completed and a TDR is due within a year or so. Significant design efforts and R\&D activities are underway including hardware prototypes of SC RF cryomodules, RF cavities, efficient klystrons (now up to 65 percent), vacuum chambers, and magnets. The magnet systems have very low fields to minimize the synchrotron radiation power.

Main advantages: Circular $e^+e^-$ colliders overall have a successful 50 year history including BEPC-II at IHEP in Beijing. Multi-ampere $e^+$ and $e^-$ beams have been demonstrated in PEP-II and KEKB. The SuperKEKB $e^+e^-$ collider in Tsukuba, now in operation, will demonstrate in the next few years nearly all the required accelerator physics techniques for CEPC, as will the future electron ring of the EIC at Brookhaven.

Main challenges: The peak luminosity drops at higher beam energies. Crab waist collision scheme with high bunch charges and mm-level vertical beam beta functions needs solid verification. SC RF cavities with multi-ampere beams with strong HOM(high order modes) damping need reliable demonstrations. Studies are ongoing for cost and power reduction.

Pre-project cost and schedule: Technically, the project is nearly ready to proceed. Pre-construction R\&D and prototyping is ongoing. Future crucial technical R\&D will concentrate on the 10 GeV SC RF cavity platforms including HOM damping with ampere level bunched beams and highly efficient RF klystrons. The project is trying to take advantage of situations uniquely available to Chinese construction. Project construction time and cost: An international collaboration is under development. Construction may start around 2026 followed by data taking perhaps starting around 2034. The project cost is projected by the proponents to be about 5 BUS (in 2021) for the Higgs based collider using Chinese accounting.

Prototypes: LEP1/2 and B-factories are considered as prototypes of CEPC. The level of achievements depend on the aspect: Beam energy: $\sim1/4$ for $t\overline t$, stored beam energy: $\sim0.013$, luminosity: $\sim1/30$.

\subsubsection{Circular Fermilab site-filler ($Z$, $W$, $ZH$) collider  proposal}

Overview: The “Site Filler” is a proposed circular $e^+e^-$ collider in a very early development stage to be located on the FNAL site \cite{bhat2022future}. A schematic is shown below. The single ring collider would operate at 46-120 GeV per beam with only a few bunches (2 to 4) colliding head-on. A luminosity at the $Z$ (91 GeV CME) would be about  from 6.3$\cdot10^{34}$ cm$^{-2}$s$^{-1}$ and at the Higgs about 1$\cdot10^{34}$ cm$^{-2}$s$^{-1}$ with the total synchrotron radiation power limited at $P_{SR}=$2$\times$50 MW. The main tunnel is $C$=16 km in length corresponding to the maximum possible circumference on the existing FNAL site with one IR. A top-up injector may be required and would be mostly new, however, with some present FNAL infrastructure gainfully repurposed. In the near term, the basic beam parameters need to be optimized and then the basic technical accelerator components will be designed. The future needed R\&D concentrates on the basic accelerator design followed by the specific component designs. Later on, R\&D will concentrate on the design of the ~12 GeV SC RF cavity system including HOM damping with ampere level bunched beams and on highly efficient RF klystrons.

Main advantages: Circular $e^+e^-$ colliders overall have a successful 50 year history. Multi-ampere beams have been demonstrated. The SuperKEKB $e^+e-$ collider in Tsukuba, now in operation, will demonstrate in the next few years nearly all the required accelerator physics techniques, as will the future EIC at Brookhaven. US circular collider experts are available. The construction time for this collider is relatively short due to the available site and smaller circumference.

Main challenges: The peak luminosity scales approximately as $P_{SR} C /E^3$ and not only drops at higher beam energies, but also is lower for smaller circumference of the “Site Filler” relative to FCCee (91 km) and CEPC (100 km). Strong synchrotron radiation reduces the available beam currents. The equilibrium beam emittances for smaller rings are large reducing the number of colliding bunches.

Pre-project cost and schedule: The pre-construction costs and schedules are smaller due to similar R\&D on lattices and hardware being done for other projects.

Project construction time and cost: Costs are under development. Construction time should take about 7 years.

\begin{table}[htbp]
\begin{tabular}{|lccc|}
\hline 
\hline
 & FCCee & CEPC & FNAL Site Filler \\
 & & & \\
Number of IPs & 4 & 2 & 1 or 2 \\
Number of bunches at Higgs & 336 & 242 & 2 \\
CM energy $\sqrt{s}$, GeV & 91/160/240/355 & 91/160/240  & 91/160/240 \\
CM energy spread at Higgs, GeV & 0.3 & 0.3 & $\sim$0.4 \\
Bunch length at Higgs, mm &2.5-4.5 & 2.3-3.9 & 2.9 \\
Main ring length, km & 91 & 100 & 16 \\
Length of new accelerators, km & 272 & 300 & 48 \\
Vert. beta at IP $\beta^{*}_y$, mm & 1 & 1 & 1 \\
Luminosity per IP at Higgs, 10$^{34}$cm$^{-2}$s$^{-1}$ & 7.7 & 8.3 & $\sim$1.3\\
RF voltage per turn at Higgs, GV & 2.2 & 2.2 & 12 \\
SR power both beams at Higgs, MW & 100 & 60(100) & 100 \\
Total facility power, MW & 290 & 340 & $O$(200) \\
Injectors/facilities & a lot reuse & all new & some reuse \\
\hline \hline
\end{tabular}
\caption{High level technical parameters for circular $e^+e^-$ colliders.} 
\label{TeeCirc}
\end{table}

\subsection{Linear RF-based $e^+e^-$ colliders}
\label{sec:eelinear}


\subsubsection{Development of the Linear Collider Designs}
The linear collider program began in the early 1980’s with the proposal of the Stanford Linear Collider (SLC).  Linear Colliders were proposed to avoid the energy loss due to synchrotron radiation which scales as $E^4$ in a storage ring.  The SLC was proposed as both a physics experiment to study the $Z_0$ boson at 92 GeV in the center-of-mass frame and as an accelerator experiment to study this new type of collider.  The SLC began colliding beams in 1988 and encountered many technical and accelerator physics challenges.  To address these challenges, many new techniques were pioneered including BNS damping of single bunch wakefields, emittance correction and preservation, polarization tuning bumps, IP deflection scans, 1st and 2nd moment beam-based adaptive feedback, and flat beam operation as well as new technologies such as a high yield positron targets, strained GaAs polarized photocathodes, high resolution single pass diagnostics, wire and laser-wire scanners.  The SLC ran for 10 years and reached the design luminosity during the final year of operation and delivered a measurement of a left-right $Z$ production asymmetry with smaller errors than any of the experiments at LEP due to the highly polarized electron beam.  Most importantly, it provided a strong basis for all of the linear collider designs going forward.

While the SLC was being constructed and commissioned, collaborations around the world began considering next generation linear colliders with higher center-of-mass energy.  This included concepts  SLAC and Japanese based on X-band RF technology, a concept from CERN using a two-beam K-band RF technology, two concepts from DESY, one based on S-band RF technology and the other on L-band SRF technology as well as others \cite{TRC1994, TRC2003}.  To support these concepts during the 1990’s a number of large-scale test facilities were constructed to verify some of the fundamental sub-systems including the FFTB (Final Focus Test Beam at SLAC, 1994), NLCTA (NLC Test Accelerator at SLAC, 1997), ATF  (Accelerator Test Facility at KEK, 1996), TTF (TESLA Test Facility at DESY, 1996), and the CTF (CLIC Test Facility at CERN, 1995).

\begin{table}
\begin{center}
\begin{tabular}{| r || r | r | r || r | r || r | r | r|| r|}
\hline
             & ILC & ILC & ILC & CLIC & CLIC & CCC & CCC & CCC & HELEN \\
\hline
CME (GeV) & 250 & 500 & 1000& 380  & 3000 & 250 & 550 & 3000 & 250\\
Peak $\mathcal{L}$ (1E34) & 2.7 & 3.6 & 5.1 & 2.3  & 5.9  & 1.3 & 2.4 & 6 & 1.35\\
Length (km)  & 20  & 31  & 40  & 11   & 54   & 8   & 8   & 27 & 7.5\\
Site power(MW)  & 138 & 215 & 300 & 110  & 580  & 150 & 175 & 700 & 110 \\
$\varepsilon_{x}$ ($\mu$m) & 5 & 10 & 10 & 0.9 & 0.66 & 0.9 & 0.9 & 0.9 & 5 \\
$\varepsilon_{y}$ (nm) & 35 & 35 & 30 & 20 & 20 & 20 & 20 & 20 & 35 \\
$N_b$ (1e9)  & 20 & 20 & 20 & 5.2 & 3.7 & 6.2 & 6.2 & 6.2 & 20\\
Bunches/train & 2625 & 2625 & 2450 & 352 & 312 & 133 & 75 & 75 & 1312\\
$f$ (Hz)   & 5 & 5 & 4 & 50 & 50 & 120 & 120 & 120 & 5 \\
$\beta^*$, hor. (mm)& 13 & 11 & 11? & 8 & 7 & 12 & 12 & 12 & 13 \\
$\beta^*$, vert. (mm)& 0.4 & 0.5 & 0.4? & 0.1 & 0.07 & 0.12 & 0.12 & 0.12 & 0.4 \\
$\sigma_{x}$ ($\mu$m) & 0.5 & 0.5 & 0.3 & 0.15 & 0.04 & 0.2 & 0.2 & 0.045 & 0.5 \\
$\sigma_{y}$ (nm) & 7.7 & 6 & 3 & 3 & 1 & 2 & 2 & 1 & 7.7 \\
\hline 
\end{tabular}
\caption{Parameters of the linear $e^+e^-$ colliders. The ILC-Higgs numbers refer to the $\mathcal{L}$-Upgrade. References: ILC \cite{Aryshev:ILC}, 
CLIC \cite{CLICreference}, 
CCC \cite{nanni2021Ccube}, HELEN \cite{belomestnykh2022higgs}.}
\label{tab:LCparams}
\end{center}
\end{table}

By the early 2000’s there were only two major concepts that were sufficiently advanced to be considered for the next large HEP facility: the US/Japan NLC/JLC X-band RF collider and the DESY SRF TESLA L-band collider.  The CERN CLIC two-beam collider at 30 GHz was not yet at the same stage of development.  After a review \cite{ITRP}, efforts focused on the International Linear Collider (ILC) SRF L-band linear collider design and a Reference Design Report and cost basis was produced in 2007.  After a Japanese site selection, these were refined with a Technical Design Report in 2012.  In parallel, in 2008, the CERN CLIC effort chose to adopt the X-band rf technology similar to that developed for the NLC/JLC and completed a Conceptual Design Report with a cost in 2012.  

\subsubsection{Status of the main designs (ILC and CLIC)}

At this time, there are two well developed design concepts, the ILC \cite{Aryshev:ILC} and CLIC \cite{CLICreference}, with detailed design reports, cost estimates, and supported with extensive beam physics simulations as well as many test facilities to verify the individual subsystems of the colliders.  Both projects have cost estimates of less than 10B ILCU and take roughly 10 years to construct.  The ILC is designed to start at 250 GeV cme and has a clear upgrade path to 1 TeV cme using the same RF technology but longer tunnels.  Additional energy or luminosity upgrades may be possible with development of more advanced RF technology.  CLIC is designed to start at 360 GeV cme and upgrade in stages to 3 TeV cme.  Detailed parameters can be found in the design documents but are summarized in Table \ref{tab:LCparams}.

The ILC has an International Develop Team (IDT) that has developed a 4-year Pre-Lab proposal that would ramp up the effort and start Final Design to allow an 8-year construction of the ILC to begin.  The Pre-Lab has not yet received funding.  The CLIC effort at CERN continues but with significantly reduced resources. It is also awaiting funding to start Final Design.

\subsubsection{Cold Copper Collider (C$^3$)}
The Cold Copper Collider (C$^3$) is a new linear collider concept that aims to achieve high-gradient acceleration in copper RF cavities through novel cavity design and cryogenic cooling at N2 temperatures~\cite{nanni2021Ccube}.

The main linac accounts for only a third of the cost of the entire machine for the 250 GeV design. C$^3$ proponents advocate for further cost reductions to other machine subsystems including particle sources, the Beam Delivery System, and support infrastructure in order to further reduce the cost~\cite{nanni2021Ccube}.

\subsubsection{Test facilities and basis of the linear collider designs}

As noted, most of the accelerator physics and technologies for the linear colliders have been demonstrated.  A linear collider can be divided into 4 main sub-systems: particle sources, damping rings to generate the desired small beam emittances, linacs to accelerate the particle beams to high energy, and final focus systems to decrease the beam size at the IP to generate high luminosity.  Dedicated facilities at the FFTB \cite{FFTB} and then the ATF2 \cite{ATF2} have verified the nonlinear focusing optics in the final focus systems required to achieve the very small beam spots.  The ATF and CESR-TA have demonstrated many of the concepts required for the damping rings and the recent generation of synchrotron light sources have shown that it is possible to routinely operate with such beam parameters. 

Dedicated RF test facilities have been used to verify the fundamentals of the linear accelerators.  For the ILC, the TESLA Test Facility was used to develop and demonstrate the 1st generation of the RF system.  The European X-FEL was later built and consisted of 100 such RF units.

Finally, the SLC provided a large-scale demonstration of many critical concepts as described above. If we compare the results of the test facilities, including the SLC and the Eu-XFEL with the requirements for ILC and CLIC we have:
\begin{itemize}
\item ILC: beam energy: $\sim$1/6, gradient: $\sim$1/1.5, luminosity: $\sim$1/1000, positron production per second: $\sim$1/30.
\item CLIC: gradient: $\sim$1, beam energy $\sim$1/2000, luminosity: 0, positron production per second: $\sim$1/20.
\end{itemize}
%


\subsection{Energy-recovery $e^+e^-$ colliders}
\label{sec:erl}


In the energy recovery linac approach the energy which beams receive from the RF field in superconducting accelerating structures is fed back to the structures by decelerating the beams on the opposite RF phase after the beams have collided at the IP. This way the virtual beam power can be much larger than the net RF power required for acceleration and the parameter optimization for an e+e- collider can be quite different from a conventional linear collider or circular collider scheme. The three concepts considered here make use of this approach and promise a very high luminosity, in absolute terms and in relation to wall plug power. This is shown in Table \ref{ERL-1}, where main parameters are listed for the example of a Higgs factory (parameters for higher or lower CM energy are also considered in the proposals). Furthermore, all schemes recycle not only the energy in the beam but also the particles by feeding them back into the accelerator by return arcs or damping rings at the low energy ends of the accelerator, therefore requiring only low-intensity beam sources for top-up of a small fraction of beam lost in the accelerate-collide-decelerate cycle. While being similar in these respects, the schemes, as described in the following, do have quite different features in their layout.
\begin{table}[h]
\centering
\begin{tabular}{|lccc|}
\hline 
\hline
 & ERLC$_{\rm CW}$ & ReLiC & CERC \\
 & & & \\
Number of IPs & 1 & 2 & 1 or more \\
CM energy $\sqrt{s}$, GeV & 250 & 240  & 240 \\
Luminosity $L$ per IP, 10$^{34}$cm$^{-2}$s$^{-1}$ & 90 & 165 & 78 \\
Wall plug power $P_{AC}$, MW & 250 & 315 & 90 \\
$L/P_{AC}$, 10$^{34}$cm$^{-2}$s$^{-1}$/MW & 0.69 & 0.68 & 0.87 \\
Bunch charge $N_e$, 10$^9$ & 0.46 & 20 & 160 \\
Beam current $I_b$, mA & 100 & 12 & 2.5 \\
R.m.s. beam size at IP $x/y$, nm & 430/6.2 & 9042/1.2 & 4633/2.2 \\
Disruption param. $D_{x,y}$  & 0.016/1.2 & 0.01/43 & 0.8/544 \\
BS spread at IP $dE_{rms}/E_0$, \% & 0.2 (equilibr.) & 0.002 (p.coll.) & 0.33 (p.coll.) \\
Accel. gradient $G$, MV/m & 20 & $\sim$20 & $\sim$16 \\
Linacs total length, km & 2$\times$15 (dual axis) & 2$\times$10 & 3.2 \\
Length of new accelerators, km & $\sim$60 & $\sim$40 & $\sim$1600 \\
Cav. quality factor $Q_0$ & 3$\cdot 10^{10}$ & $> 10^{10}$ & $10^{11}$ \\
Cryo coefficient $P_{AC} / P_{cryo}$, kW/W & 0.22 $@$ 4.5K & 1.25 $@$ 1.8K& 1.25 $@$ 1.8K \\

\hline \hline
\end{tabular}
\caption{ Main parameters of ERL-based $e^+e^-$ collider proposals at CM energy of Higgs factory. For ReLiC the total luminosity doubles for two IPs. For CERC the luminosity is shared for multiple IPs.} 
\label{ERL-1}
\end{table}

\subsubsection{ERLC}

{\em Basic layout:}
The ERLC concept \cite{ERL1}  foresees two “twin” (or “dual axis”) linacs for acceleration and deceleration of electrons and positrons. The electron (positron) beam is accelerated in one branch of twin linac 1 (2). After collision it is decelerated in the opposing twin linac 2 (1), where it transfers its energy to the positron (electron) beam, which is accelerated in the other branch of the twin linac 2 (1). After deceleration the electron (positron) beam is recycled to the entrance of twin linac 1 (2), which closes the loop. Energy transfer requires strong RF coupling which is to be achieved by using dual-axis SRF cavities. The return transfer lines include damping wigglers at an energy of 5 GeV, which radiate about 0.5\% of the beam energy, so that the damping time corresponds to about 200 revolutions. The beams have similar emittances and IP beam sizes as in a linear collider (ILC), but with much smaller bunch charge. The latter strongly reduces beamstrahlung (BS), which is essential to avoid a strong build-up of energy spread, which has to be counteracted by the damping in the return transfer lines. The energy spread shown in Table \ref{ERL-1} shows the equilibrium energy spread resulting from BS and radiation damping by the wigglers (the single collision energy spread from BS is 0.002\%). With deceleration from 125 GeV to 5 GeV the relative energy spread increases 25-fold. By de-compressing the bunches the energy spread in the return arc is reduced for beam dynamics reasons, re-compression takes place before re-injection into the linac. The high luminosity in this scheme results from the high circulating beam current, obtained by filling every RF bucket of the 1.3GHz Nb3Sn SRF linac. The power consumption is strongly dominated by cryogenic load from RF dissipation and higher order mode (HOM) losses, despite a very optimistic assumption on the overall cryogenic efficiency. The net RF power is comparatively small and due to energy losses in the damping wigglers and to higher order modes in the accelerating structures. RF power for stabilization of the RF field is not yet included.

The power consumption is relatively moderate for a 250GeV collider with very high luminosity, leading to a high ratio of luminosity to AC power, more than an order of magnitude higher than that of the ILC. Power contributions from RF stabilization, magnets, injectors and infrastructure are not yet included here. The ERLC study also includes the option of operating the collider in a slow pulsed mode (on a scale of seconds) to reduce the power consumption, at the expense of luminosity. Such a scheme could be applied if SRF technology with low RF dissipation does not become available, but it creates several challenges regarding particle sources and the injection scheme, as well as regarding the ramp-up and down of the beam current in the ERL linacs, and is not considered as the preferred solution.

{\em R\&D and further studies}:
Due to the low average accelerating gradient, the ERLC is rather long. Furthermore, the need to have four linacs (or two dual axis linacs) for two beams is a challenge for limiting the construction costs. In principle, the ERLC consists of two large 5 GeV storage rings with very uncommon insertions made up of 120 GeV accelerating and decelerating linacs, very strongly RF-coupled to each other, and final focus systems with interaction point. The properties of such a machine regarding stability, tolerances, equilibrium emittances, etc. are not easy to evaluate and detailed studies of beam dynamics are required to assess the viability of the scheme. From a technology point of view, the most demanding R\&D item is the dual-axis SRF linac structure. Design, prototyping and full test of such devices require a significant effort before it can become clear whether this approach is suitable for a large-scale accelerator facility. The power efficiency of the concept assumes to strongly benefit from improved Q0 in the SRF cavities and from the successful development of Nb3Sn technology operating at 4.5K instead of 1.8K. The RF field stabilization in dual-axis cavities is another issue which has to be addressed in the R\&D program. R\&D is also needed on superfast injection/extraction kickers and polarization control.
 
\subsubsection{ReLiC}

{\em Basic layout}:
The ReLiC concept \cite{ERL2}  has a 10km long SRF linac on either side of the interaction region, and the transfer of energy between the decelerated and accelerated electron and positron beams takes place in the same structures. So, in total four beams are propagating in the same linac, one electron and positron beam being accelerated and one counter-propagating electron and positron beam being decelerated. In steady state the net beam current is zero and ideally no RF power is necessary for beam acceleration, except compensating for energy loss from BS and HOM, and for RF field stabilization. In order to avoid collisions of the counter-propagating short bunch trains (10 bunches per train), beam separation chicanes are foreseen at every 200m in the linacs. A combination of magnetic and electrostatic deflectors has to be used to separate electron (positron) bunches from electron (positron) bunches as well as electron from positron bunches. The timing of the bunch trains is carefully chosen such that no parasitic collisions occur in the linac sections between the chicanes. The decelerated beams are stored in (in total four) damping rings at 2 GeV  for a few damping times before being injected again into the linacs. This “resets”the transverse and longitudinal phase distribution in the bunches after each cycle, which permits a much stronger perturbation of the beam by the beam-beam interaction. Consequently, the bunch charge can be higher and the luminosity very large. The BS effect is kept small due to the large horizontal beam size,the energy spread per collision quoted in Table \ref{ERL-1} has been obtained here from a simple analytical approximation (this figure is not specified in the ReLiC paper). The BS energy spread is increased 60-fold with deceleration. If necessary, bunch decompression can be used prior to injection into the damping rings. The issue of energy spread can become more demanding for higher CM energy variants of the concept. The linacs are assumed to operate with relatively low-frequency (500MHz) SRF cavities in order to reduce HOM power.

{\em R\&D and further studies}: 
The ReLiC proposal is at a very early conceptual stage and important design aspects have to be further worked out. With the relatively high bunch charge collective effects have to be carefully investigated. Furthermore, the high beam current requires a very large RF power in the damping rings and the large number of bunches which have to be stored likely also require a large ring circumference. Regarding the SRF linac technology, the necessary R\&D can build on already ongoing developments of strongly HOM-damped RF structures and on existing or planned ERL test facilities. Successful R\&D on Nb$_3$Sn SRF technology seems to be highly desirable for the  benefit of power efficiency.
 
\subsubsection{CERC}

{\em Basic layout}:
The CERC approach \cite{ERL3} combines ERL and recirculating linac concepts. In a circular arrangement both electron and positron beams pass four times through the same SRF linac structures, thus providing energy transfer between decelerated and accelerated beams as well as reducing the required accelerating voltage by a factor of four. Due to synchrotron radiation the recirculation beamlines for the accelerated and decelerated beams have to be different, leading to in total 16 beamlines which are assumed to be installed in a 100km circumference tunnel, including the two final focus systems and interaction region. Merging and separating sections are needed to guide the beams through the same linac sections. Energy loss from synchrotron radiation as well as BS and HOM is compensated by additional net RF power. The decelerated beams are stored in two damping rings before being re-injected for the next cycle, very similar to the ReLiC scheme. However, the bunch charge is higher and the BS effect is much stronger than for ReLiC, whereas the beam current is much smaller. This yields a lower luminosity, but also a lower power consumption.

{\em R\&D and further studies}:
Conceptually, the CERC scheme allows to save on the cost of the RF linacs, because the total installed accelerating voltage is 30GV instead of 240GV in a linear collider. Still, staying at or below the cost of FCC-ee is a challenge. Regarding bunch de-compression and damping ring issues, this scheme has significant challenges, because the BS relative energy spread is amplified by a factor of 60 with deceleration to the damping ring energy of 2 GeV. Bunch de-compression schemes and damping rings with high energy acceptance including the corresponding beam dynamics have to be carefully studied. On the other hand, due to the much lower beam current, the demands for the damping rings regarding number of bunches stored and RF power are more relaxed. Additional design challenges are related to the in total 1600km of beam lines, including sophisticated schemes for merging and separation of the 16 beams. A full analysis of the beam dynamics is required, including imperfections, to assess whether the assumed high beam quality can be achieved in the recirculation process. As for the linac technology, further R\&D can be based on already existing developments and use existing or planned ERL test facilities. SRF R\&D could benefit the power efficiency of the scheme, but due to the much shorter linac the power saving effect would be smaller – as shown in Table \ref{ERL-1}, the power consumption is already rather moderate.    A recent analysis of the state-of-the-art for an ERL can be found in Ref. \cite{CERNLDG:2022}.

\subsection{Muon Colliders}
\label{sec:mumu}

The concept of a muon collider has been developing in the HEP community for decades as an alternative to hadron or electron accelerators.  As with most proposed future HEP colliders, the physics case for the muon collider is largely centered on search for and study of new physics such as electroweak symmetry breaking, dark matter, and the naturalness of the weak scale. It may also serve as a dedicated facility for further exploration and characterization of the Higgs bosons and it’s interaction with other fundamental particles and forces.
 
Muons pose a few distinct advantages that make them attractive compared with hadrons and electrons, even for the same physics goals. Specifically, they carry the key advantage of electrons, without the primary weakness: They serve as high precision probes that can release all of their energy upon collision, but by virtue of their large mass compared with electrons, they do not suffer from the same limiting phenomenon of synchrotron radiation and thus can reach high energies in circular colliders with anticipated higher luminosity-to-power (ab$^{-1}$/TWh) ratio than circular and linear e+e- colliders above 1 TeV cme.  Also, Higgs production is greatly enhanced for muons over electrons; there is practically no bremsstrahlung in muon collisions.  Additionally, for the same physics reach, it can be shown that the required muon collider center of mass energy is an order of magnitude lower than the equivalent hadron collider \cite{ali2021muon}, giving muon colliders the potential to be more compact and cost effective than their hadron counterparts \cite{long2021muon}.   Despite these compelling arguments, there are also significant technical challenges primarily associated with the lifetime of the muon have limited their realization thus far.

Concepts for a muon collider typically contain conventional accelerating and magnet technologies, along with the requirement to address a few key elements specific to muons, namely:\\
1.     Generation of a muon beam.  A muon injector is envisioned consisting of a $\sim$(4-8) GeV proton driver and target to generate pions which decay into muons with both positive and negative charges. The opposite charges are then separated into positive and negative muon beams for acceleration and collision.\\
2.     Cooling of the muons. The initial muon beams must undergo cooling by orders of magnitude prior to injection into an accelerating system.  Some schemes for ionization cooling (recently demonstrated in MICE \cite{mice2020demonstration}) have been envisioned but are still in the early stages of development. The cooling will likely require high field SC magnets.\\
3.     Acceleration.  Schemes such as a large rapid-cycling booster ring or recirculating linear accelerator requiring high field SC magnets have been proposed.\\ 
4.     A collider ring with dedicated interaction points and several hundred turns, with average luminosity proportional to the field in the SC dipole magnets.\\

\begin{table}[h]
\centering
\begin{tabular}{|lccc|}
\hline 
\hline
 & Higgs Factory & MC-3TeV & MC-10TeV \\
 & & & \\
CM energy $\sqrt{s}$, TeV & 0.125 & 3  & 10 \\
Luminosity per IP, 10$^{34}$cm$^{-2}$s$^{-1}$ & 0.008 & 2.3 & 20 \\
Collider circumference, km & 0.3 & 4.5 & 10 \\
Number of IPs & 1 & 2 & 2 \\
Number of bunches & 1 & 1 & 1 \\
Repetition rate, HZ & 15 & 5 & 5 \\
Bunch charge $N_{\mu}$, 10$^{12}$ & 4 & 2.2 & 1.8 \\
Bunch length, mm & 63 & 5 & 1.5 \\
Bet-function at IP $\beta^*$, mm & 17 & 5 & 1.5 \\
R.m.s. beam size at IP, $\mu$m & 75 & 3 & 0.9 \\
Avg. power to beams, MW & 0.05 & 10.5 & 28.8 \\
\hline \hline
\end{tabular}
\caption{Main parameters of Muon Colliders.} 
\label{MCtab}
\end{table} 
 
While much of this can be considered conventional accelerator componentry, the unstable nature of the muon introduces a significant complications . Namely, the entire scheme (1)-(4) must be completed within the lifetime of the muon, which is dilated in the lab frame according to the muon energy throughout the process, but is still quite short ($\sim$20 ms at TeV) for the processes proposed.  The cooling step must be faster than the muon rest frame lifetime (2.2$\mu$s or about 700m), but requires several cooling stages to achieve the required phase space reduction.  In addition, the collisions must occur before significant decay of the muons, which places constraints on the size, accumulation, and storage times for acceleration and collision.
 
Another significant challenge stemming from muon decay is the impact of the by-products. Electrons from the decay will indirectly result in background in the detectors.  Additionally, the decays will produce an intense flux of neutrinos that will interact with surrounding earth to produce ionizing radiation.  Some mitigating scenarios have been considered, for instance adjusting the orbits to spread the impact, or reducing the muon beam intensity for an equivalent brightness through further reduction in emittance, but a clean solution is not yet obvious.
 
A considerable amount of R\&D is still required for the muon collider to overcome these challenges.  Some of the R\&D such as high field SC magnets are proceeding according to the needs of other proposed collider schemes such as the FCChh.  On the other hand, R\&D efforts such as the cooling scheme and the target development are more specific to the muon collider and thus required a dedicated program.  Overall, the R\&D effort is likely in the 10+ year time range.  Construction time would be on scale with conventional large-scale facilities at 10-15 years (for a 10-14 TeV cme muon collider), with some possible overlap between these two.  A muon collider is clearly a multibillion dollar facility, but exact cost has not yet be estimated.    A recent analysis of the state-of-the-art can be found for muon colliders can be found in Ref. \cite{CERNLDG:2022}.

\subsection{Hadron and hadron-lepton colliders}
\label{sec:hhhl}


Circular hadron collider proposals include FCC-hh from CERN, SPPC from China, a possible Fermilab site-filler $pp$-collider and the Collider-under-the-sea from the USA. Main design parameters appear in Table \ref{Hadrontab} below.

\begin{table}[h]
\begin{tabular}{|lcccc|}
\hline 
\hline
 & FCC-hh & SPPC & FNAL Site Filler & Coll. under sea \\
\hline
CM energy $\sqrt{s}$, TeV & 100 & 125  & 24 & 500\\
Injection energy, TeV & 3.3 & 3.2  & & 50\\
Perimeter, km & 91.2 & 100  & 16 & 1,900\\
Number of arcs & 8 & 8  & &\\
Arc length, km & 9.8 & 10.5  & &\\
Number of IP & 4 & 2  & 2 & 2\\
LSS length at IP, km & 1.4 & 1.25  & &\\
LSS length at tech site, km & 2.134 & 1.25-3.4 & &\\
Bending field, T & 17 & 12 - 20  & 24.4 & 3.5 \\
Peak lumi./IP (multipl.), $10^{34}cm^{-2}s^{-1}$ & 5-30 (1000) & 10  & 3.5 & 50\\
Beam current, A & 0.5 & 0.19  & 0.45 & \\
Bunch population $1\times 10^{11}$ & 1 & 0.4  & .93 & \\
Bunch spacing, ns & 25 & 25  & 25 & 30\\
Normalized emittance, $\mu m$ & 2.2 & 1.2  & 1.5 &\\
Synchrotron radiation/ring, MW & 2.7 & 2.2  & 0.04 & 18\\
Stored energy per beam, GJ & 7.8 & 4.0  & .29 & \\
Total power consumption, MW & 560 & 400  & 200 - 300 & 200,000\\
\hline \hline
\end{tabular}
\caption{Main parameters of hadron collider proposals.} 
\label{Hadrontab}
\end{table} 
Hadron-lepton collider proposals include LHeC and FCC-eh from CERN, and CEPC+SPPC from China: all are considered as incremental projects with respect to the corresponding hadron colliders.  

\subsubsection{Future Circular Collider FCC-hh}
FCC-hh is proton and ion collider, with a center-of-mass energy of 100 TeV, aiming at an integrated luminosity of the order of 20 inv-ab in each of its two main detectors over 25 years of operation (proton-proton) \cite{FCC-hh}. In its latest version, it would be installed in a 91.2 km circumference quasi-circular tunnel in the Geneva basin, next to CERN. Four long straight sections of 1.4 km each will house the insertions for experiments, and another four of length 2.143 km each will be devoted to injection, extraction, RF and collimation, leaving 76.9 km for the arcs. The lattice in the arcs consists in FODO cells of length 213 m with a phase advance of 90°, composed of six 14 m long twin-aperture dipoles between quadrupoles. The bending field of about 17 T (16 T in the case of combined-function magnets) calls for high-field superconducting magnets using advanced superconductors such as Nb3Sn, operated in superfluid helium below 2 K. Injection would re-use the existing chain of CERN machines – Linac4, PS, PSB, SPS and LHC operated at 3.3 TeV – as pre-accelerators. Direct injection from a new superconducting synchrotron at 1.2 TeV to be installed in the SPS tunnel is considered as a possible alternative.

The main technical challenges requiring further studies and R\&D arise from the high energy of the beams, the intense flux of collision debris from the high-luminosity experiments, and the large synchrotron radiation power in the arcs. The large-scale production of high-field superconducting accelerator magnets requires the development of Nb3Sn superconductor with adequate current density, filament size and cost, the control of strain in the coils and their insulation, their protection in case of resistive transitions and the implementation of reproducible construction techniques and reliable quality assurance. A robust and efficient collimation system, as well as a fault-tolerant beam extraction are required to protect the magnets from the huge energy stored in the circulating beams. The cryogenic beam vacuum system, using helium-cooled beam screens must limit beam-gas scattering, remove the largest fraction of synchrotron radiation heat load and avoids beam instabilities and electron cloud. International collaborations are under way on all these studies and developments.

Prototypes: LHC is considered as a prototype of FCChh. The level of achievements depend on the aspect: Beam energy: $\sim1/7$, magnetic field: 1/2.

\subsubsection{Super Proton-Proton Collider SPPC}

SPPC is a proton-proton circular collider to be built in China, with a circumference of 100 km and centre-of-mass energy of 125 TeV. with an intermediate run at 75 TeV \cite{SPPC}. The integrated luminosity goal is 30 inv-ab assuming two interaction points and 10-15 years operation. The machine layout is composed of eight identical arcs and eight long straight sections housing the detectors, injection and extraction, RF stations and a complex collimation system. The twin-aperture dipole magnets would use iron-based high-temperature superconductors to produce an initial field of 12 T. In a second phase, following development of this class of superconducting materials, new magnets would allow reaching 20-24 T bending field. The operation temperature of the magnets is not defined yet, and so is the cryogenic system. The machine also features a cryogenic beam screen and vacuum system. A totally new injector chain, composed of a 1.2 GeV proton linac, 10 GeV rapid-cycling synchrotron, 180 GeV medium-stage synchrotron and 3.2 TeV final-stage superconducting synchrotron would serve the main ring collider. With the addition of a heavy-ion linac and rapid-cycling synchrotron, the collider could also be operated with ions.
The main challenges facing the SPPC project are similar to those of FCC-hh. A roadmap has been established in China for the development of high-field magnets, particularly using iron-based high-temperature superconductors which could be easier to implement on a large scale than the demanding Nb3Sn technology. As in FCC-hh, beam collimation, machine protection, management of the synchrotron radiation and other beam-induced effects are critical issues to be studied.   

In the present thinking, both the FCC-hh and SPPC hadron colliders would be installed in the same tunnels, respectively, as the FCC-ee and CEPC lepton colliders after exploitation of the latter for physics. A difference between these projects, though, resides in the fact that CEPC and SPPC could be accommodated simultaneously in their tunnel, while FCC-ee and FCC-hh would sequentially occupy the tunnel at CERN, smaller in cross-section, thus requiring removal of FCC-ee for installing FCC-hh (in the same way as LEP was removed to make space for the LHC).  A consequence of this approach is that the quoted incremental costs of the hadron colliders do not include the cost of the reused tunnels, of other civil works in underground and at ground level, and of basic infrastructure. Still, in the present exercise and for the sake of comparison, we have considered the complete “green field” cost estimates for both projects.

Prototypes: LHC is considered as a prototype of SPPC. The level of achievements depend on the aspect: Beam energy: $\sim 1/9$, magnetic field: 1/2.4.

\subsubsection{FNAL site-filler hadron collider}

There are several proposals put forward for accelerators on the Fermilab site \cite{bhat2022future}. Among them is an ambitious idea to build a $pp$ collider that will fit on the existing site. The Tevatron and the existing injector complex would serve as the entire injector chain. However, only a 16 km circumference would be allowed. So, reaching an energy of twice the LHC would require dipoles operating at approximately 24 T, 50 \% higher than the quite challenging 16 Tesla dipoles proposed for the FCC-hh. Such magnets would require the use of high temperature superconductors, a technology that has yet to be proven. 

\subsubsection{Collider-under-the-Sea}

A novel concept proposes to build a 100 TeV collider housed in a ring pipeline submerged in the Gulf of Mexico. It could then later be used as an injector for a 500 TeV machine. The main goal is to eliminate the tunnel cost and allow use of low field (3.5 Tesla), low-cost magnets. A 1,900 km circumference would provide a C0M energy of 500 TeV. The 100 TeV collider would have a circumference of 300 km with the same dipole field. The collider detectors would be housed in a bathysphere approximately the size of the CMS detector at the LHC. The beam dynamics is dominated by synchrotron damping that sustains the luminosity for > 10 hours. Another novel aspect of the concept is to use non-insulated (NI) Cu-clad REBCO tapes operating at around 25 K using either He vapor or liquid hydrogen. There are several challenges associated with this concept. The more obvious is installation of the ring and detectors and maintaining alignment against fluctuations due to ocean currents. One particularly significant hurdle would be the estimated 200 GW power requirement. 

\subsubsection{Hadron-lepton colliders}

LHeC and FCC-eh [5] consist in the addition of multi-pass energy-recovery linacs (ERL) to bring intense, high-energy electron beams in collision with the protons of the main LHC and FCC-hh rings \cite{LHeC}. LHeC aims for an integrated luminosity of (1) inv-ab at a TeV center-of-mass energy, achieved by colliding 50 GeV electrons with the 7 TeV proton beam of the LHC. Two superconducting linacs of about 900 m each, placed opposite to each other, accelerate electrons by 8.1 GeV per pass. Six recirculating arcs complete the three-turn racetrack configuration of the machine, with perimeter an integral fraction (1/5) of that of the LHC. For FCC-eh, a similar ERL would accelerate electrons to 60 GeV before colliding them with the 50 TeV proton beam of FCC-hh, yielding 3.5 TeV centre-of-mass energy. Each linac is composed of 112 four-cavity cryomodules operating c.w. at 802 MHz to apply an accelerating gradient of 19.73 MV/m onto the 20 mA electron beam (500 pC charge at 40 MHz bunch frequency). The electrical power consumption has been constrained to 100 MW.

The main technical challenges of LHeC and FCC-eh are the high-power ERLs and the emittance preservation of the high-brightness electron beams in the recirculating arcs. The ERL development facility PERLE, to be built at Orsay (France), will address these issues by accelerating similar (500 pC at 40 MHz) electron beams at 500 MeV in three passes through two cryo-modules.

Little information exists on the CEPC+SPPC hadron-lepton collider. Different than the projects above, the co-existence of both CEPC and SPPC in their large common tunnel does not require an additional electron-accelerating complex: the 120 GeV electron beam of CEPC will collide with the 62.5 TeV proton beam of SPPC to yield a centre-of-mass energy of 5.48 TeV. A main challenge will be the electrical power consumption of the two large rings and their injectors operating simultaneously: the proponents of the project have quoted 400 MW.

\subsection{Advanced wakefield acceleration based $e^+e^-$ colliders}
\label{sec:aac}

Advanced collider designs are based on high peak-gradient ($>1$~GV/m) acceleration in wakefields. The wakefields are excited by drivers, that can either be particle bunches or laser pulses and are sustained by either plasmas or dielectric-based structures. Particle acceleration in wakefields allows to greatly increase the peak and average accelerating gradient compared to conventional metallic cavities that are powered by radio-frequency waves. Collider designs can therefore be more compact than conventional linear collider designs. 

There are several wakefied collider designs, depending on the particle collision energies (1-15 TeV), the particle species (electron-positron or gamma-gamma), the wakefield drivers (lasers or electron beams) and the medium sustaining the wakefields (plasma or dielectric-based structures).

Plasma is an ionized medium allowing to overcome the electrical breakdown limiting the accelerating gradient of RF cavities. Plasmas have demonstrated accelerating gradients in excess of 50 GeV/m and plasma based colliders are designed with average gradients exceeding 1 GeV/m. Structure wakefields accelerators make use of the short lifetime of the fields in the structures which allows them to increase the average accelerating gradient to >250 MV/m without electrical breakdown.
Due to the nature of the wakefield acceleration process, particle bunches that are accelerated in wakefields are short ($\ll$ pico-seconds) allowing to potentially suppress beamstrahlung at the collider interaction point.

The Beam Delivery System (BDS) will be a crucial part of future linear colliders with extremely high center-of-mass energy $\mathcal{O}(10~\mbox{TeV})$. For these high energy collisions, beamstrahlung will have a significant effect on the luminosity spectrum and also lead to large backgrounds. While this may be considered a drawback for performing  precision physics normally associated with linear colliders, an ultra-high energy linear collider can still operate as a discovery machine with a reach that is competitive with FCC-hh and SPPS. By converting the normally flat-beam BDS system to a round-beam delivery system, the luminosity-per-power figure of merit increases substantially. This can be observed as the upward inflection of the Advanced Linear Collider curves in Figure~\ref{fig:lumi_pow}. 

There are unresolved questions regarding positron acceleration in plasma. A $\gamma\gamma$ collider would avoid positron beams and still achieve high luminosities with round beam focusing.
\\
\\
Advanced collider proposals based on the following concepts have been submitted: 

\subsubsection{Laser-driven plasma wakefield $e^+e^-$ colliders for 1, 3, 15 TeV c.o.m. energies }
Laser driven plasma wakefield colliders can be all optical (no drive beams or conventional accelerators required) and consist of a sequence of approximately meter-length 5-10 GeV laser plasma wakefield acceleration stages \cite{SCHROEDER2016113} . In each stage, a short intense laser pulse drives wakefields in a plasma. Between two stages, the depleted drive pulse is coupled out and a fresh drive pulse is coupled in compactly by plasma mirrors.

\subsubsection{Beam-driven plasma wakefield $e^+e^-$ colliders for 1, 3, 15 TeV c.o.m. energies}
Beam driven plasma wakefield colliders consist of a sequence of ~25 m long 25 GeV plasma acceleration modules that include a 3.3 m long electron beam driven plasma section, beam injection and extraction system as well as beam transfer optics \cite{PWFA} . The drive beams are generated by a CW SRF recirculating linac.

\subsubsection{Beam-driven structure wakefield $e^+e^-$ colliders for 1, 3, 15 TeV c.o.m. energies}
In structure wakefield colliders, witness bunches are accelerated in dielectric-based structures that are driven by electron beams. Proposed collider schemes consist of a sequence of 150 GeV modules that share one drive beam \cite{Jing:2013uxa}. Each module is made up of 50 discrete 3 GeV (15m long) sections sharing one drive bunch. Every 3 GeV module contains 38 0.3 m-long two beam accelerator pairs based on dielectric structures.

\subsubsection{R\&D requirements}
Current wakefield collider proposals are strawperson designs based on modeling and provide estimates of collider performance.  Integrated design studies are needed and are the next step toward advancing these proposals.  All concepts have demonstrated proof-of-principle single stage high-gradient particle acceleration of electrons. R\&D efforts are required to demonstrate key components of the collider designs, such as e.g:  high-quality positron acceleration in plasma, high-efficiency staging of multiple plasma sections, and design of a beam delivery system. Laser-driven wakefield accelerators will also require extensive R\&D to develop high-average power laser systems at high efficiency. Dielectric-based structure accelerators will require R\&D efforts for the development of short pulse high gradient structures, wakefield damping and high charge beam transportation.   A recent analysis of the state-of-the-art for the plasma collider concepts can be found in Ref. \cite{CERNLDG:2022}.

\begin{table}[!ht]
\small
    \centering
    \begin{adjustbox}{angle=90}
    \begin{tabular}{|l|l|l|l|l|l|l|l|l|l|}
    \hline
        Technology & PWFA & PWFA & PWFA & SWFA & SWFA & SWFA & LWFA & LWFA & LWFA \\ \hline
        Aspect Ratio & Flat & Flat & Round & Flat & Flat & Round & Flat & Flat & Round \\ 
        CM Energy & 1 & 3 & 15 & 1 & 3 & 15 & 1 & 3 & 15 \\ 
        Single beam energy (TeV) & 0.5 & 1.5 & 7.5 & 0.5 & 1.5 & 7.5 & 0.5 & 1.5 & 7.5 \\ 
        Gamma & 9.78E+05 & 2.94E+06 & 1.47E+07 & 9.78E+05 & 2.94E+06 & 1.47E+07 & 9.78E+05 & 2.94E+06 & 1.47E+07 \\ 
        Emittance X (mm mrad) & 0.66 & 0.66 & 0.1 & 0.66 & 0.66 & 0.1 & 0.1 & 0.02 & 0.1 \\ 
        Emittance Y (mm mrad) & 0.02 & 0.02 & 0.1 & 0.02 & 0.02 & 0.1 & 0.01 & 0.007 & 0.1 \\ 
        Beta* X (m) & 5.00E-03 & 5.00E-03 & 1.50E-04 & 5.00E-03 & 5.00E-03 & 1.50E-04 & 2.50E-02 & 1.40E-02 & 1.50E-04 \\ 
        Beta* Y (m) & 1.00E-04 & 1.00E-04 & 1.50E-04 & 1.00E-04 & 1.00E-04 & 1.50E-04 & 1.00E-04 & 1.00E-04 & 1.50E-04 \\ 
        Sigma* X (nm) & 58.07 & 33.53 & 1.01 & 58.07 & 33.53 & 1.01 & 50.55 & 9.77 & 1.01 \\ 
        Sigma* Y (nm) & 1.43 & 0.83 & 1.01 & 1.43 & 0.83 & 1.01 & 1.01 & 0.49 & 1.01 \\ 
        N\_bunch (num) & 5.00E+09 & 5.00E+09 & 5.00E+09 & 3.13E+09 & 3.13E+09 & 3.13E+09 & 1.20E+09 & 1.20E+09 & 7.50E+09 \\ 
        Freq (Hz) & 4200 & 14000 & 7725 & 11000 & 36000 & 19800 & 46856 & 46856 & 3435 \\ 
        Sigma Z (um) & 5 & 5 & 5 & 40 & 40 & 40 & 8.4 & 8.4 & 2.2 \\ 
        Beamstrahlung parameter & 15 & 78 & 6590 & 1 & 6 & 515 & 2 & 37 & 22466 \\ 
        $n_{\gamma}$ & 1.5 & 1.5 & 5.7 & 2.2 & 2.2 & 8.4 & 0.8 & 1.5 & 5.7 \\ \hline
        Single Beam Power (MW) & 1.7 & 16.8 & 46.4 & 2.8 & 27.0 & 74.4 & 4.5 & 13.5 & 31.0 \\ 
        Two Beam Power (MW) & 3.4 & 33.6 & 92.8 & 5.5 & 54.1 & 148.7 & 9.0 & 27.0 & 61.9 \\ \hline
        Geometric Lumi (cm\^-2 s\^-1) & 1.01E+34 & 1.01E+35 & 1.50E+36 & 1.03E+34 & 1.01E+35 & 1.51E+36 & 1.05E+34 & 1.13E+35 & 1.50E+36 \\ 
        Beamstrahlung lumi & 1.99E+34 & 1.99E+35 & 1.52E+36 & 2.03E+34 & 2.00E+35 & 1.52E+36 & 2.09E+34 & 2.17E+35 & 1.52E+36 \\ 
        Wall plug to drive laser/beam eff & 0.4 & 0.4 & 0.4 & 0.774 & 0.774 & 0.774 & 0.4 & 0.4 & 0.5 \\ 
        Laser/beam drive to main eff & 0.375 & 0.375 & 0.375 & 0.42 & 0.42 & 0.42 & 0.2 & 0.2 & 0.12 \\ 
        Wall plug to main beam eff & 0.15 & 0.15 & 0.15 & 0.32508 & 0.32508 & 0.32508 & 0.08 & 0.08 & 0.06 \\ 
        Site power Wall to main only (MW) & 22 & 224 & 619 & 17 & 166 & 457 & 113 & 338 & 1032 \\ 
        Lumi/Power (1e34/MW) & 0.04 & 0.04 & 0.08 & 0.06 & 0.06 & 0.11 & 0.01 & 0.03 & 0.05 \\ \hline
        GUINEA-PIG Total Lumi & 1.83E+34 & 1.85E+35 & 4.2E+37 & 2.08E+34 & 2.13E+35 & 4.2E+36 & 1.53E+34 & 2.58E+35 & 6E+36 \\ 
        GUINEA-PIG Lumi 1\% (20\%) & 6.86E+33 & 6.23E+34 & 5E+35 & 8.49E+33 & 6.14E+34 & 5E+35 & 1.03E+34 & 8.72E+34 & 5E+35 \\ 
        GP Total Lumi/Power & 0.08 & 0.08 & 6.79 & 0.12 & 0.13 & 0.92 & 0.01 & 0.08 & 0.58 \\ 
        GP Lumi 1\%/Power (20\%) & 0.03 & 0.03 & 0.08 & 0.05 & 0.04 & 0.11 & 0.01 & 0.03 & 0.05 \\ \hline
        Length of 2 Linacs (km) & 1 & 3 & 14 & 5 & 15 & 75 & 0.44 & 1.3 & 6.5 \\
        Length of Facility & 14 & 14 & 14 & 8 & 18 & 90 & 3.5 & 4.5 & 9.5 \\ \hline
    \end{tabular}
    \end{adjustbox}
\caption{Parameters of the advanced WFA-based colliders.}
\end{table}

\clearpage
\newpage
\subsection{TRL (technical readiness level) definitions}
\label{TRLAppendix}

\begin{table}[h!]
\small
\begin{tabular}{|p{0.3cm}|p{8cm}|p{8cm}|}
\hline 
\hline
1 & Basic principles observed and reported.
& Lowest level of technology readiness. Scientific research begins to be translated into applied research and development. Examples might include paper studies of a technology’s basic properties.\\
2 & Technology concept and/or application formulated.
& Invention begins. Once basic principles are observed, practical applications can be invented. Applications are speculative and there may be no proof or detailed analysis to support the assumptions. Examples are limited to analytic studies.\\
3 & Analytical and experimental critical function and/or characteristic proof of concept.
& Active research and development is initiated. This includes analytical studies and laboratory studies to physically validate analytical predictions of separate elements of the technology. Examples include components that are not yet integrated or representative.\\
4 & Component and/or breadboard validation in laboratory environment.
& Basic technological components are integrated to establish that they will work together. This is relatively “low fidelity” compared to the eventual system. Examples include the integration of “ad hoc” hardware in the laboratory.\\
5 & Component and/or breadboard validation in relevant environment.
& The Fidelity of breadboard technology increases significantly. The basic technological components are integrated with reasonably realistic supporting elements so it can be tested in a simulated environment.\\
6 & System/subsystem model or prototype demonstration in a relevant environment. 
& A representative model or prototype system, which is well beyond that of TRL 5, is tested in a relevant environment. Represents a major step up in a technology’s demonstrated readiness.\\
7 & System prototype demonstration in an operational environment.
& Prototype near, or at, planned operational system. Represents a major step up from TRL 6, requiring the demonstration of an actual system prototype in an operational environment such as an aircraft, vehicle, or space.\\
8 & Actual system completed and qualified through test and demonstration.
& Technology has been proven to work in its final form and under expected conditions. In almost all cases, this TRL represents the end of true system development. Examples include developmental test and evaluations of the system in its intended weapon system to determine if it meets design specifications.\\
9 & Actual system has proven through successful mission operations.
& The actual application of the technology in its final form and under mission conditions, such as those encountered in operational test and evaluation. Examples include using the system under operational mission.\\
\hline \hline
\end{tabular}
\caption{ Technical Readiness Level (TRL) values used in the risk evaluation -- see, e.g., \cite{heder2017nasa} and references therein.}
\end{table}